\documentclass[final]{svjour3}            
\usepackage{amsmath}
\usepackage{amssymb}
\smartqed  
\usepackage{graphicx}
\usepackage{latexsym}
\journalname{Journal of Economic Interaction and Coordination}
\begin{document}
\title{Response of double-auction markets to instantaneous Selling-Buying signals with stochastic Bid-Ask spread}
\titlerunning{Response of double-auction markets}

\author{Takero Ibuki  \and Jun-ichi Inoue}

\authorrunning{T. Ibuki and J. Inoue} 

\institute{Takero Ibuki and Jun-ichi Inoue \at
              Graduate School of 
              Information Science and Technology, Hokkaido University, 
              N14-W9, Kita-ku, Sapporo 064-0814, Japan \\
              Tel.: +81-11-7067225\\
              Fax: +81-11-7067391\\
   \email{\{ibuki, j$\underline{\,\,\,}$inoue\}@complex.ist.hokudai.ac.jp}
          }

\date{Received: date / Accepted: date}
\maketitle
\begin{abstract}
Statistical properties of order-driven double-auction markets with Bid-Ask spread are investigated 
through the dynamical quantities such as 
response function. We first attempt to utilize the so-called {\it Madhavan-Richardson-Roomans 
model} (MRR for short) to simulate the stochastic process of 
the price-change in empirical data sets  (say, 
EUR/JPY or USD/JPY exchange rates) 
in which the Bid-Ask spread fluctuates in time.
We find that the MRR theory apparently fails to 
simulate so much as the qualitative behaviour  (`non-monotonic' behaviour) of the response function 
$R(l)$ ($l$ denotes the difference of times at which the response function is evaluated) 
calculated from the data.  
Especially, we confirm that 
the stochastic nature of the Bid-Ask spread causes 
apparent deviations from a linear relationship between the $R(l)$ and the 
auto-correlation function $C(l)$, namely, 
$R(l) \propto -C(l)$. 
To make the microscopic model of double-auction markets having stochastic Bid-Ask spread, 
we use the minority game with a finite market history length and 
find numerically that appropriate extension of the game shows quite similar 
behaviour of the response function to the empirical evidence. 
We also reveal that the minority game modeling with the 
 adaptive (`annealed') look-up table reproduces the 
 non-linear relationship $R(l) \propto -f(C(l))$ ($f(x)$ stands for a non-linear function 
 leading to `$\lambda$-shapes') 
more effectively than the fixed (`quenched') look-up table does. 
\end{abstract}
\keywords{
Double-auction \and Bid-Ask spread \and 
Response function \and Minority game \and Stochastic process \and Non-equilibrium phenomena 
\and Agent-based simulations \and Econophysics
}
\PACS{89.65.Gh \and 02.50.-r \and 05.40.-a \and 05.10.Gg \and 02.50.Cw}
\section{Introduction}
\label{sec:Intro}
How a specific trading mechanism 
effects on the price formation is one of the essential queries to understand 
the process and outcomes of exchanging assets under 
a given concrete rule. 
To investigate the issue, 
a lot of studies concerning the micro-structure of markets have been done 
in various research fields \cite{Ohara}.

Recently, lots of on-line trading services on 
the internet were constructed by several major banks such as the Sony Bank \cite{Sony}.  
As the result, one can gather a lot of trading data sets 
to investigate the statistical properties extensively. 
As such studies, several authors focused on the fact that the Sony Bank uses a trading system in which 
foreign currency exchange rates change according to 
a first-passage process (FPP) \cite{Redner,Kappen,Montero}. 
Automatic FOREX trading systems such as the Sony Bank 
are now popular in Japan where many investors
use a scheme called {\em carry trade} by 
borrowing money in a currency 
with low interest rate and lending 
it in a currency offering higher interest rates. 
With these demands in mind, 
several studies have been done 
to investigate the stochastic process 
and made a model of it to reproduce the FPP in order to 
provide useful information for customers 
\cite{Sazuka2007a,Sazuka2007,Inoue2007,SazukaInoue2007,SIS,Inoue2008,Inoue2010,IHSS}. 

The data sets of the Sony Bank rate \cite{Sony} are composed of time index and trading rate at that time. 
As we explained, a huge number of market data  are reduced to a small amount of it, namely, the number of 
the Sony bank rate is reduced by the first-passage process 
and unfortunately, the market rates behind the Sony Bank rate are not available for us. 

As well-known, there are several data sets whose price are determined by 
the so-called {\it double-auction} system. 
In the double-auction market, 
each trader (investor) posts his (or her) selling price or buying price in market order  
of a specific commodity with its volume to the market.
Then, the market maker determines the minimum price of  buying orders, 
what we call {\it Ask}, and the maximum price of selling orders, the so-called {\it Bid} at each 
trading time and discloses these prices to the public. 
Then, the difference between the Bid and the Ask is referred to as {\it spread} or {\it Bid-Ask spread}.
In market rates available for traders (on the web for instance), 
there are two types of Bid-Ask spread, that is, `constant' or `distributed', 
and which type of spread is disclosed depends on the market makers (securities companies).

Results of market making, especially, statistical properties of the Bid-Ask spread might have an impact on the market and 
several studies have been done to reveal the relationship between the properties of 
Bid-Ask spread and behaviour of the market \cite{Ohara,Elliott,MRR,Bouchaud1,Bouchaud2,Ponzi}. 
For instance, Madhavan, Richardson and Roomans \cite{MRR}
proposed a phenomenological model to 
explain the price dynamics of double-auction market in market order, 
however, their model is apparently limited to the case in 
which the Bid-Ask spread remains constant during the price dynamics. 
Therefore, much more extensive studies including empirical data analysis 
seem to be needed to investigate to what extent the model proposed by  Madhavan, Richardson and Roomans 
can explain the behaviour of market with {\it stochastic} Bid-Ask spread through some relevant quantity.   

In this paper, we investigate statistical properties of double-auction markets 
with Bid-Ask spread through the dynamical quantities such as response function. 
We first attempt to examine  the so-called {\it Madhavan-Richardson-Roomans 
model} (MRR for short) to simulate the stochastic process of 
the price-change in empirical data sets  (say, 
EUR/JPY or USD/JPY exchange rates) 
in which the Bid-Ask spread fluctuates in time.
We find that the MRR theory apparently does not  
simulate so much as the qualitative behaviour  (`non-monotonic' behaviour) 
of the response function calculated from the data sets.  
It is possible for us to show that 
a linear relationship $R(l) \propto -C(l)$ 
between auto-correlation function $C(l)$ and response function $R(l)$ 
holds for the MRR model. 
Namely, these two macroscopic quantities 
are related each other and the relationship should 
be explained from the microscopic view point as 
statistical physics provides the microscopic foundation of 
thermodynamics. 
Moreover, we find that 
the linear relationship $R(l) \propto -C(l)$ 
is apparently broken down in order-driven 
double-auction markets with fluctuating Bid-Ask spread. 
This fact tells us that 
on the analogy of physics, 
the phenomenological MRR model for the constant spread 
might be regarded as 
`thermodynamics'  
which usually deals with 
the macroscopic quantities such as price, 
auto-correlation,  response functions and the relationship between them. 
It does not need to consider the detail behaviour of 
microscopic ingredients 
such as traders. 
In this paper, we show that the phenomenological model is apparently limited and 
fails to reproduce the dynamical quantities $C(l),R(l)$ efficiently. 

Hence,  here we attempt to construct 
a kind of `statistical mechanics' in finance, 
which provides a microscopic foundation of phenomenological theory  such as the MRR model.  
For this end, we utilize the minority game with a finite market history length having 
the distributed Bid-Ask spread to reproduce similar 
behaviour of macroscopic dynamical quantities as the empirical evidence shows.

In our minority game modeling, we first fix each decision component 
(buying: $+1$, selling:  $-1$) in 
their look-up tables before playing the game 
(in this sense, the decision components are `quenched variables' 
in the literature of disordered spin systems such as spin glasses \cite{Mezard}).  
We also consider the case in which a certain amount of traders update their 
decision components according to the macroscopic 
market history (they `learn' from the behaviour of markets) 
so as to be categorized into two groups with a finite probability
(in this sense, the components are now regarded as `annealed variables').  
Namely, at each round of the game, if the number of sellers is smaller/greater than 
that of buyers, a fraction of  traders, what we call {\it optimistic group}/{\it pessimistic group}, 
 is more likely to rewrite their own decision components from $-1$/$+1$ to $+1$/$-1$. 
 We find that the minority game modeling with the 
 adaptive look-up table reproduces the 
 non-linear relationship $R(l) \propto -f(C(l))$ ($f(x)$ stands for a non-linear function 
 leading to `$\lambda$ -shapes') 
 more effectively than fixed (frozen) look-up table does. 

This paper is organized as follows. 
In the next section \ref{sec:data}, 
we explain our data sets and their format. We investigate their statistical properties. 
In the next section \ref{sec:data_analysis}, we evaluate 
two relevant quantities, 
namely, 
the auto-correlation function 
$C(l)$ and the response function $R(l)$, 
which are 
our key quantities to discuss the double-auction markets, for our data sets. 
We find that a linear relationship 
$R(l) \propto -C(l)$ holds 
for the data sets having a constant Bid-Ask spread, however, 
the relation is broken for 
the data with a stochastic spread. 
In section \ref{sec:MRR}, we introduce the so-called MRR model 
as a phenomenological model and 
derive the $C(l)$ and $R(l)$. 
The difference between the MRR theory and the empirical evidence, the origin of the difference is discussed. 
In section \ref{sec:MG}, we introduce and modify the minority game with 
a finite market history length and apply to explain the typical behaviour 
of the response function for the data with stochastic Bid-Ask spread. 
In section \ref{sec:adapt}, 
we reveal that the minority game modeling with the 
 adaptive (`annealed') look-up table reproduces the 
 non-linear relationship $R(l) \propto -f(C(l))$ ($f(x)$ stands for a non-linear function 
 leading to `$\lambda$-shapes') 
more effectively than fixed (`quenched') look-up table does. 
In section \ref{sec:discuss}, we comment on the possible extensions of our approach. 
The last section is summary.   
\section{Statistical properties of data sets}
\label{sec:data}
In order to check the validity of our modeling of markets, 
we gathered data sets of double-auction markets 
from the web site {\sf http://www.metaquotes.net/} \cite{Meta} 
by using the free software {\it MetaTrader4}. 
We shall explain the data format of the {\it MataTrader 4}. 
We used the script which is available on the web \cite{Meta}. 
By using the script, the data sets are stored as the 
following format: 
\mbox{}\\
\begin{center}
\begin{verbatim}
2009/12/24,17:17:40,131.053,131.092 
2009/12/24,17:17:41,131.053,131.088
2009/12/24,17:17:41,131.052,131.088
2009/12/24,17:17:43,131.048,131.071
2009/12/24,17:17:44,131.043,131.076
..................................
..................................
\end{verbatim}
\end{center}
\mbox{}\\
From the far left column 
to the far right, transaction time (Year/Month/Day, hour:min:sec), Bid, Ask 
are shown. 
For instance, the first line denotes 
the Bid is 131.053 and Ask is 131.092 on 
24th December 2009 at 17:17:40. 
In this paper, we treat the data set 
written by the above format. 
Among data sets concerning 
various different financial assets, 
we shall use here specific four data sets, namely, 
USD/JPY exchange rates (23rd-28th October 2009), 
EUR/JPY exchange rates (22nd-28th November 2009), 
Nasdaq100 (22nd-31st October 2009) and 
price of gold (28th-30th October 2009). 
Each data set contains $10^{5}$-data points. 

In the conventional (standard) data for continuous-time 
double-auction markets,  
we usually use the data having 
transactions (buying or selling price and
the transaction time) including 
the quote (Bid and Ask prices posted to the market with the time). 
However, unfortunately, 
the data set provided by the {\it MetaTrader4} 
does not contain any information about the transaction. 
Namely, the `Bid and Ask values' we mentioned above 
are the best selling price and the best buying price, 
and the time at which the transaction takes place. 
Therefore,  the price itself is not available for these data sets. 

Hence, here we assume that the mid point $m_{t}$ of 
the Bid $b_{t}$ and Ask $a_{t}$ at time $t$, that is,  
$m_{t}=(a_{t}+b_{t})/2$ 
is a sort of buying or selling 
price when the transaction is approved.  
Then, we shall define the sign of 
 the `return'  of the mid points (the difference between successive mid points) 
 as a {\it Selling-Buying signal} $\epsilon_{t}$, namely, 
$\epsilon_{t}={\rm sgn}(m_{t+1}-m_{t})$. 
Of course, these definitions of 
`prices' and the `Selling-Buying signals' are 
different from the conventional one, 
however, we shall try to investigate the behaviour of the system having such a 
slightly different definition of the prices and signals in limited data sets. 

\begin{figure}[ht]
\begin{center}
\includegraphics[width=6cm]{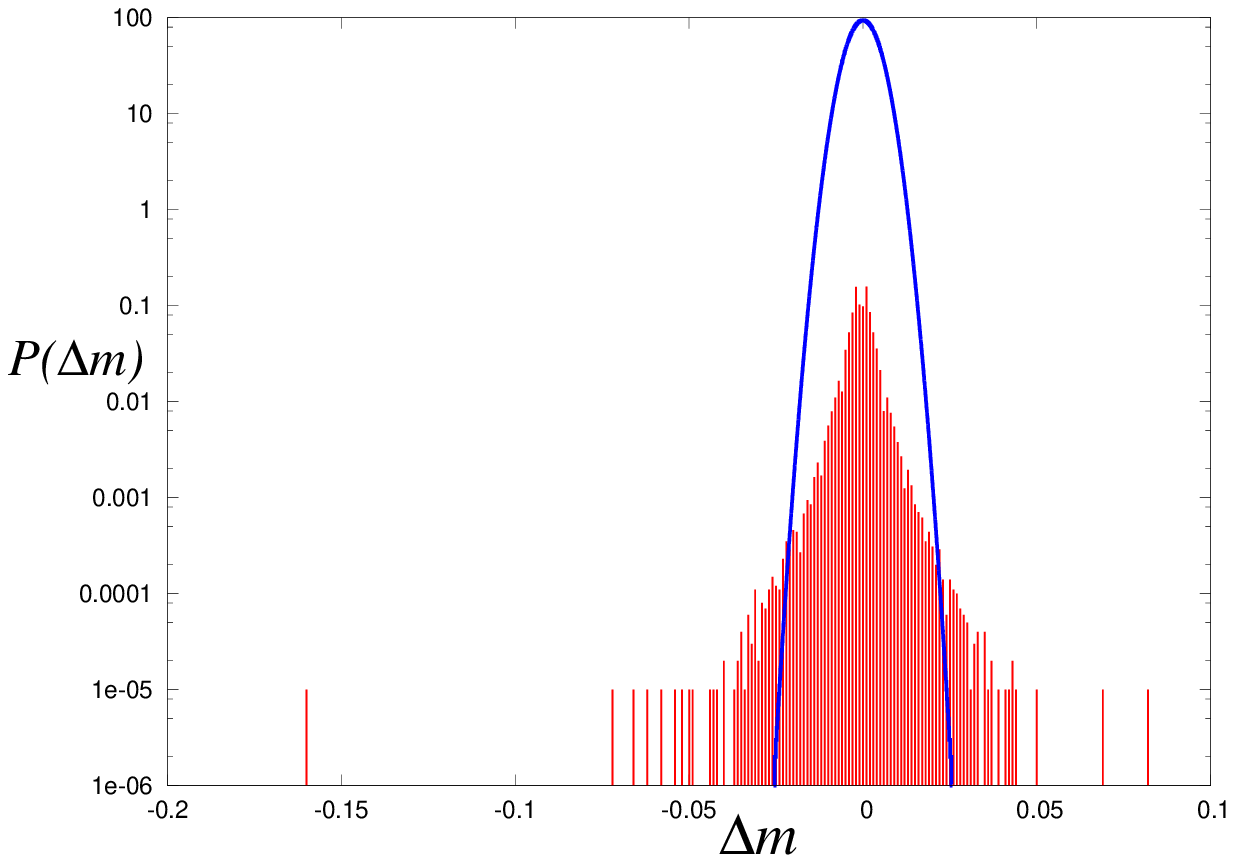}
\includegraphics[width=6cm]{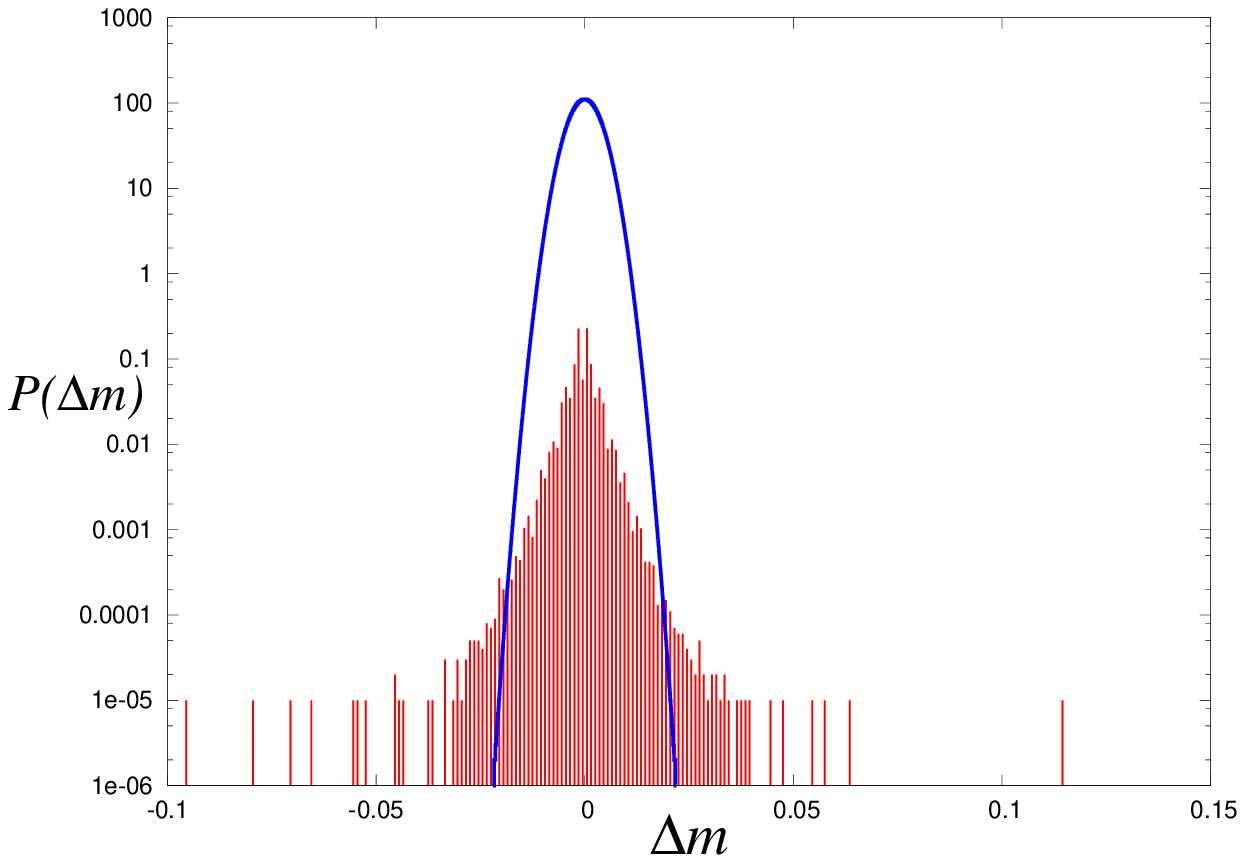} \\
\includegraphics[width=6cm]{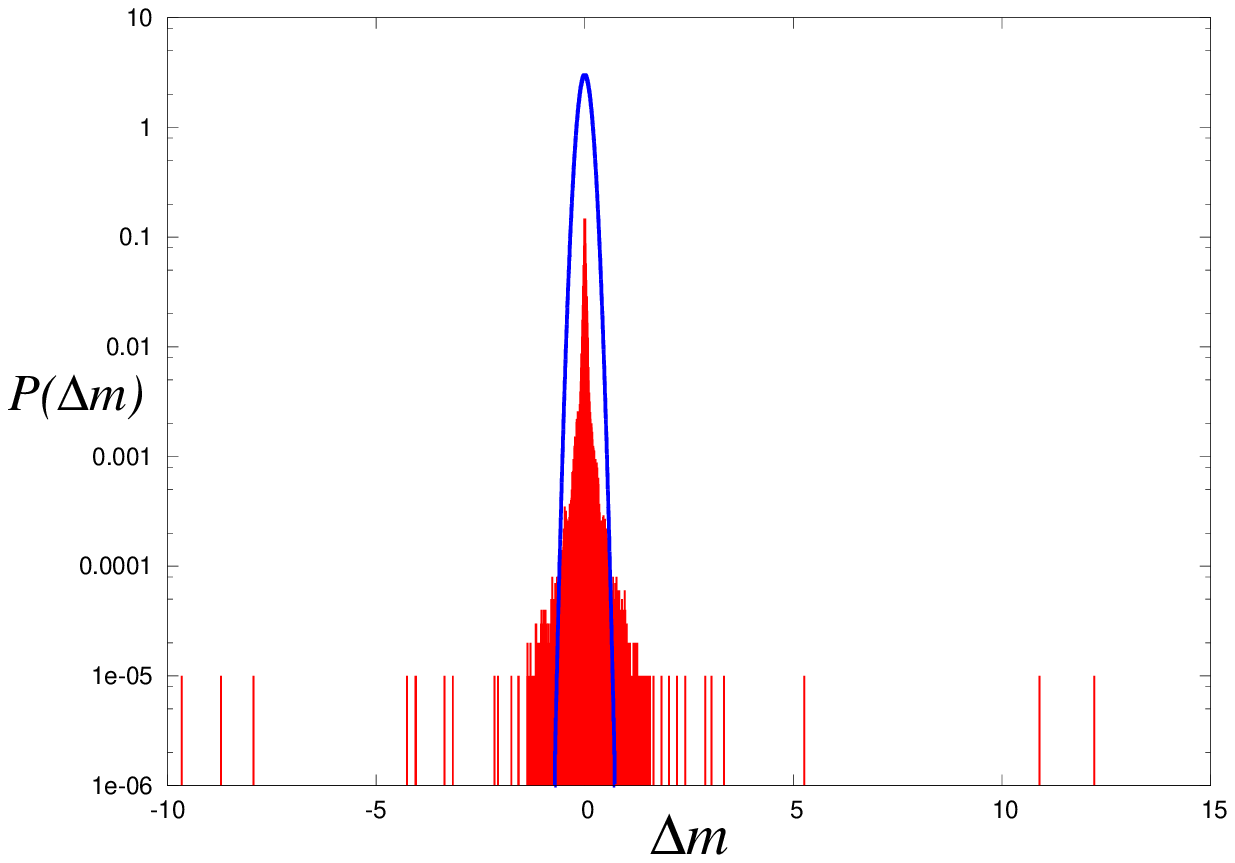}
\includegraphics[width=6cm]{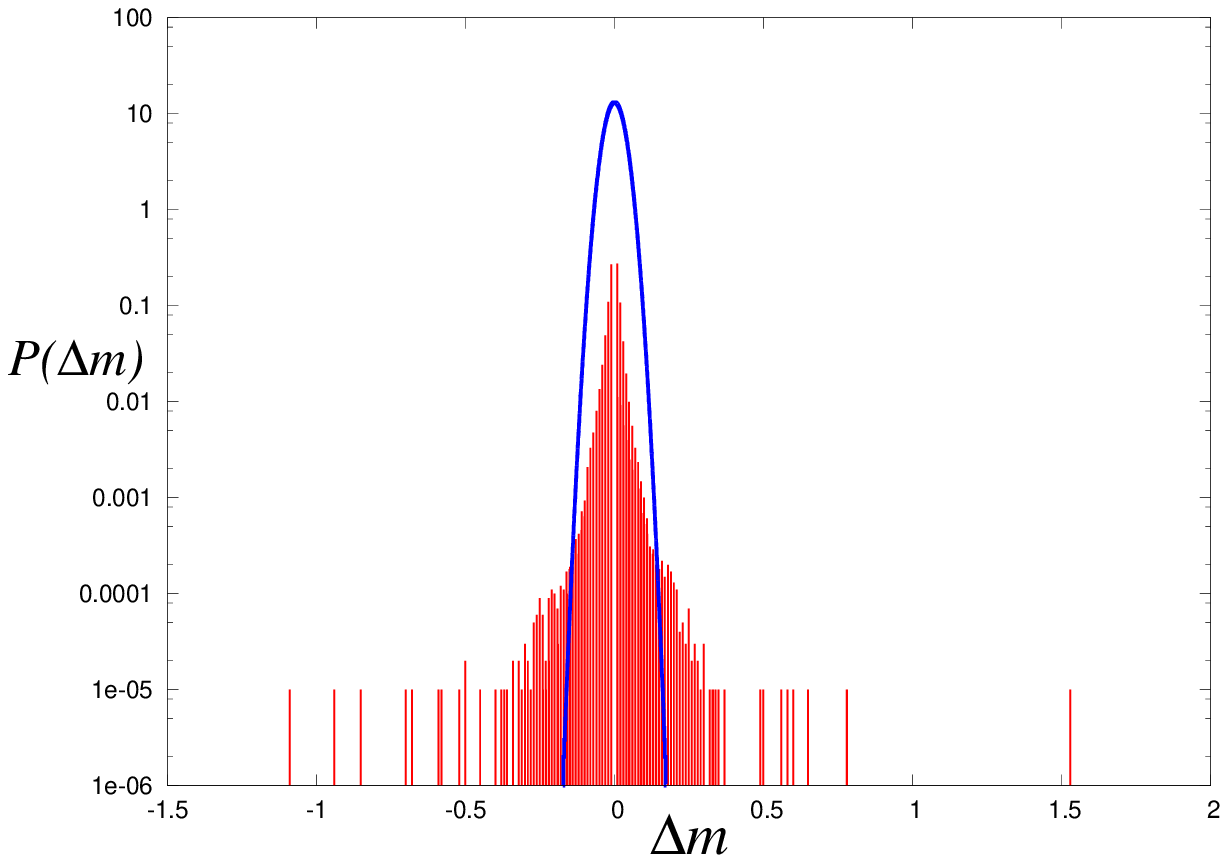}
\end{center}
\caption{\footnotesize 
Empirical distributions of `return' 
(the difference between successive mid points)
$\Delta m_{t} \equiv m_{t+1}-m_{t}$ 
for our four kinds of data sets. 
From the upper left to the lower right, 
$P(\Delta m_{t})$ for 
EUR/JPY exchange rates, 
USD/JPY exchange rates, Nasdaq100 and 
price of gold are plotted. 
 To compare the results with normal Gaussians, 
 we calculate the empirical mean 
 $ \overline{\Delta m} \equiv (1/T)\sum_{t=0}^{T-1}
 \Delta m_{t}$ and the empirical variance 
 $\overline{\sigma}_{\Delta m}^{2} \equiv (1/T) \sum_{t=0}^{T-1}
 (\Delta m_{t}-\overline{\Delta m})^{2}$ for each data and 
 plot the Gaussian ${\cal N}(\overline{\Delta m},\overline{\sigma}_{\Delta m})$ in the same panel. 
 From the upper left to the lower right, 
 these normal Gaussians are 
 ${\cal N}(0.000012, 0.000018), 
 {\cal N}(-0.000007, 0.000013), 
 {\cal N}(-0.000726, 0.017722)$ and 
 ${\cal N}(0.000080, 0.000922)$, respectively. 
 }
\label{fig:fg1}
\end{figure}

In this section, we first calculate the histogram of 
the return of the mid point 
 $\Delta m_{t} \equiv m_{t+1}-m_{t}$. The results are shown in Fig. \ref{fig:fg1}. 
 From these panels, we clearly find that the return of the mid point 
 is distributed with `heavy tails' as the conventional return of the price has \cite{Bouchaud}. 
 To compare the results with Gaussians, 
 we calculate the empirical mean 
 $ \overline{\Delta m} \equiv (1/T)\sum_{t=0}^{T-1}
 \Delta m_{t}$ and the empirical variance 
 $\overline{\sigma}_{\Delta m}^{2} \equiv (1/T) \sum_{t=0}^{T-1}
 (\Delta m_{t}-\overline{\Delta m})^{2}$ for each data and 
 plot the corresponding Gaussian ${\cal N}(\overline{\Delta m},\overline{\sigma}_{\Delta m})$ in the same panel. 
\mbox{}

We next focus on the Bid-Ask spread 
$S_{t}=a_{t}-b_{t}$ 
which is one of the key values in this study. 
We are confirmed that the above four data sets 
are classified into 
two types according to each statistical property of 
the spread. 
Namely, 
the spread of USD/JPY or EUR/JPY exchange rates 
is time-dependent and fluctuates, 
whereas, the spread of Nasdaq100 or price of gold 
is a time-independent constant. 
\begin{figure}[ht]
\begin{center}
\includegraphics[width=4cm]{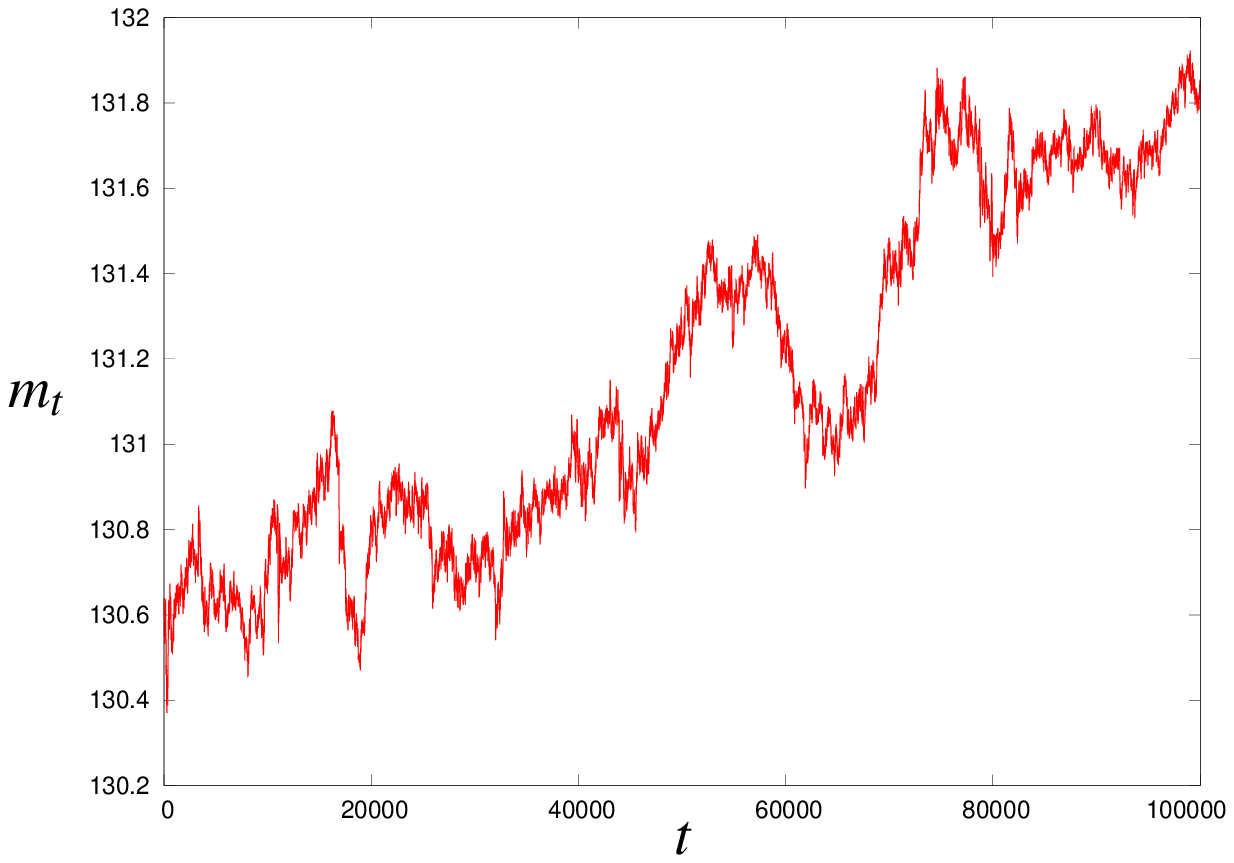}
\includegraphics[width=4cm]{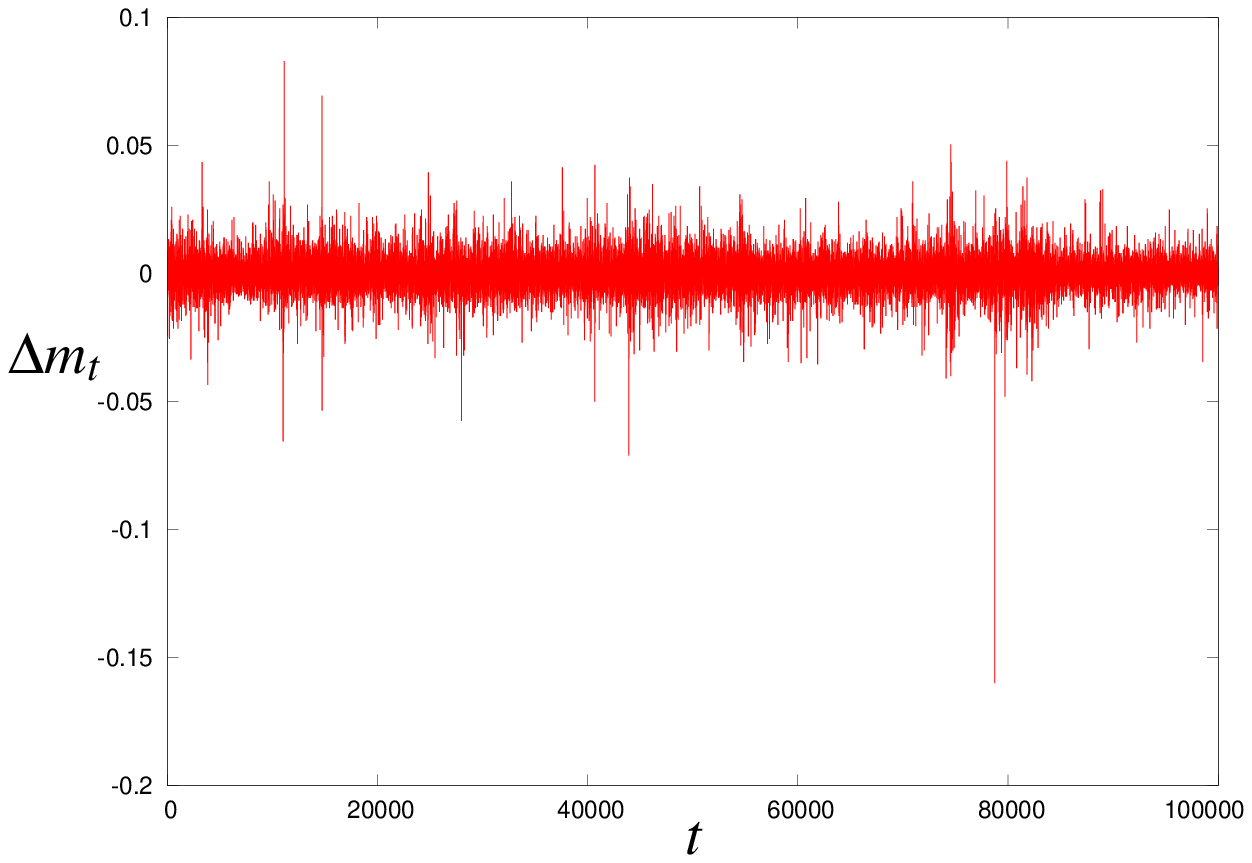}
\includegraphics[width=4cm]{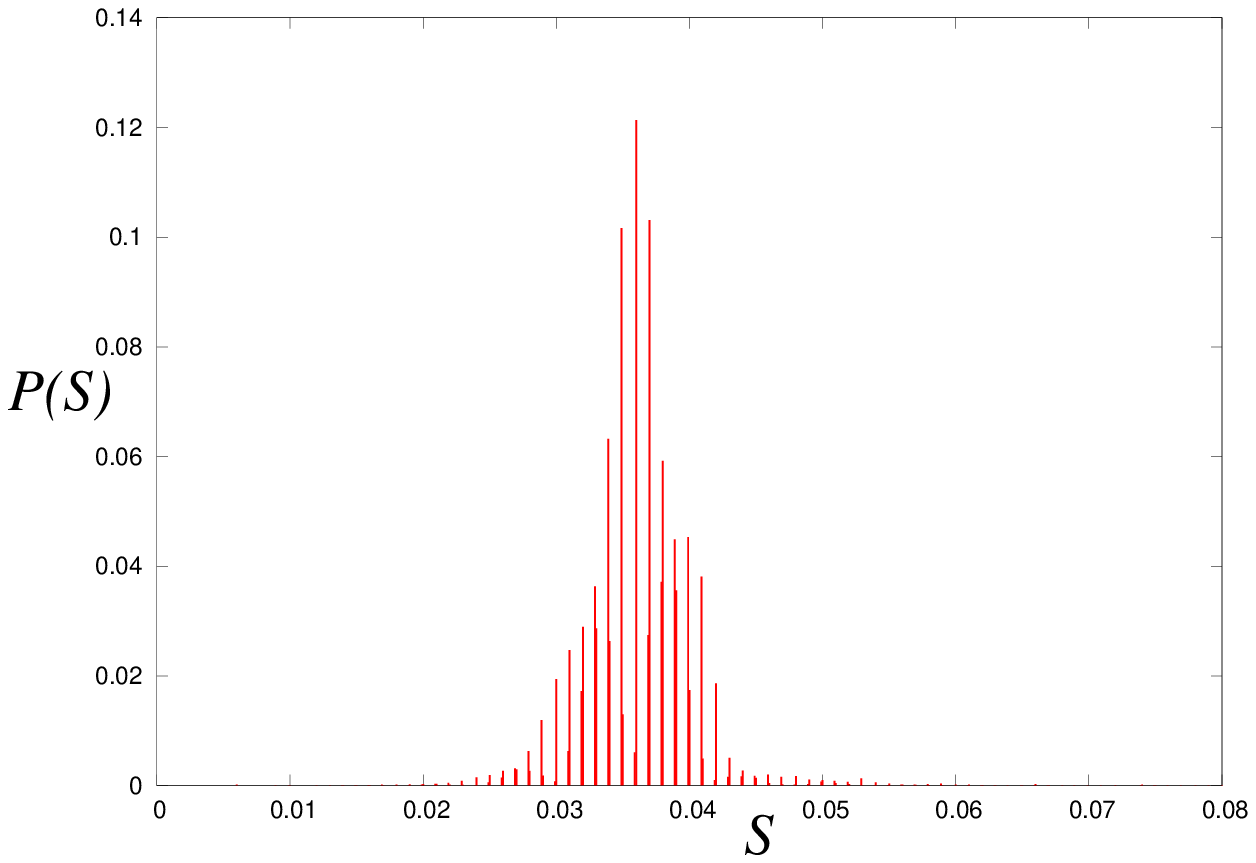} \\
\includegraphics[width=4cm]{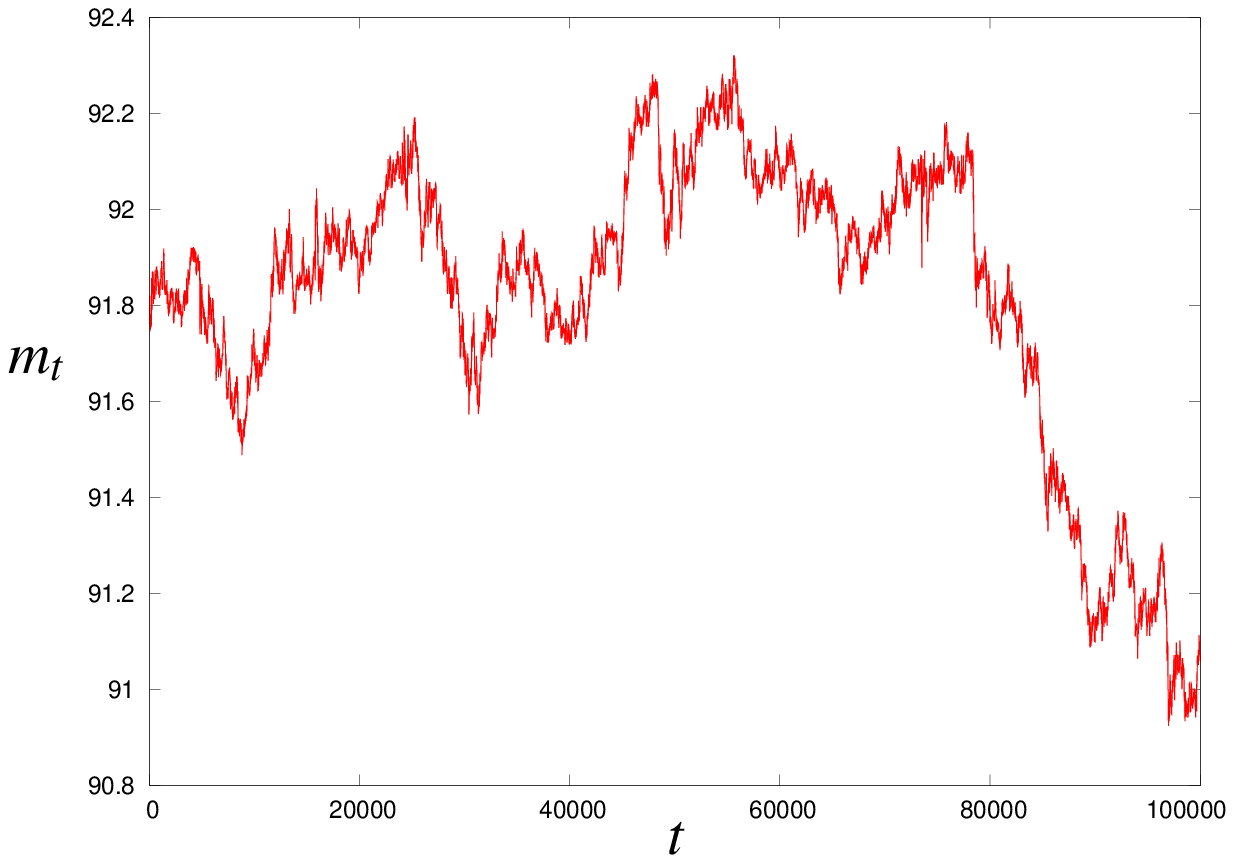}
\includegraphics[width=4cm]{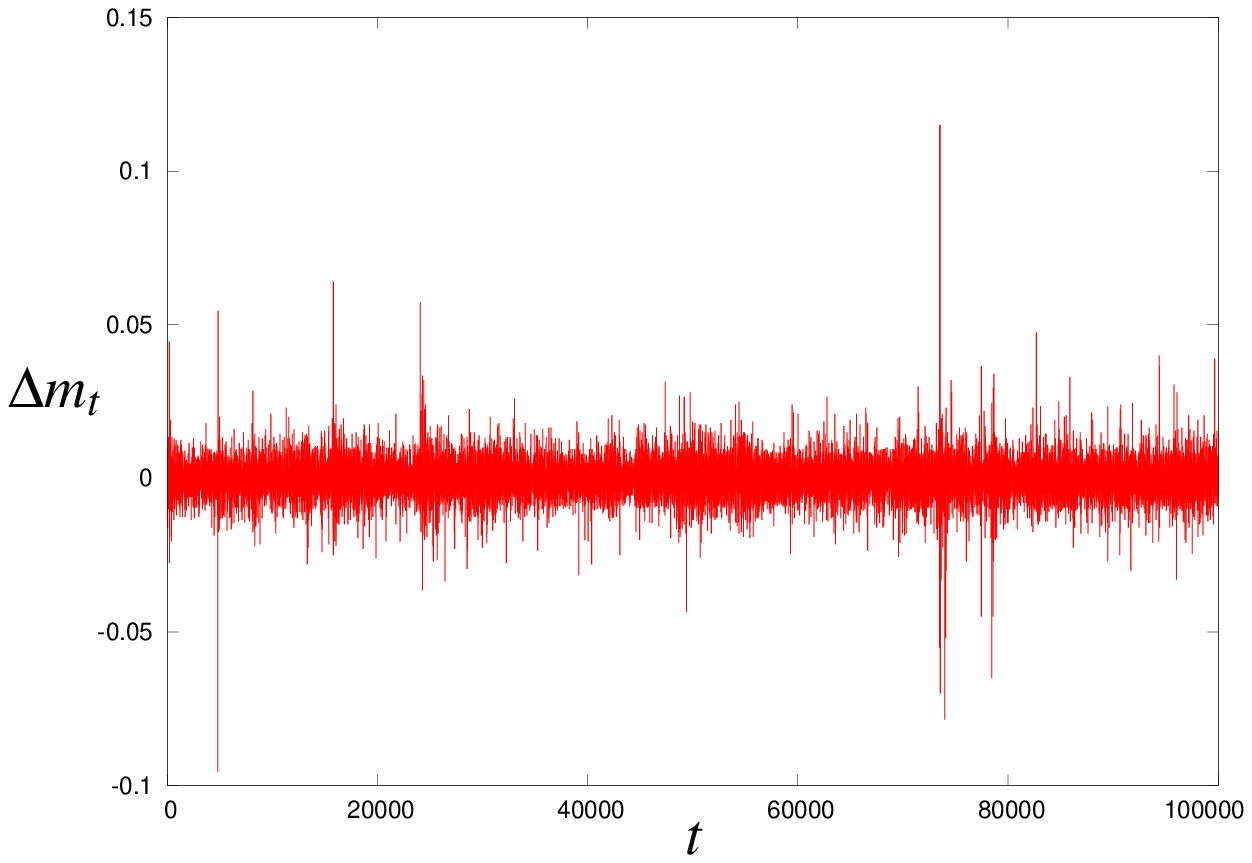}
\includegraphics[width=4cm]{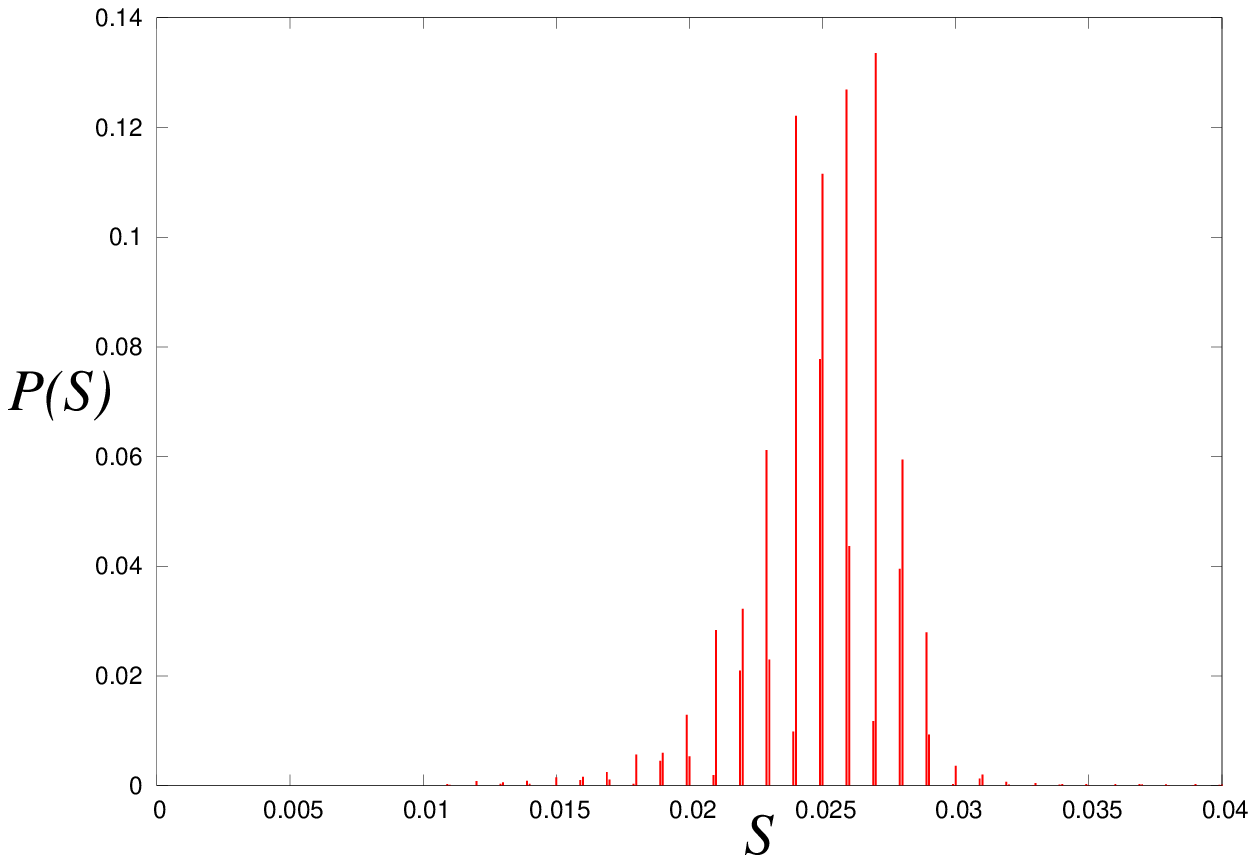} 
\caption{\footnotesize 
Statistical properties of 
the data set in which the spread fluctuates in time, 
EUR/JPY exchange rates (the upper panels) and 
USD/JPY exchange rates (the lower panels) are shown. 
From left to right, the mid point $m_{t}$, 
the return of the mid point $\Delta m_{t} \equiv m_{t+1}-m_{t}$ 
as a function of $t$, and 
the distribution of the spread $P(S)$ are plotted. }
\label{fig:fg_usd1}
\end{center}
\end{figure}
\begin{figure}[ht]
\begin{center}
\includegraphics[width=4cm]{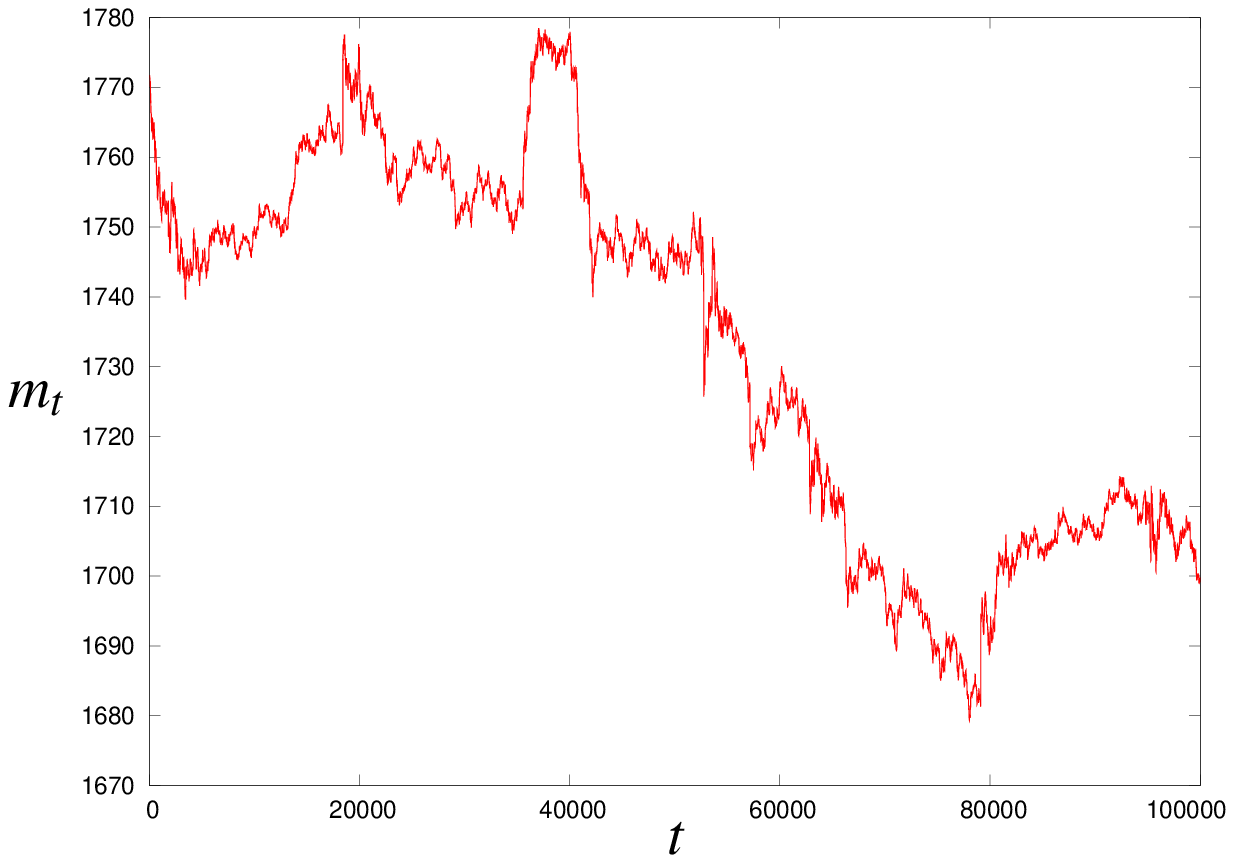}
\includegraphics[width=4cm]{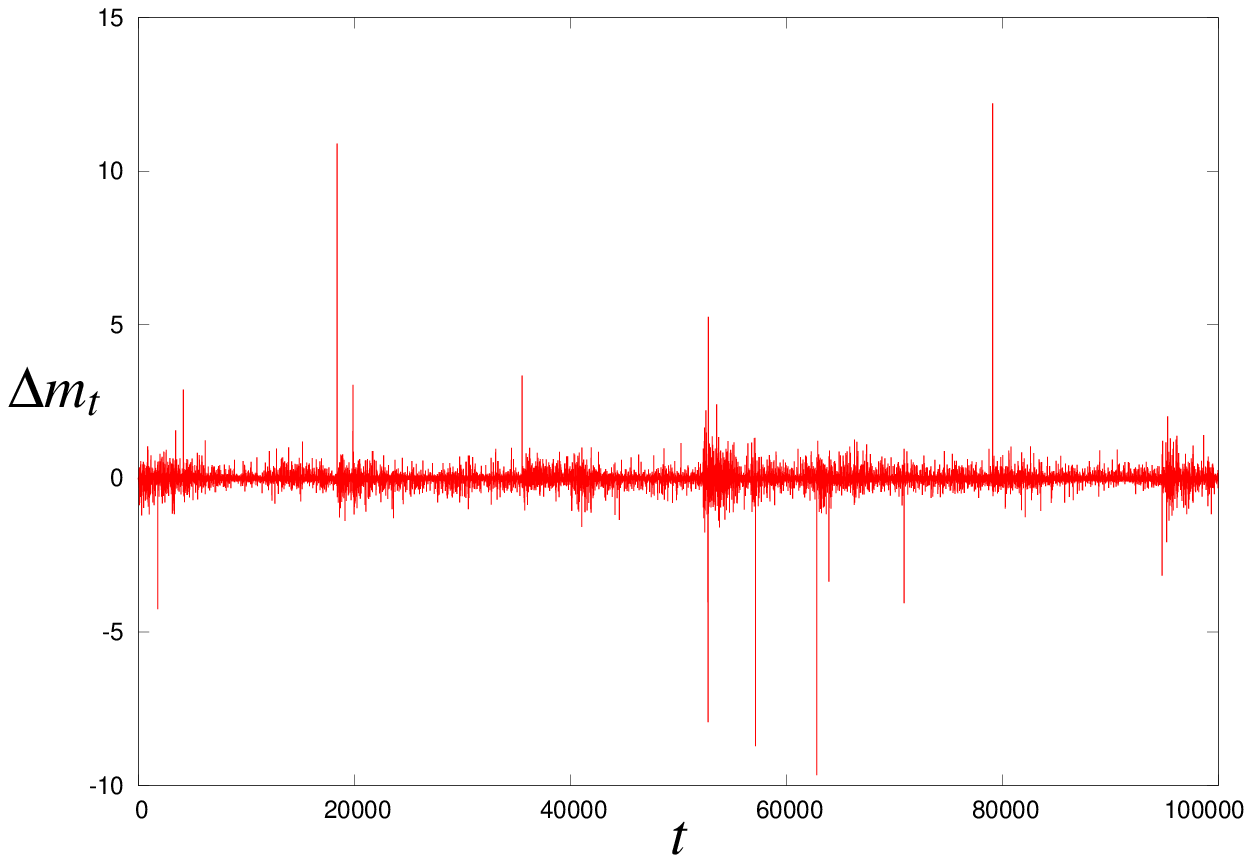}
\includegraphics[width=4cm]{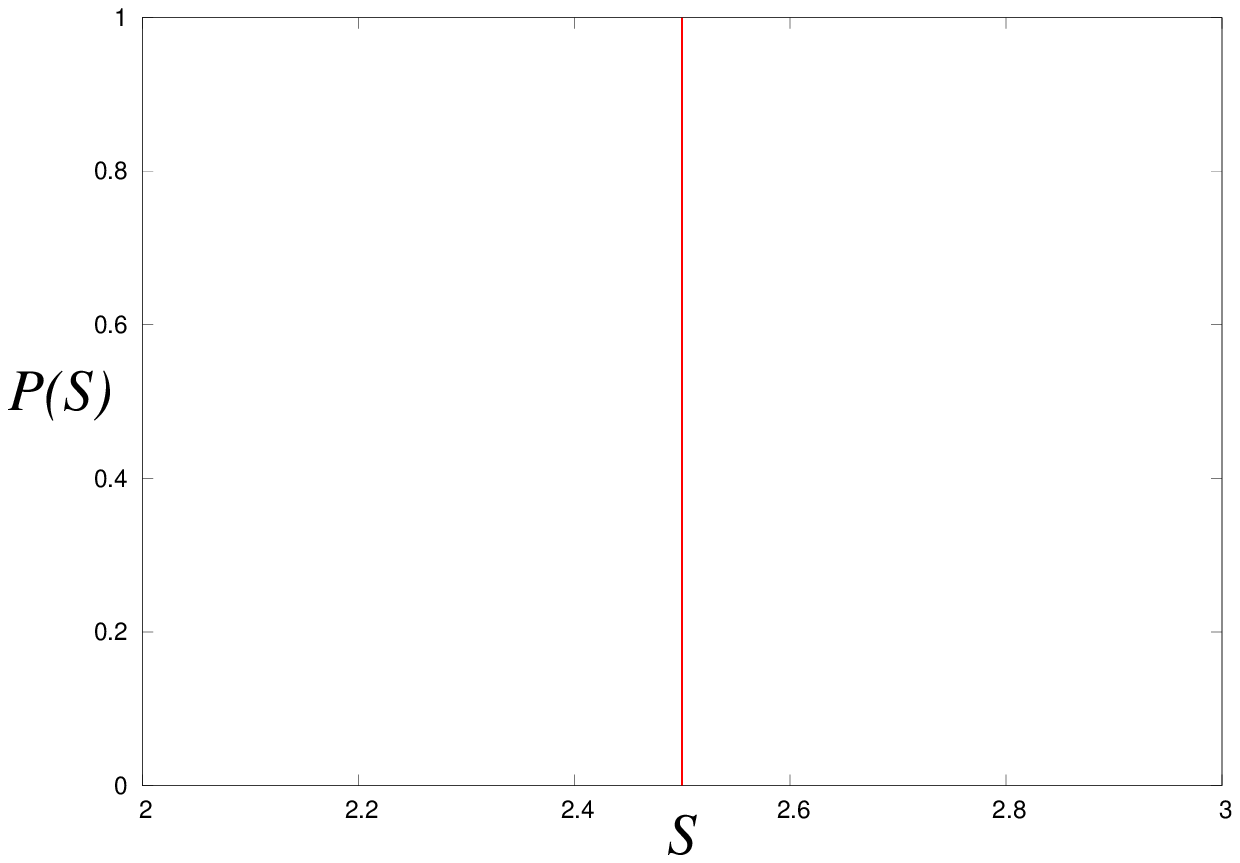} \\
\includegraphics[width=4cm]{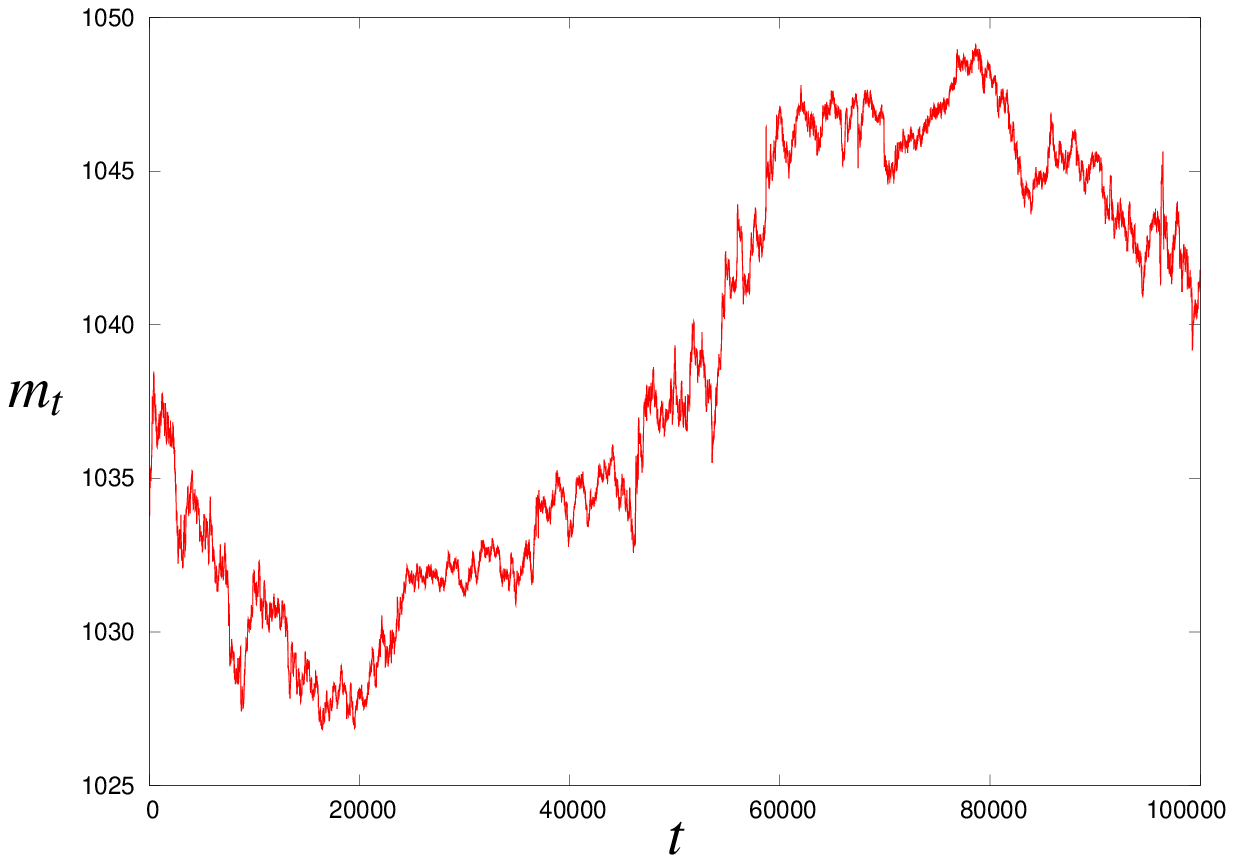}
\includegraphics[width=4cm]{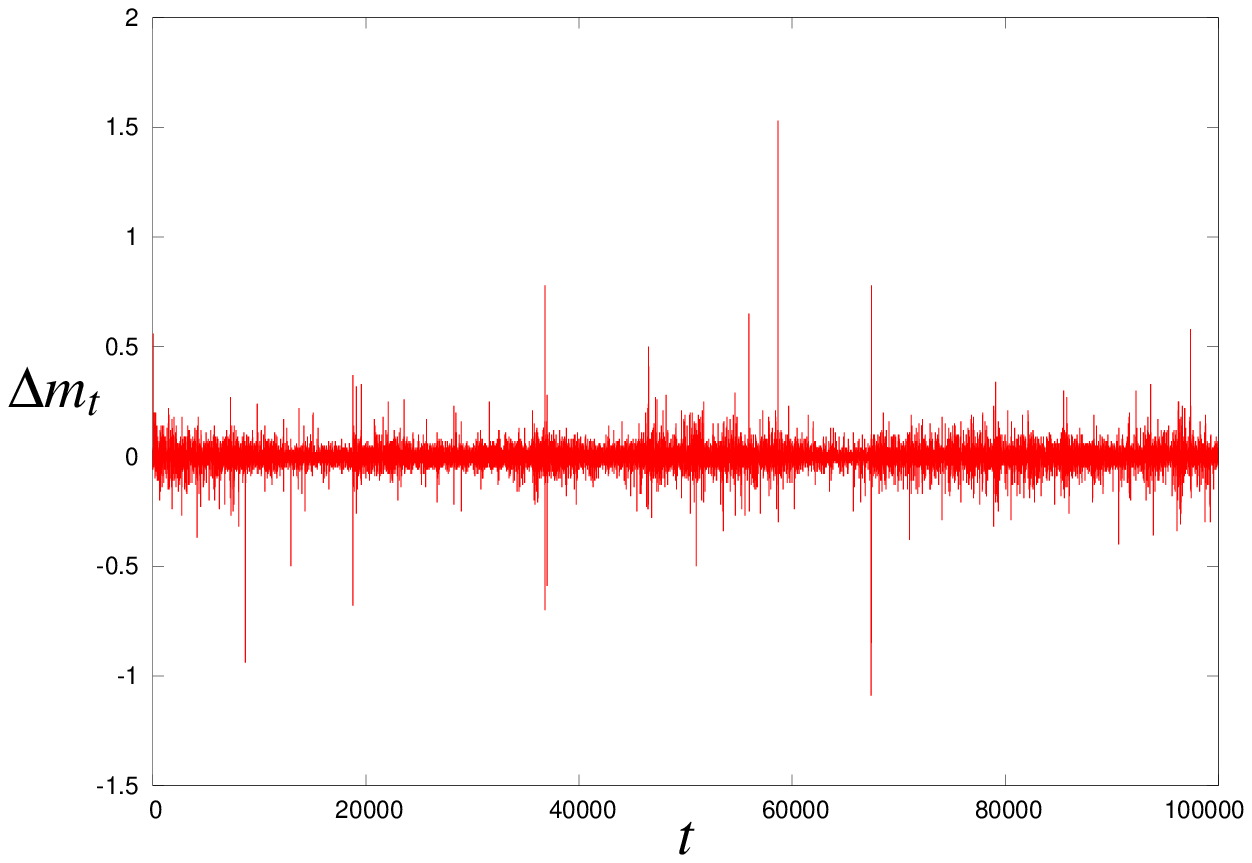}
\includegraphics[width=4cm]{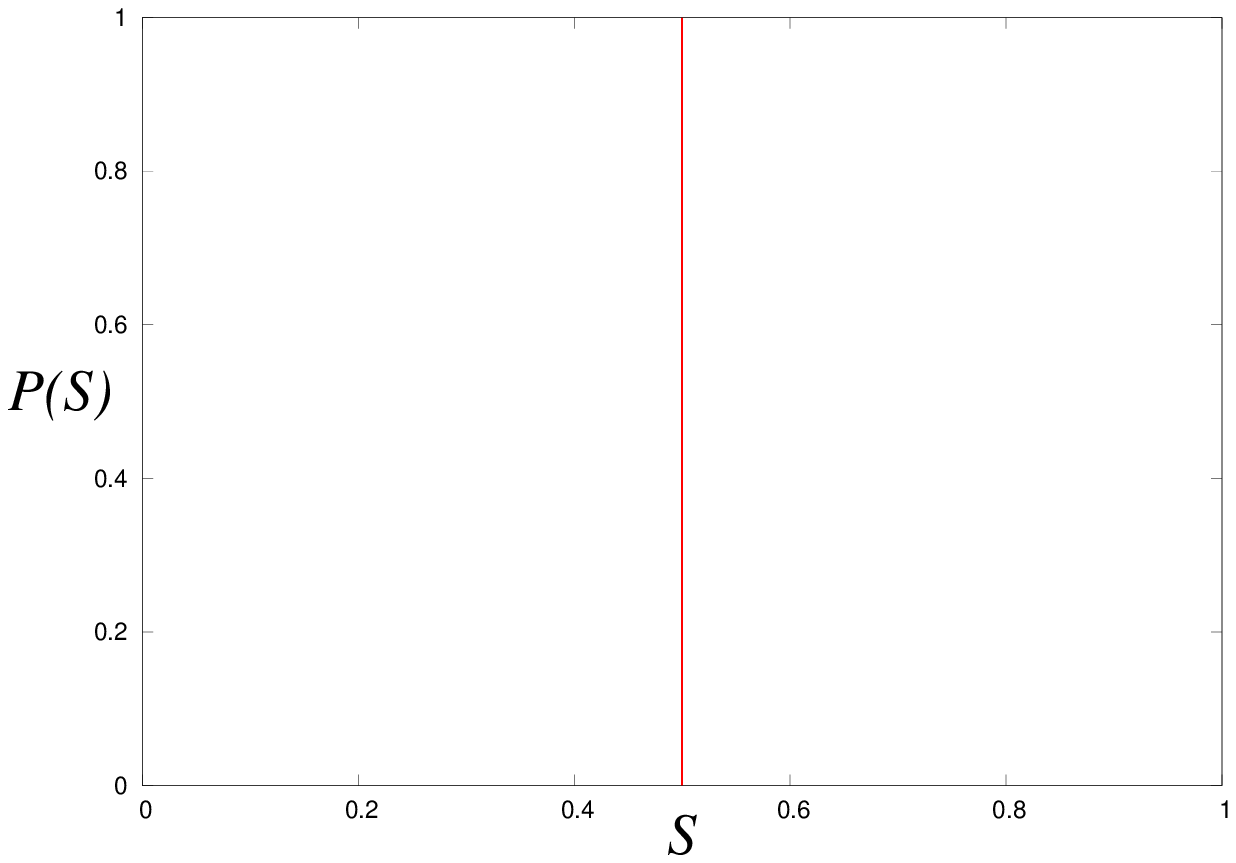}
\caption{\footnotesize 
Statistical properties of 
the data set in which the spread is a time-independent constant, 
Nasdaq100 (the upper panels) and 
price of gold (the lower panels) are shown. 
From left to right, the mid point $m_{t}$, 
the return of the mid point $\Delta m_{t} \equiv m_{t+1}-m_{t}$ 
as a function of $t$, and 
the distribution of the spread $P(S)$ are plotted. 
}
\label{fig:fg_nas1}
\end{center}
\end{figure}
\mbox{}
In Fig. \ref{fig:fg_usd1} for 
EUR/JPY  and USD/JPY exchange rates, 
and in Fig. \ref{fig:fg_nas1} for Nasdaq100 and price of gold, 
we plot the mid point $m_{t}$, 
the return of the mid point $\Delta m_{t} \equiv m_{t+1}-m_{t}$ 
as a function of $t$ and 
the distribution of the Bid-Ask spread $P(S)$. 
From these figures, we clearly find that 
the Bid-Ask spread for 
the exchange rates apparently fluctuates, whereas 
the spread for Nasdaq100 or price of gold is a constant 
leading up to a single delta peak in the empirical distribution. 
From now on, 
the data in which the spread fluctuates is refereed to as 
{\it data with stochastic Bid-Ask spread}, whereas, 
the data in which the spread is constant is 
called as  {\it data with constant Bid-Ask spread}.  
One of the main goals of this paper is to 
reveal the relationship 
between the statistical properties of Bid-Ask spread and 
the behaviour of auto-correlation and response functions for double-auction markets. 
\section{Empirical data analysis}
\label{sec:data_analysis}
In this section, we evaluate two macroscopic dynamical quantities, 
namely, 
auto-correlation and 
response functions 
by making use of empirical data analysis. 
These two relevant quantities 
are explicitly defined by 
\begin{eqnarray}
C(l) & = & 
\lim_{T \to \infty}
\frac{1}{T}
\sum_{t=0}^{T-1}
\epsilon_{t} \epsilon_{t+l}
\label{eq:empiricalC} \\
R(l) & = & 
\lim_{T \to \infty}
\frac{1}{T}
\sum_{t=0}^{T-1}
\epsilon_{t}(
m_{t+l}-m_{t}).
\label{eq:empiricalR} 
\end{eqnarray}
In order to evaluate these functions, we need 
the information about Selling-Buying signal 
$\epsilon_{t}$. However, as we mentioned in the previous section, 
the data gathered through the {\it  MetaTrader4} \cite{Meta} does not contain any information about it explicitly. 
To overcome this problem, we here assume 
that $\epsilon_{t}$ is given in terms of `return' of the mid point:  
\begin{eqnarray}
\epsilon_{t} & = & 
{\rm sgn}(m_{t+1}-m_{t}).
\end{eqnarray}
Namely, we assume when 
the mid point increases at the instant $m_{t+1}>m_{t}$, 
the number of traders who 
posted their own buying signal to the market also increases. 
As the result, the Selling-Buying signal 
$\epsilon$ is more likely to 
take buying $+1$ at that instant $t$. 

Under the above assumption, 
for our four data sets, 
namely, EUR/JPY, USD/JPY exchange rates, 
Nasdaq100 and price of gold, 
we calculate the auto-correlation function $C(l)$ 
and the response function $R(l)$ via (\ref{eq:empiricalC}) 
and (\ref{eq:empiricalR}), respectively. 
\subsection{Auto-correlation function}
\begin{figure}[ht]
\begin{center}
\includegraphics[width=6cm]{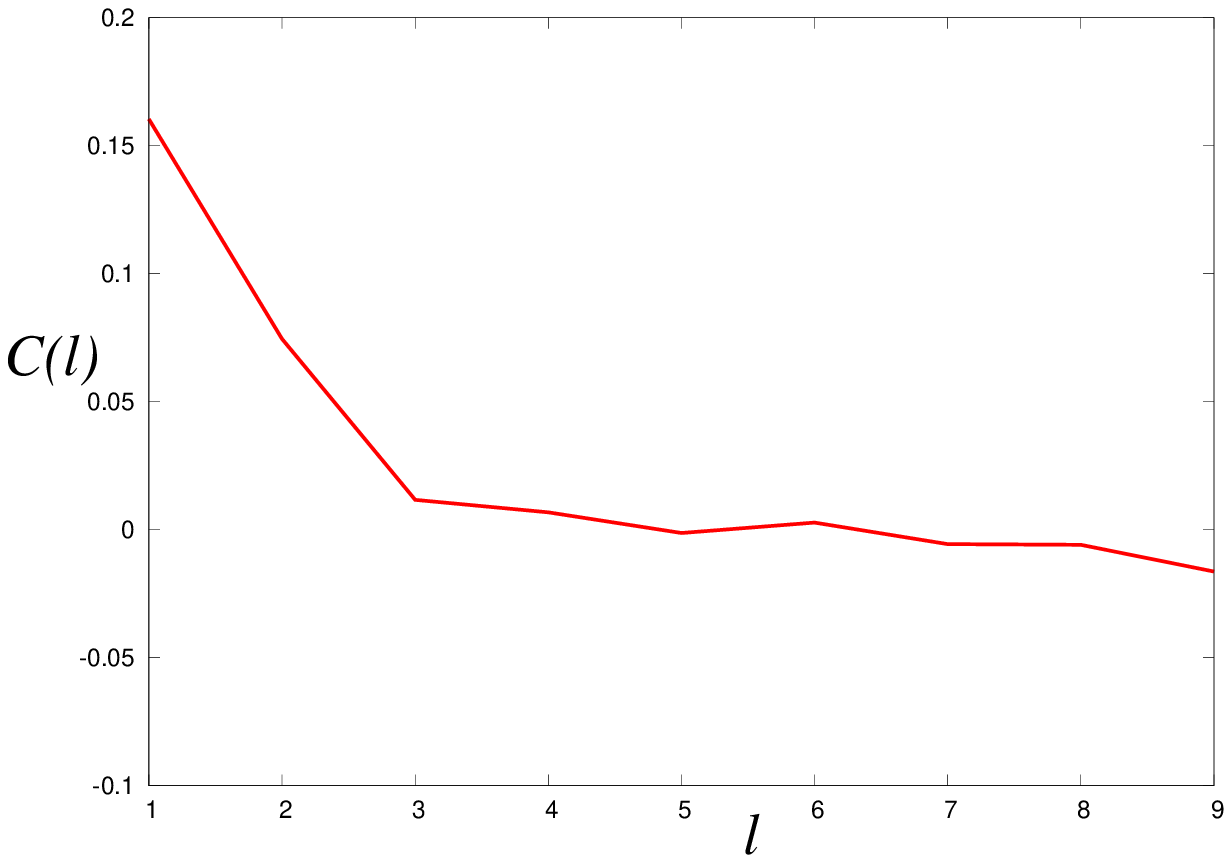}
\includegraphics[width=6cm]{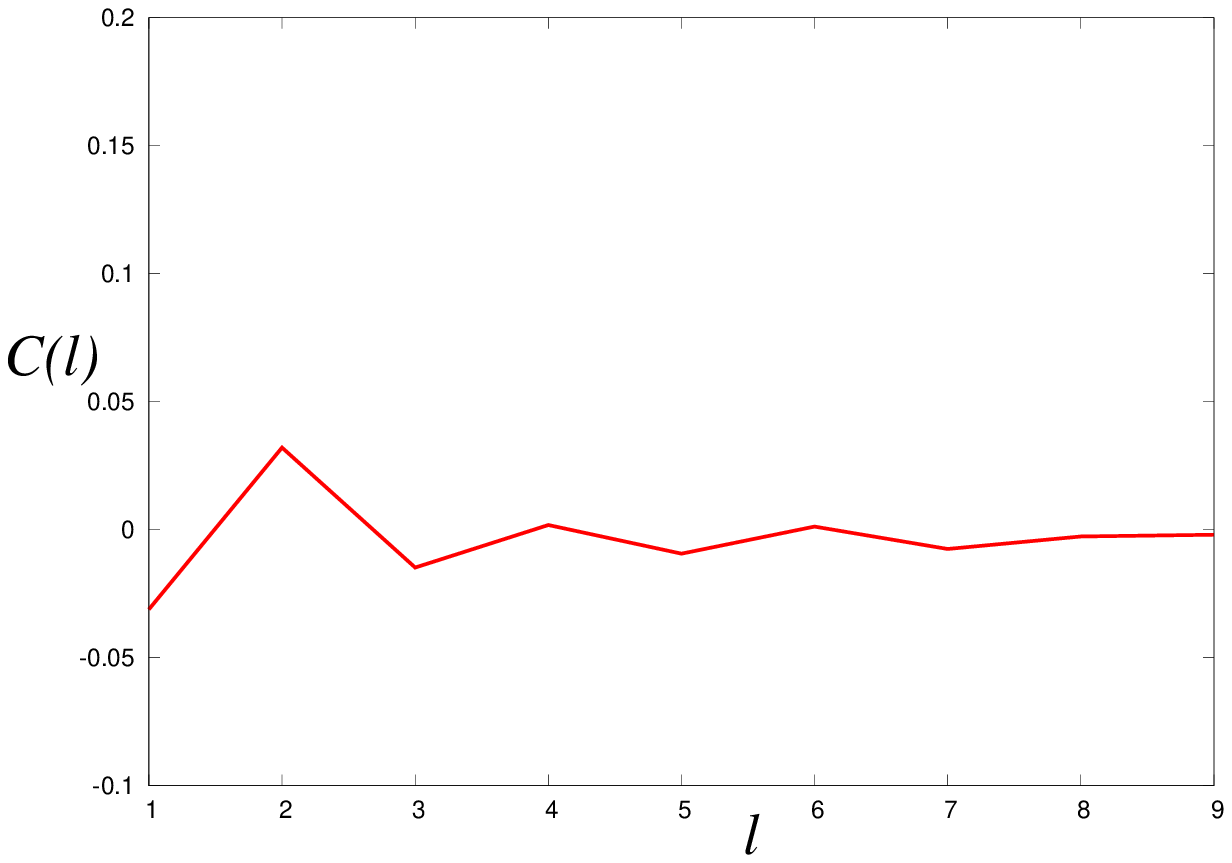} \\
\includegraphics[width=6cm]{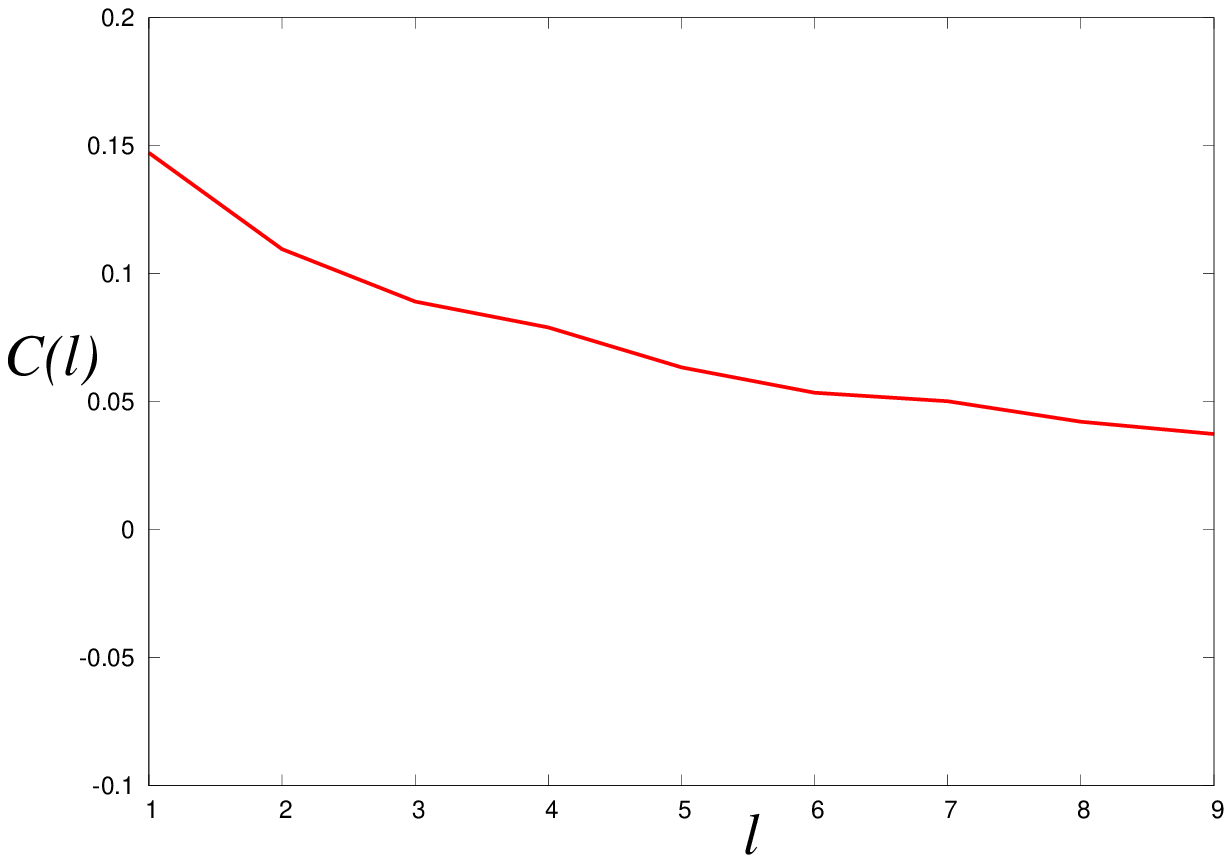}
\includegraphics[width=6cm]{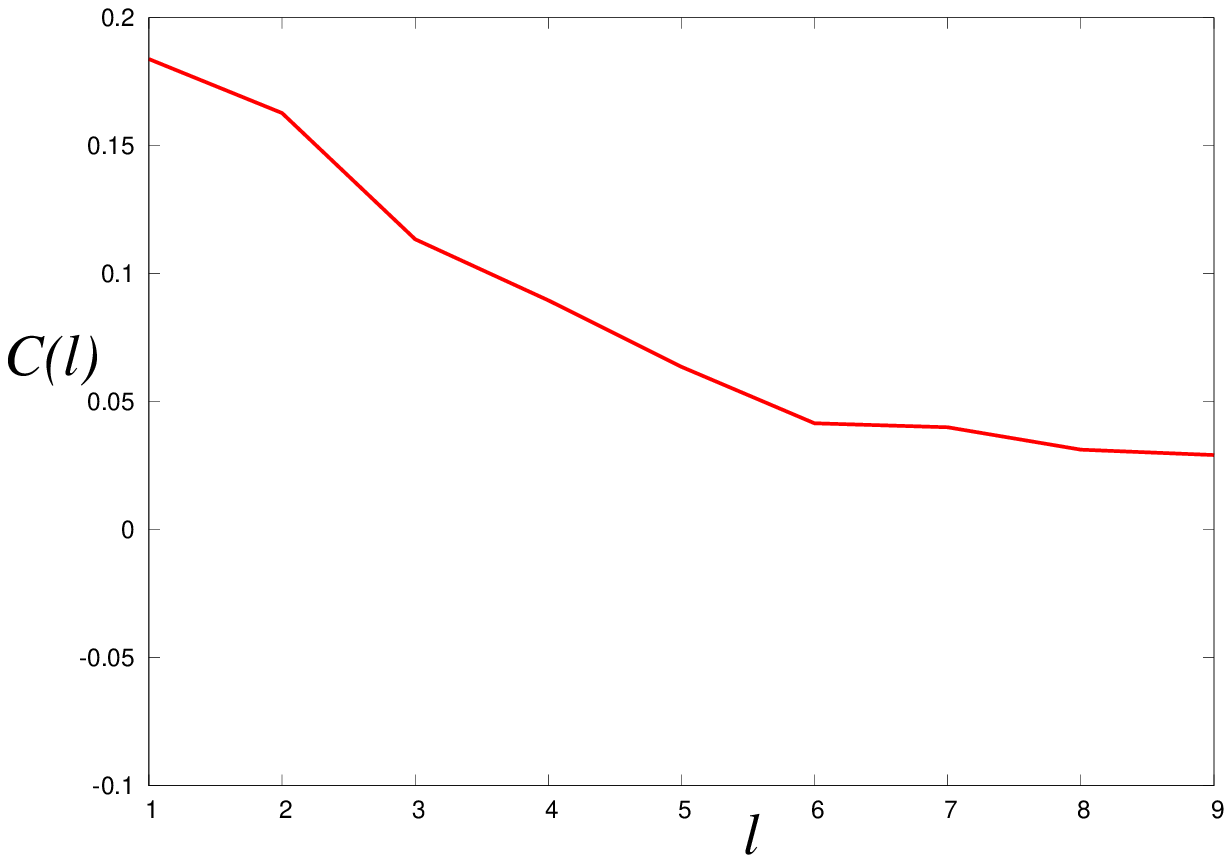}
\end{center}
\caption{\footnotesize
Typical behaviour of 
the auto-correlation function $C(l)$. 
From the upper left to the lower right, 
we plot 
EUR/JPY exchange rates (the estimated $\rho=0.24$), 
USD/JPY exchange rates (the estimated $\rho=0.28$), 
Nasdaq100 (the estimated $\rho=0.47$), 
price of gold (the estimated $\rho=0.51$). }
\label{fig:rho_eur}
\end{figure}

We first plot the auto-correlation function 
for the above four data sets in Fig.  \ref{fig:rho_eur}. 
From the upper left to the lower right, 
we plot 
EUR/JPY exchange rates (under the assumption on the asymptotic form: 
$C(l) \sim \rho^{l},\, l \gg 1$, the estimated $\rho=0.24$), 
USD/JPY exchange rates (the estimated $\rho=0.28$), 
Nasdaq100 (the estimated $\rho=0.47$ ), 
price of gold (the estimated $\rho=0.51$). 
From these panels, we find that 
the correlation in the Selling-Buying signals decreases in 
the time difference $l$ although the result for USD/JPY exchange rate 
possesses the negative correlation in $l=1$ and converges to zero with a slight oscillation. 
\subsection{Response function}
\begin{figure}[ht]
\begin{center}
\includegraphics[width=6cm]{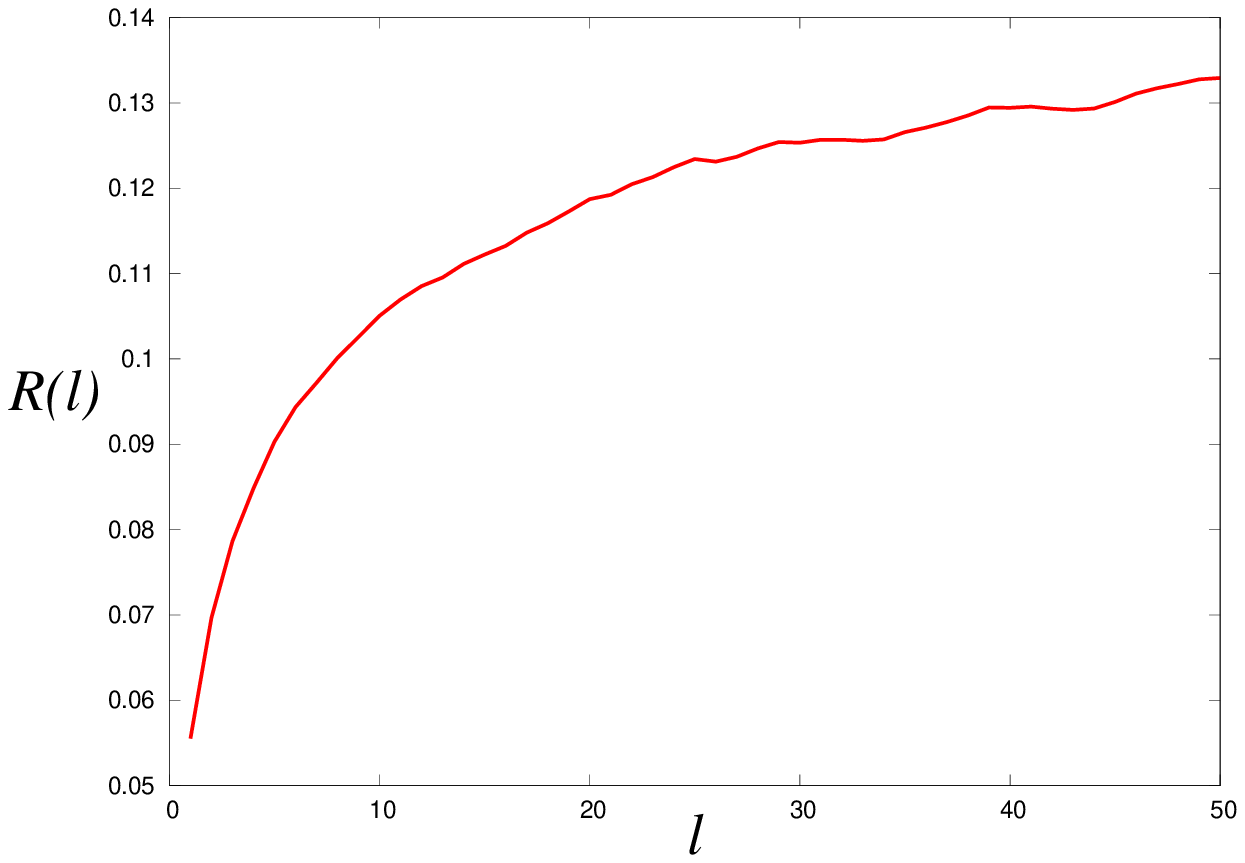}
\includegraphics[width=6cm]{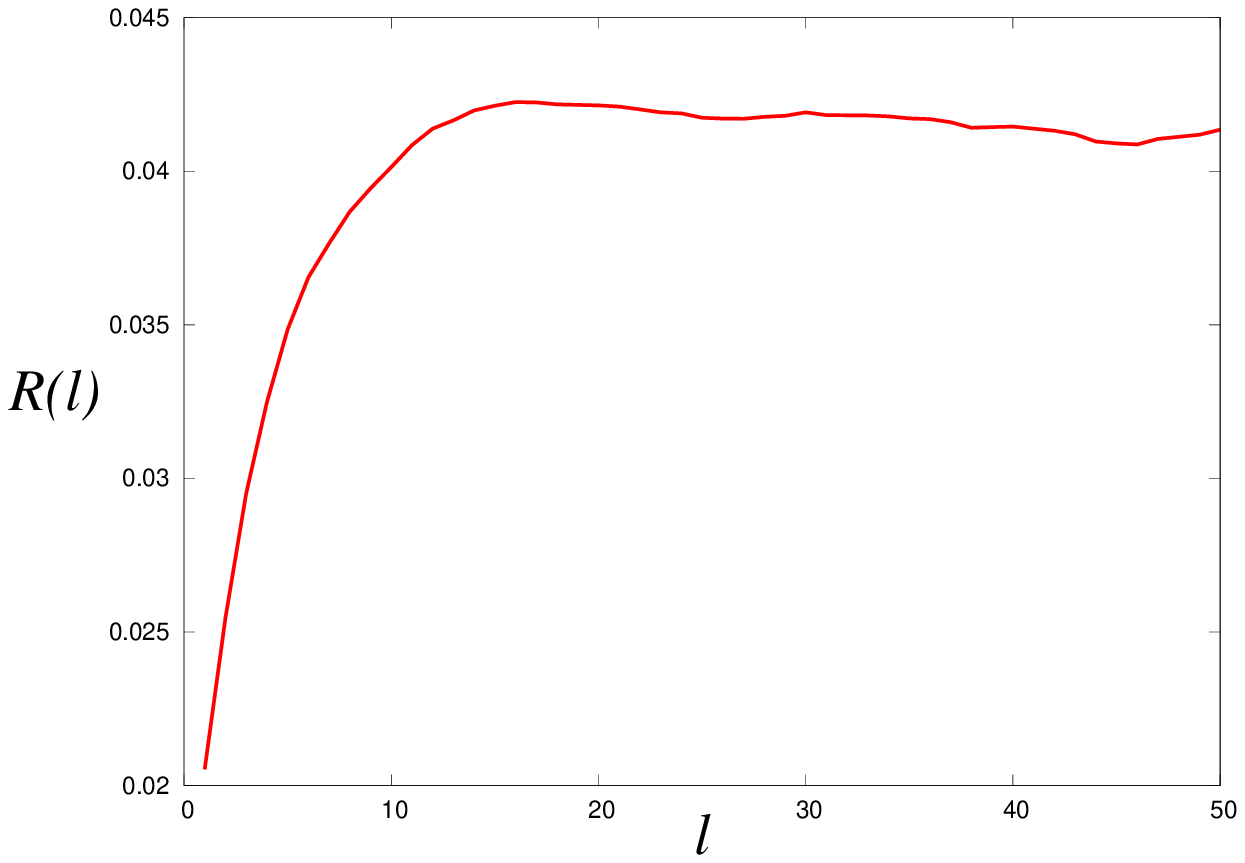} \\
\includegraphics[width=6cm]{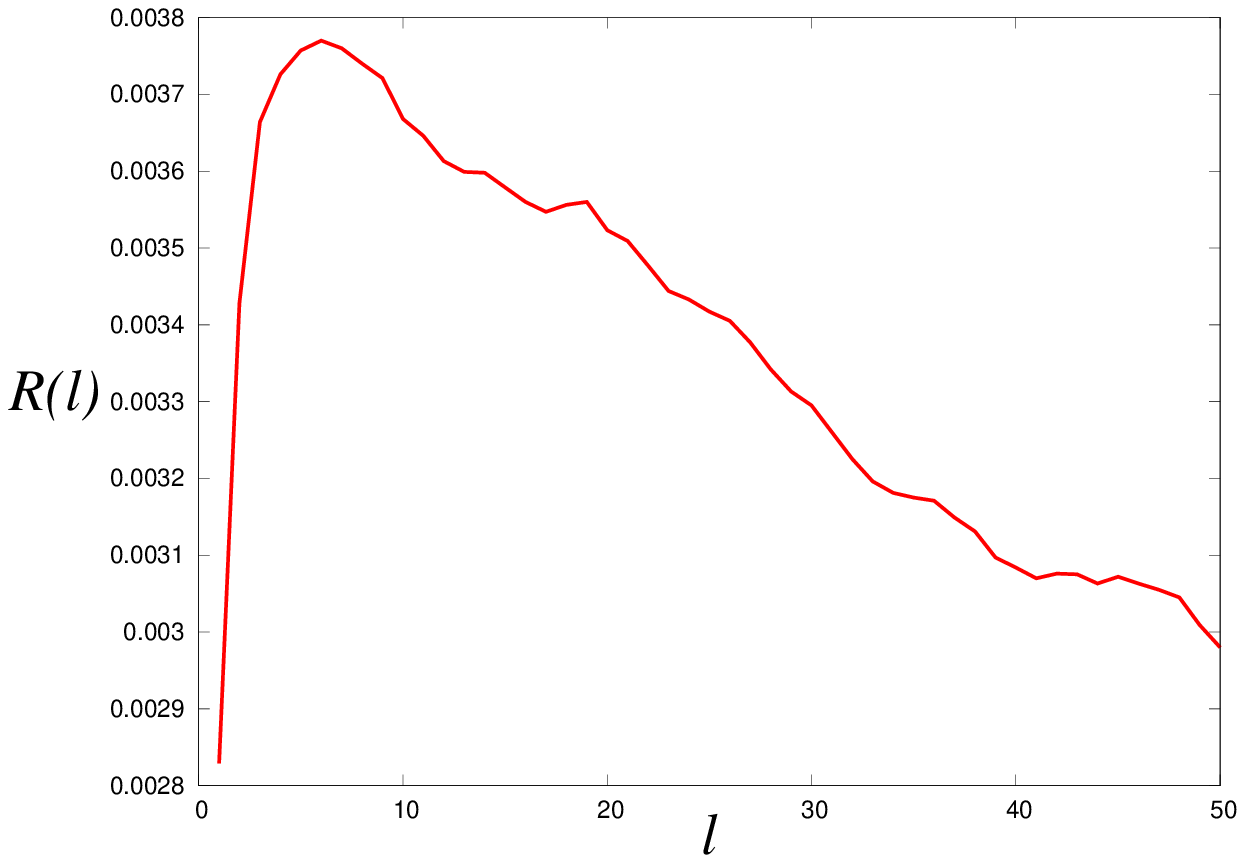}
\includegraphics[width=6cm]{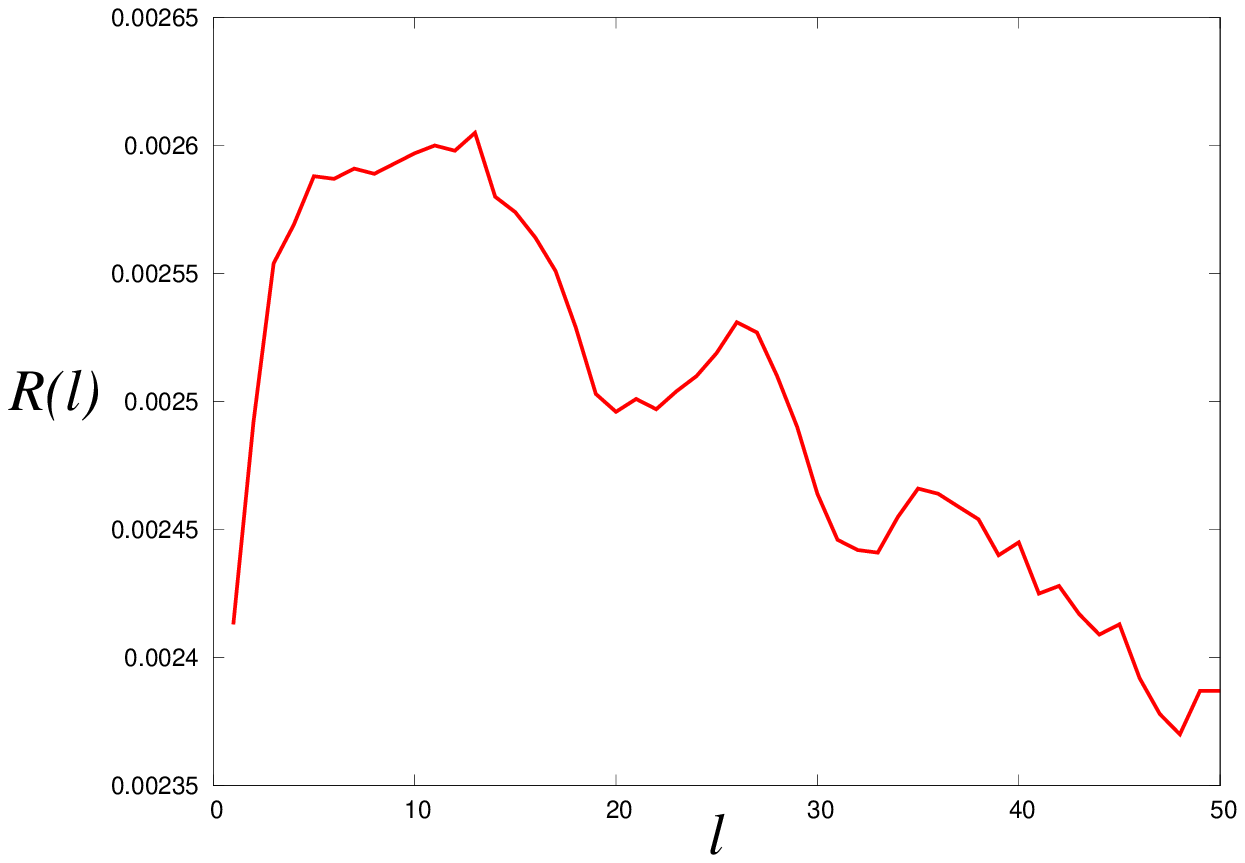}
\end{center}
\caption{\footnotesize 
Typical behaviour of the response function 
for our empirical data: 
from the upper left to the lower right, 
the results for 
Nasdaq100, price of gold, 
EUR/JPY exchange rate and 
USD/JPY exchange rate are shown.}
\label{fig:fg_empirical}
\end{figure}

We next evaluate the response function 
for our four data sets. 
The results are shown in Fig. \ref{fig:fg_empirical}. 
In this figure, 
we plot the response function 
for the data with stochastic Bid-Ask spread (the lower panels) and 
for the data with a constant Bid-Ask spread 
(the upper panels). 
From these panels, we find that some `non-monotonic' 
behaviour in $R(l)$ appears for the data set with stochastic Bid-Ask spread.  
\subsection{Relationship between $C(l)$ and $R(l)$}
From the definitions, 
both auto-correlation and response functions 
are functions of the time-difference $l$. 
In the previous subsections, we investigated their behaviour 
independently. However, 
it might be assumed that these two quantities are related each other. 
Therefore, 
it is useful for us to 
make `scatter plots' to reveal the dynamical relationship 
underlying these two quantities. 
\begin{figure}[ht]
\begin{center}
\includegraphics[width=6cm]{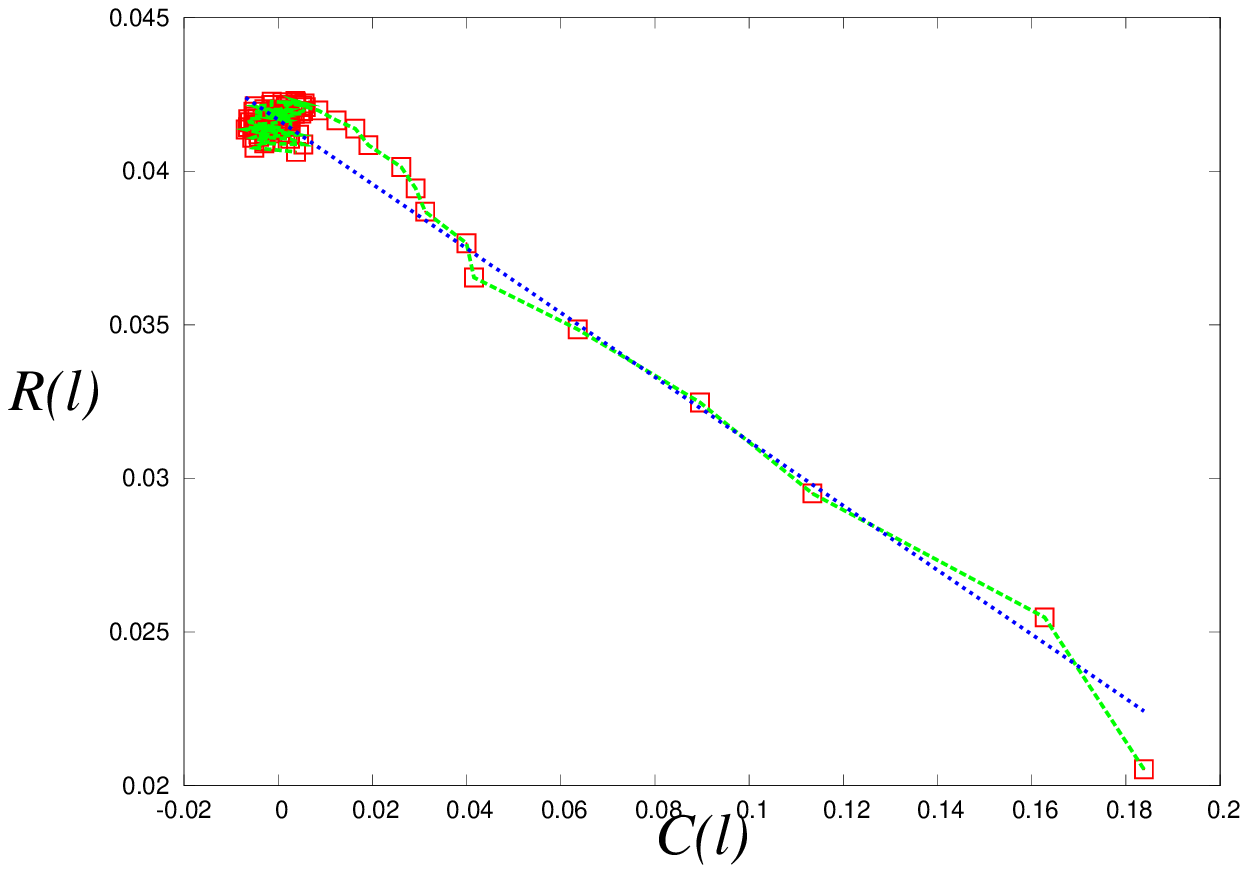}
\includegraphics[width=6cm]{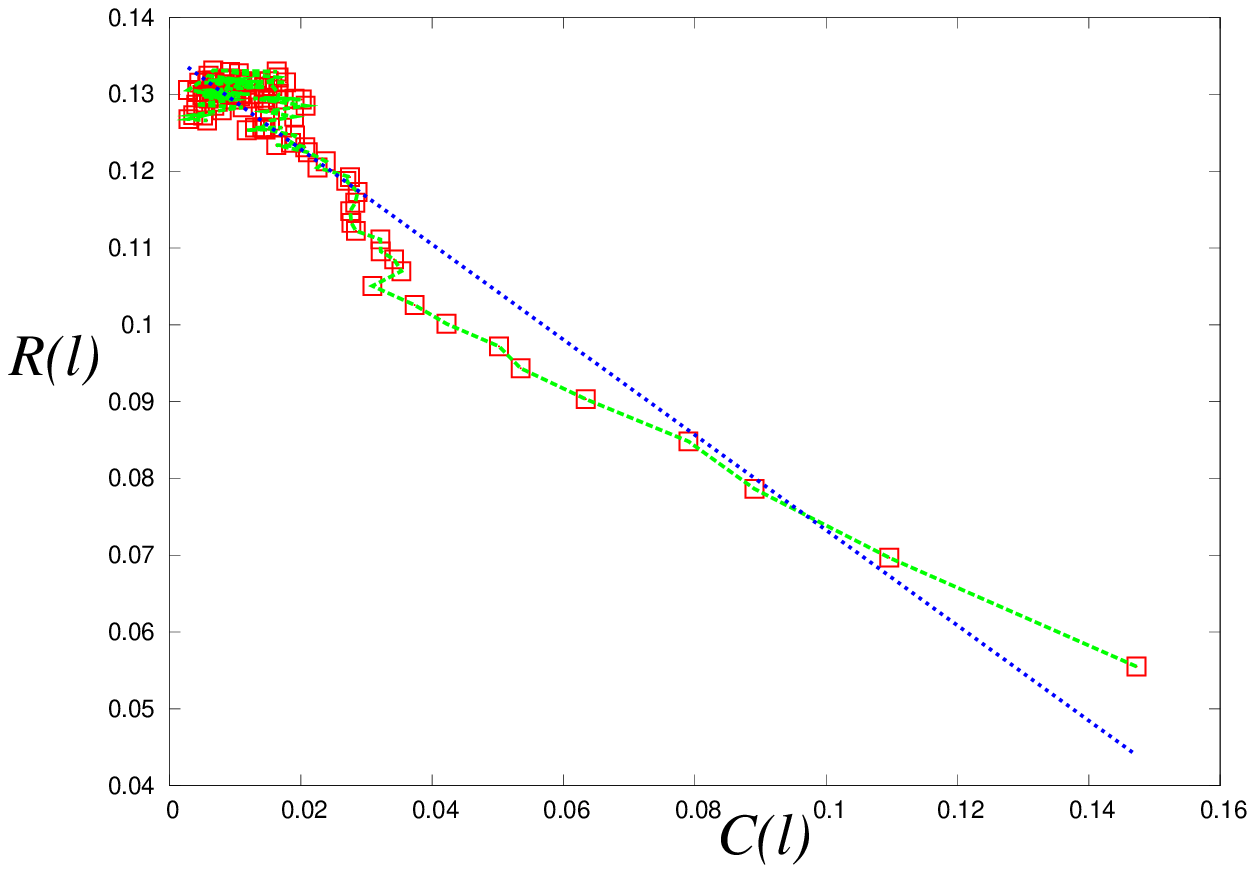} \\
\includegraphics[width=6cm]{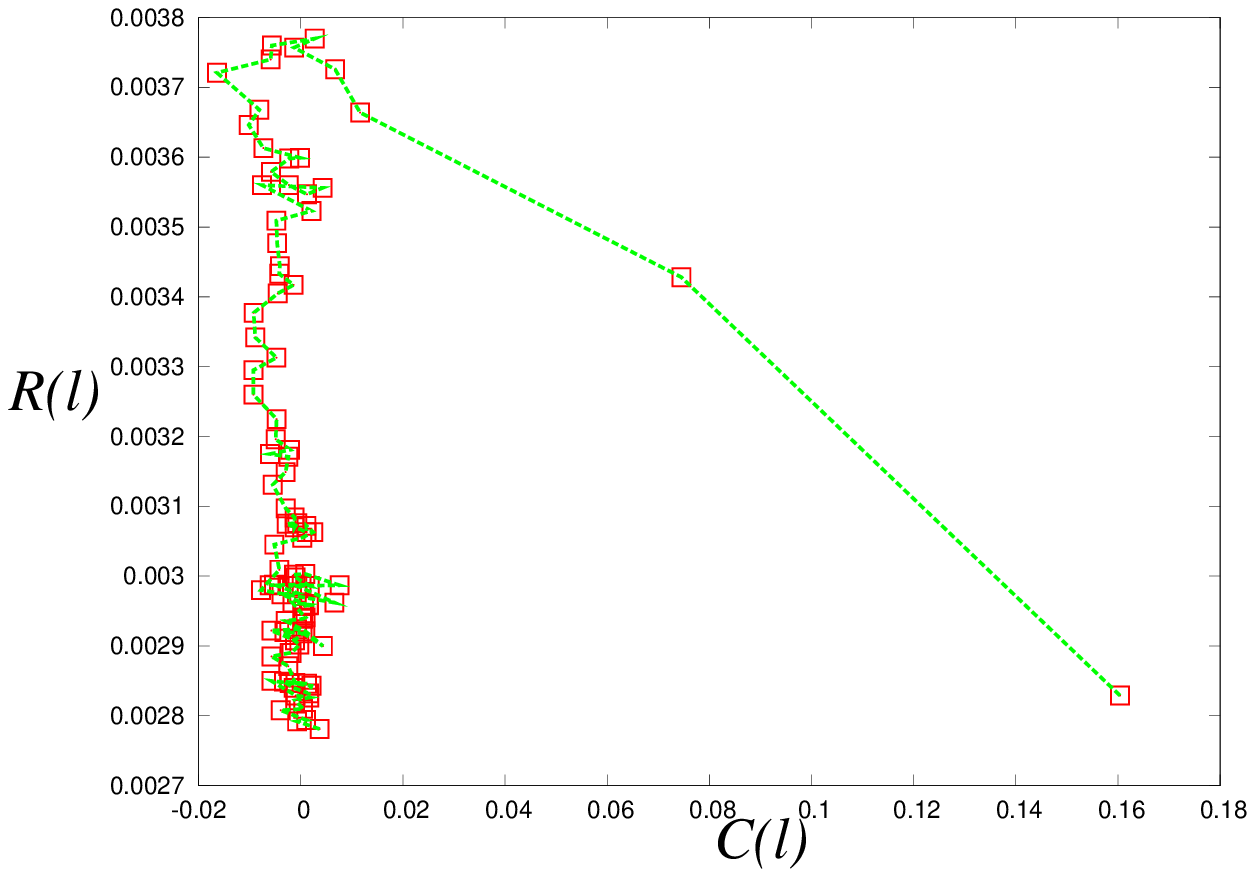}
\end{center}
\caption{\footnotesize 
The relationship between $R(l)$ and $C(l)$. 
The upper two panels are results for the 
gold and Nasdaq100 with constant spreads, whereas the lower panel denotes 
the result for the EUR/JPY exchange rate having 
fluctuating spreads.
We find that the linear relationship 
$R(l) \propto -C(l)$ 
holds for the data having a constant spread, whereas the linear relationship is apparently broken down in 
the EUR/JPY exchange rate  
which possesses a fluctuating Bid-Ask spread. 
}
\label{fig:CR_real}
\end{figure}
\mbox{} 
In Fig. \ref{fig:CR_real}, 
we plot 
the relationship between $R(l)$ and $C(l)$ by scatter plots.  
The upper two panels are results for the gold and 
Nasdaq100 with constant spreads, whereas the lower panel
denotes the result for the EUR/JPY exchange rate having 
fluctuating spread.
We find that the linear relationship 
$R(l) \propto -C(l)$ is apparently broken down in 
the EUR/JPY exchange rate  
which possesses a fluctuating Bid-Ask spread. 

In the next section, we examine a phenomenological 
model to explain the non-linear relationship 
$R(l) \propto -f(C(l))$ 
($f(x)$ denotes a non-linear function) theoretically. 
\section{A phenomenological approach}
\label{sec:MRR}
In order to explain the behaviour of 
the auto-correlation and response functions, 
we examine a phenomenological approach based on 
the so-called 
{\it Madhavan-Richardson-Roomans 
model} (MRR for short) \cite{MRR,Bouchaud1} to simulate 
the stochastic process of price-change in empirical data sets. 
In the MRR model, the price $p_{t}$ updates according to 
the following rule. 
\begin{eqnarray}
p_{t+1} & = &  
p_{t} +
\theta (\epsilon_{t}-\rho \epsilon_{t-1})+ \xi_{t} 
\label{eq:updateP}
\end{eqnarray}
where 
$\xi_{t}$ denotes a noise in the 
market satisfying 
$\langle \xi_{t} \rangle=0$ and 
$\langle \xi_{t} \xi_{t^{'}} \rangle = 
\delta_{t,t^{'}}\Sigma^{2}$.  
$\theta$ is a constant value to control 
the slope of instantaneous price change.
The label $\epsilon_{t}$ means 
a {\it Selling-Buying signal} to represent  
$\epsilon_{t}=+1$ for a buying signal and 
$\epsilon_{t}=-1$ vice versa.

Behaviour of the above update rule is dependent on 
the statistical properties of Selling-Buying signals $\epsilon_{t}$. 
$\rho$ is a correlation factor and in the MRR theory, 
we assume that $\epsilon_{t}$ follows a simple Markovian process, namely, 
\begin{eqnarray}
\sum_{\epsilon_{t}=\pm 1}
\epsilon_{t} 
P(\epsilon_{t}|\epsilon_{t-1}) & = & 
\rho \epsilon_{t-1}
\label{eq:Markov}.
\end{eqnarray}
\mbox{}

The price value of $p_{t+1}$ provided that 
the Selling-Buying signal in the previous time step 
is $\epsilon_{t}=+1$ 
should be the Ask $a_{t}$ and 
the price $p_{t+1}$ provided that 
the signal is  $\epsilon_{t}=-1$ should be 
the Bid $b_{t}$. Hence, we naturally define the 
time-dependence of Ask and Bid as follows. 
\begin{eqnarray}
a_{t} & = & 
p_{t} + \theta (1 - \rho \epsilon_{t-1}) + \phi 
\label{eq:a_i} \\
b_{t} & = & 
p_{t} + \theta (-1 -\rho \epsilon_{t-1}) -\phi 
\label{eq:b_i}
\end{eqnarray}
where $\phi$ denotes a kind of transaction cost and the value itself is 
set to a constant in the MRR model. 
From these rules, we easily find 
the Bid-Ask spread at time $t$ as 
\begin{eqnarray}
S_{t} & = & a_{t}-b_{t}=
2 (\theta + \phi).
\end{eqnarray}
Namely, in the MRR model, the spread is a time-independent constant 
during the dynamics. 

On the other hand, the mid point of the Bid and Ask is given by 
\begin{eqnarray}
m_{t} & = &  
\frac{1}{2}
(a_{t}+b_{t}) = 
p_{t}-\theta \rho \, \epsilon_{t-1}.
\label{eq:pm}
\end{eqnarray}
Therefore, for the parameter choice $\theta=0$ or $\rho=0$, 
the mid point $m_{t}$ is identical to the price $p_{t}$. 
For the above update rules of 
price, Bid, Ask, spread and mid point, 
we investigate the macroscopic properties of double-auction markets 
through the auto-correlation function and the response function. 
\subsection{Auto-correlation function}
From the definition of Markovian process (\ref{eq:Markov}), 
the auto-correlation function is given by 
\begin{eqnarray}
C (l) & = & 
\langle \epsilon_{t}
\epsilon_{t+l} 
\rangle  \equiv  
\sum_{\epsilon_{t}=\pm 1}
\cdots 
\sum_{\epsilon_{t+l}=\pm 1}
P(\epsilon_{t},\cdots, \epsilon_{t+l})
\epsilon_{t}
\epsilon_{t+l}  = 
\rho^{l}. 
\label{eq:corr}
\end{eqnarray}
We should keep in mind that 
the auto-correlation function is originally 
defined by (\ref{eq:empiricalC}). However, in the limit of $T \to \infty$, 
one can replace the time-average by 
the average over the joint probability 
of the stochastic variables 
$\epsilon_{t},\cdots,\epsilon_{t+l}$ as 
$(1/T)\sum_{t=0}^{T-1}(\cdots)
=
\sum_{\epsilon_{t}=\pm 1}
\cdots 
\sum_{\epsilon_{t+l}=\pm 1}
(\cdots)P(\epsilon_{t},\cdots,\epsilon_{t+l})$ 
according to the law of large number.

The correlation factor $\rho$ should be 
$|\rho| \leq 1$.  
For a positive $\rho$, the correlation function decays exponentially as 
$C(l) = {\rm e}^{-l \log (1/\rho)}$. 
\subsection{Response function}
We next consider the response function of the market, that is defined by 
\begin{eqnarray}
R(l) & = & 
\langle 
\epsilon_{t} \cdot 
(m_{t+l}-m_{t})
\rangle
\label{eq:lag_func}
\end{eqnarray}
where the bracket $\langle \cdots \rangle$ has the same meaning as that in 
(\ref{eq:corr}) has. 

From  the above response function, one obtains some information 
about the response of the market 
at time $t+l$ to the Selling-Buying signal 
at arbitrary time $t$. 
Namely, 
the response function 
measures to what extent 
the mid point increases (decreases) on average 
for interval $l$ when a buying (selling) signal is posted 
to the market $l$ steps before we observe the mid point.   
After simple algebra, we easily obtain the relationship between 
the response function $R(l)$ and the correlation function $C(l)$  for 
the MRR model as follows. 
\begin{eqnarray}
 R(l) & = & \theta (1-C(l))
\label{eq:RC}
\end{eqnarray}
As we saw before, for a positive correlation factor $\rho>0$, 
the $C(l)$ monotonically 
decreases as $C(l) = {\rm e}^{-l \log (1/\rho)}$. 
Hence, the response function also behaves monotonically and converges to $\theta$ as 
$R(l)=\theta (1-{\rm e}^{-l \log (1/\rho)}) \to \theta $ ($l \to \infty$). 
It should be noticed that 
the linear relationship between $C(l)$ and $R(l)$ holds from (\ref{eq:RC}). 

From equation (\ref{eq:RC}), we also find 
$R(1)=\theta (1-\rho)$ and 
this fact tells us that  
$R(\infty)/R(1) = (1-\rho)^{-1}$ holds. 
The volatility defined by 
\begin{eqnarray}
\sigma^{2} (l) & \equiv & 
\frac{1}{l} \langle 
(m_{t+l}-m_{t})^{2}
\rangle
\end{eqnarray}
also reads 
\begin{eqnarray}
\sigma^{2} (l) & = &  
\Sigma^{2}
+
\theta^{2} 
(1-\rho)^{2}
\left\{
1+ 
\frac{2\rho (1-\rho^{l-1})}
{1-\rho}
\right\}
\end{eqnarray}
and 
$\sigma^{2} (1) = 
\Sigma^{2}+\theta^{2}(1-\rho)^{2}$, 
$\sigma^{2} (\infty)=\Sigma^{2}+
\theta^{2}(1-\rho^{2})$. 
We may use the above rigorous equations 
to check the validity of our computer simulations. 

\begin{figure}[ht]
\begin{center}
\includegraphics[width=6cm]{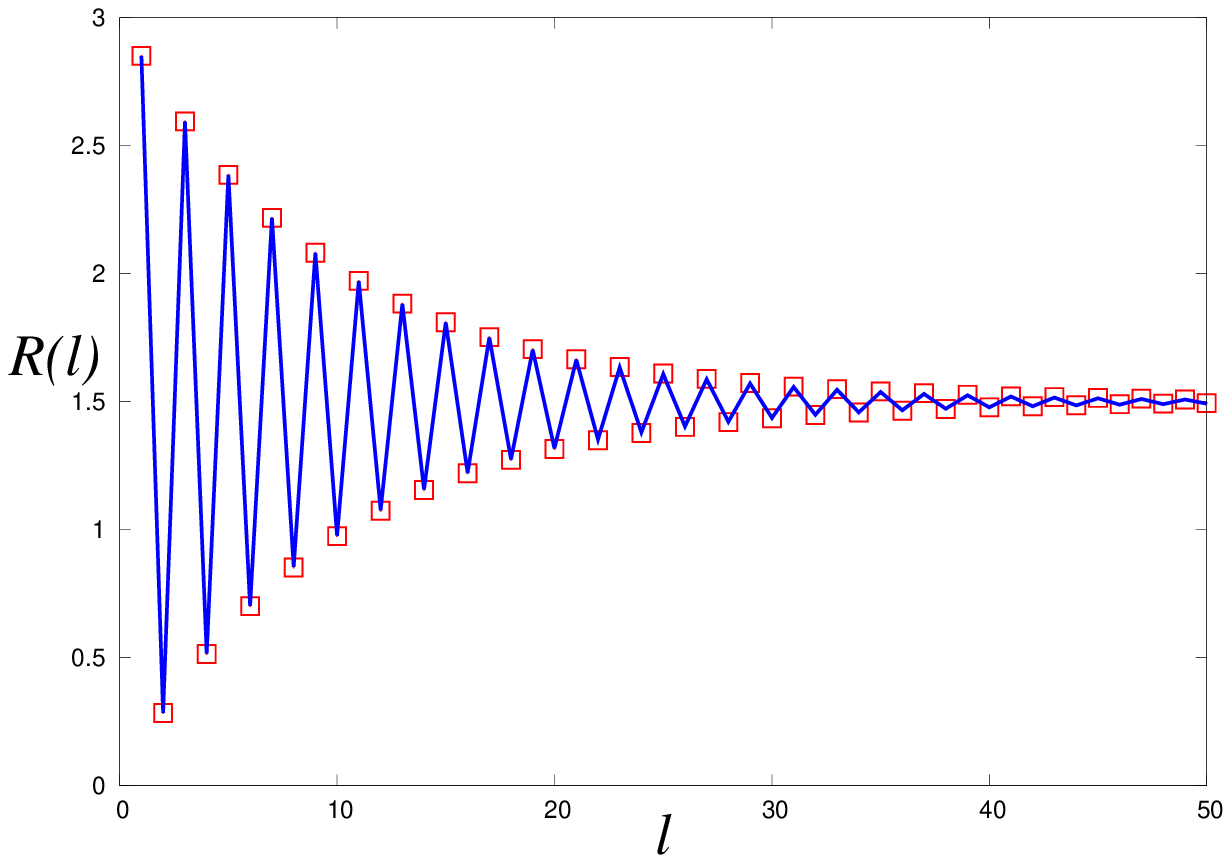}
\includegraphics[width=6cm]{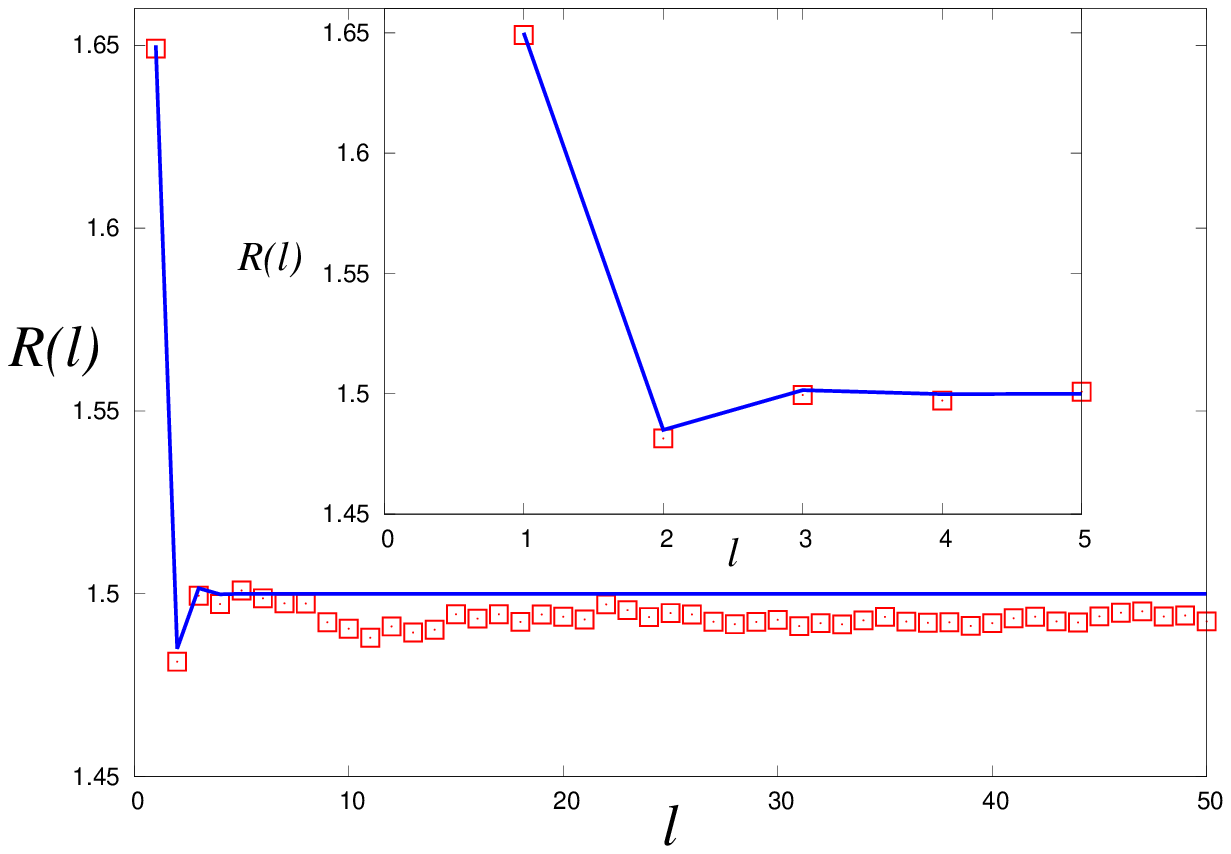} \\
\includegraphics[width=6cm]{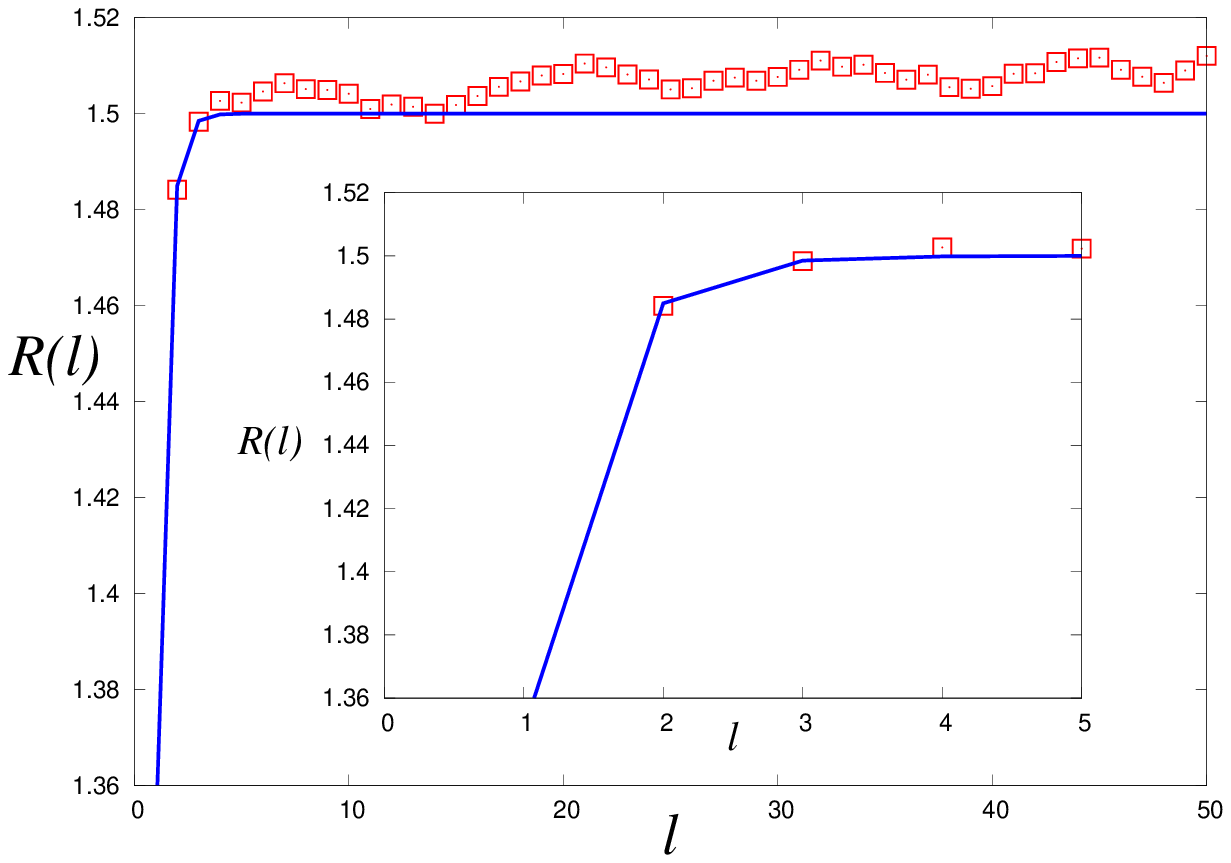}
\includegraphics[width=6cm]{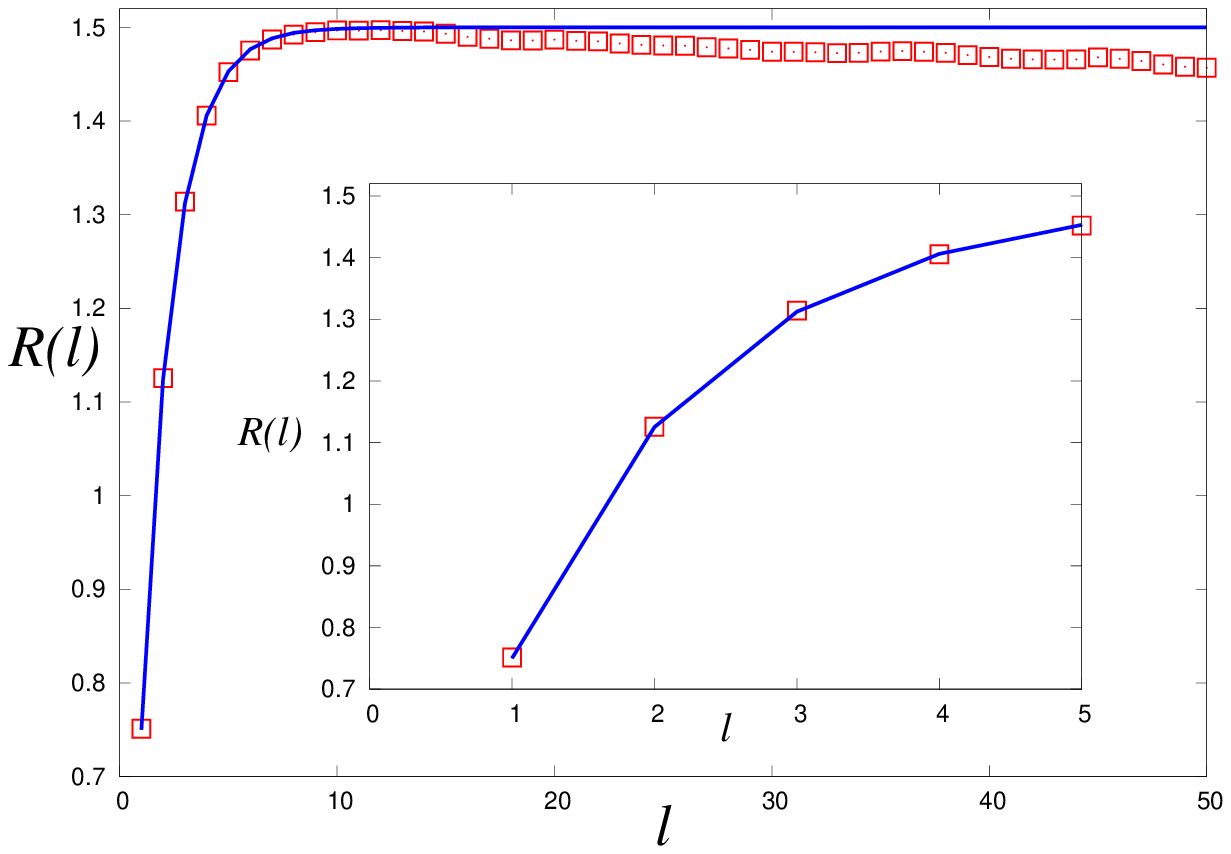}
\end{center}
\caption{\footnotesize 
Typical behaviour of the response function 
for the MRR model. 
From the upper left to the lower right, we plot the 
response function for 
$\rho=-0.9,-0.1,0.1,0.5$. 
The solid lines are theoretical predictions 
by $R(l)=\theta (1-\rho^{l})$. The boxes are 
obtained by numerical simulations for a finite $T=10^{5}$. }
\label{fig:fg3}
\end{figure}

In Fig. \ref{fig:fg3}, we show the typical behaviour of response function for 
several choices of $\rho$. 
From these panels, we find that for a positive 
correlation factor, the response function 
monotonically converges to the value $\theta$. 
In this figure, we also show the results obtained by simulating the update equation 
for the price (\ref{eq:updateP}) and the mid point (\ref{eq:pm}), 
and calculating the response function numerically by making use of 
(\ref{eq:empiricalR}). 
We find that the both theoretical prediction (solid lines) and 
the simulation (boxes) are in good agreement. 
Moreover, for the choice of 
positive correlation factors, 
say, $\rho=0.1$ and $0.5$, 
the response function 
increases monotonically and converges to $\theta (=1.5)$ 
as the MRR theory predicted. 

\begin{figure}[ht]
\begin{center}
\includegraphics[width=8cm]{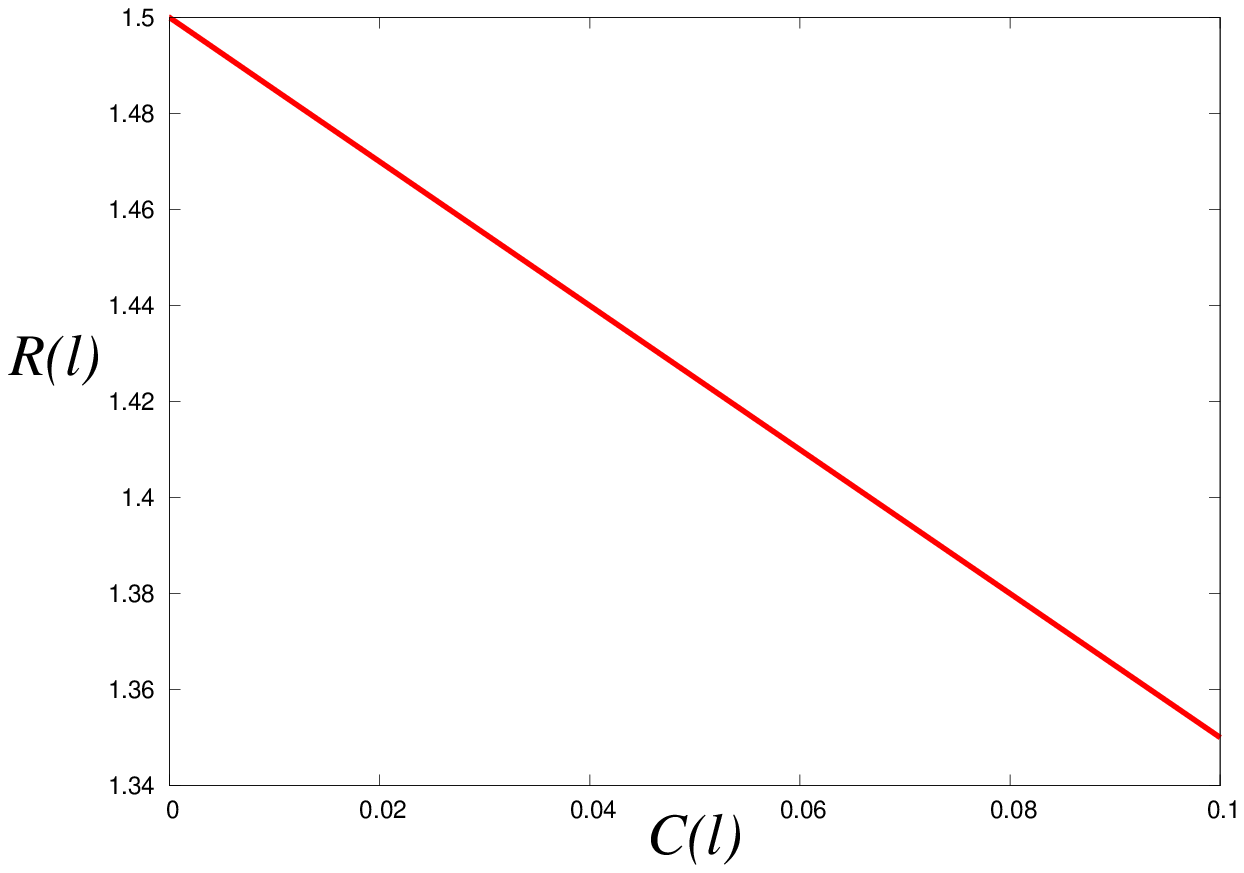} 
\end{center}
\caption{\footnotesize 
The relationship between $R(l)$ and $C(l)$ 
for the MRR model. 
We find that the linear relationship 
$R(l)  = \theta (1-C(l)) \propto -C(l)$ 
holds. 
}
\label{fig:CR_MRR}
\end{figure}

In Fig. \ref{fig:CR_MRR}, 
we show the relationship between $R(l)$ and $C(l)$ for the MRR model. 
We clearly find that the linear relationship 
$R(l)  = \theta (1-C(l)) \propto -C(l)$ actually holds. 
We should notice that 
this result is completely different from 
the relationship shown in the lower panel of 
Fig. \ref{fig:CR_real} 
which is result for the data having fluctuating Bid-Ask spread. 
This failure of the MRR model 
to simulate  the `$\lambda$-shape' of 
the $C(l)$-$R(l)$ scatter plot 
for the data with stochastic Bid-Ask spread is 
obviously due to the assumption of constant Bid-Ask spread, 
namely, $S_{t}=a_{t}-b_{t}=2(\theta + \phi)$ on the MRR theory. 
We also conclude that the breaking of the linear relationship 
$R(l) \propto -C(l)$ 
is macroscopically due to the fact that 
the response function 
evaluated on the basis of the MRR theory 
behaves `monotonically' and converges to a finite value $\theta$. 

This fact is one of the limitations of the MRR theory to 
explain the empirical data for double-auction markets. 
To make the efficient model for the double-auction market with 
stochastic Bid-Ask spread, we use a game theoretical approach 
based on the so-called {\it minority game}.
\section{A minority game modeling of double-auction markets}
\label{sec:MG}
In order to make a model to simulate the 
`non-monotonic' behaviour of the response function 
for the financial data with stochastic Bid-Ask spread, 
we start our argument from standard minority game \cite{Arthur,Challet1997,Challet,Coolen} 
with a finite market history length. 
\subsection{General set-up}
In our computer simulations for the minority game, 
at each round $t$ (time step) of the game, 
each trader $i$ 
($i=1,\cdots,N$: $N$ should be an odd number 
to determine the `minority group') decides his (or her) decision: 
${\cal B}_{i}(t)=+1$ ({\it buy}) or ${\cal B}_{i}(t)=-1$ ({\it sell}) 
to choose the minority group. 
Then, we evaluate the total decision of the traders: 
\begin{eqnarray}
A(t) & = & \frac{1}{\sqrt{N}} 
\sum_{i=1}^{N}{\cal B}_{i}(t)
\label{eq:total_bid}
\end{eqnarray}
for each round $t$.  
It should be noted that the 
factor $N^{-1/2}$ is needed to make the 
$A(t)$ of order $1$ object (independent of the size $N$). 
From the definition (\ref{eq:total_bid}), 
the market is {\it seller's market} 
for $A(t)>0$, whereas 
the market behaves as {\it buyer's market} 
for $A(t)<0$.  
The $A(t)$ follows complicated 
stochastic process 
and one might consider that the price $p$ is updated 
in terms of the $A(t)$ as follows.  
\begin{eqnarray}
p(t+1) & = & 
p(t) +\beta \{
A(t) + \psi \, {\rm sgn}[A(t-1)]
\}
\label{eq:MGp}
\end{eqnarray}
where $\beta$ and $\psi$ are positive constants. 
The above update rule means that 
the price increases if the `buying group' is majority 
and decreases vice versa. This setting of the game seems to be naturally accepted. 
A bias term $\psi$ appearing in (\ref{eq:MGp}) plays an important 
role to simulate the auto-correlation function as we will see later on. 

To decide `buy' or `sell', each trader uses the following information vector defined by 
\begin{eqnarray}
\mbox{\boldmath $\lambda$}(A,Z : t) & = & 
\left(
\begin{array}{c}
{\rm sgn}[(1-\zeta)A(t-1)+\zeta Z(t,1)]  \\
{\rm sgn}[(1-\zeta)A(t-2)+\zeta Z(t,2)]  \\
\cdots \\
{\rm sgn}[(1-\zeta)A(t-M)+\zeta Z(t,M)] 
\end{array}
\right)
\end{eqnarray}
where ${\rm sgn}(x)$ denotes a sign function and 
$Z(t,\xi), 
\xi =1,\cdots, M$ is a white noise defined by 
\begin{eqnarray}
\langle 
Z(t,\xi) 
Z(t^{'},\xi^{'}) 
\rangle & = & 
\delta_{t,t^{'}}
\delta_{\xi, \xi^{'}}.
\end{eqnarray}
Therefore, 
each trader uses the information 
of market through 
the up-down configuration 
of the return $A$ (with some additive noise $Z$) 
back to the previous $M$-steps. If the $\zeta$ is close to 
$1$, the `real' market history through 
the $A$ is hided by the `fake' market 
history through the noise $Z$. 

For a given information vector 
$\mbox{\boldmath $\lambda$}(A,Z : t) =
\mbox{\boldmath $\lambda$}$ chosen 
from all possible $2^{M}$ candidates,  
each trader $i$ decides her (or his) 
action at round $l$ by the following 
strategy vector: 
\begin{eqnarray}
\mbox{\boldmath $r$}_{\Lambda}^{i} & = & 
(r_{\Lambda}^{i1}, 
\cdots, r_{\Lambda}^{is}),\,\,\, \Lambda =1,\cdots, 2^{M}
\end{eqnarray}
where 
we defined $\Lambda$ as the index (entry) of the selected 
information vector $\mbox{\boldmath $\lambda$}$ and 
$s$ stands for the number of 
the possible strategies for each trader. 
Each component of the above strategy vector 
$\mbox{\boldmath $r$}_{\Lambda}^{i}$ takes 
$+1$ (buy) or $-1$ (sell).  
Therefore, each trader has her (his) own look-up table 
which is defined by a matrix with size $s \times 2^{M}$ as 
\begin{eqnarray}
\mbox{\boldmath $R$}^{i} \equiv 
\left(
\begin{array}{c}
\mbox{\boldmath $r$}_{1}^{i} \\
\mbox{\boldmath $r$}_{2}^{i} \\
\cdot \\
\cdot \\
\cdot \\
\mbox{\boldmath $r$}_{2^{M}}^{i}
\end{array}
\right) & = & 
\left(
\begin{array}{cccc}
r_{1}^{i1} & r_{1}^{i2} & \cdots & r_{1}^{is} \\
r_{2}^{i1} & r_{2}^{i2} & \cdots & r_{2}^{is} \\
\cdots & \cdots & \cdots & \cdots \\
\cdots &  \cdots & \cdots & \cdots \\
\cdots & \cdots & \cdots & \cdots \\
r_{2^{M}}^{i1} & r_{2^{M}}^{i2} & \cdots & r_{2^{M}}^{is} 
\end{array}
\right).
\end{eqnarray}
Each component of the above look-up table 
$r_{\Lambda}^{is}=\{+1,-1\}$ is 
fixed (`quenched') before playing the game. 
However, in the next section, 
we consider the case in which 
the components of look-up tables are rewritten 
during the game. 
In this paper, we concentrate ourselves 
to the simplest case of two strategies $s=2$ and 
$\zeta=0$ (`real' market history). 
Then, the trader $i$ changes 
her/his own pay-off value ${\cal P}_{ic}$ by the following 
update rule: 
\begin{eqnarray}
{\cal P}_{ic}(t+1) & = & 
{\cal P}_{ic}(t) - 
\frac{1}{\sqrt{N}}
{\cal B}_{i}(t) A(t) 
\label{eq:pa} \\
{\cal B}_{i}(t) & = & 
\sum_{\Lambda=1}^{2^{M}}
\delta (\Lambda,
\mbox{\boldmath $\lambda$}(A,Z:t))
r_{\Lambda}^{i \tilde{c}_{i}(t)}
\label{eq:bit2}
\end{eqnarray}
where $\delta (x,y)$ denotes the Kronecker's delta and 
$\tilde{c}_{i}(t)$ means the {\it optimal 
strategy} in the sense that $\tilde{c}_{i}(t)$ is 
given by 
\begin{eqnarray}
\tilde{c}_{i}(t) & = & 
{\arg\max}_{c} [{\cal P}_{ic}(t)].
\end{eqnarray}
The meaning of the update rule (\ref{eq:pa}) is given as follows. 
If $A(t)>0$ and the majority group consists of traders 
who post their decisions $+1$ to the market, 
the trader $i$ attempts to post her/his decision as 
an opposite sign of $A(t)$, namely, 
${\cal B}_{i}(t)=-1$. Thus, the trader $i$ acts so as to 
satisfy the condition ${\cal B}_{i}(t) A(t) <0$ which leads to 
increase of her/his pay-off value ${\cal P}_{ic}(t+1)$. 

By taking into account the fact that 
we are dealing with the case of $s=2$ 
($c=1,2$), 
we rewrite the equation 
(\ref{eq:pa}) as 
\begin{eqnarray}
q_{i}(t+1) & = & q_{i}(t) - 
\frac{1}{\sqrt{N}}
\sum_{\Lambda=1}^{2^{M}}
\delta (\Lambda, \mbox{\boldmath $\lambda$}(A, Z : t)) 
\eta_{\Lambda}^{i} A(t)
\label{eq:ql}
\end{eqnarray}
by means of 
\begin{eqnarray}
q_{i}(t) &  = &  
\frac{1}{2} ({\cal P}_{i1}(t)-{\cal P}_{i2}(t)),\,\,\,\,
\eta_{\Lambda}^{i}  =  
\frac{1}{2} 
(
r_{\Lambda}^{i1}
-
r_{\Lambda}^{i2}
).
\end{eqnarray}
Substituting 
(\ref{eq:bit2}) into the definition of 
the total bit $A(t)$, we have 
\begin{eqnarray}
A(t) & = & 
\frac{1}{\sqrt{N}}
\sum_{i=1}^{N}
{\cal B}_{i}(t) = 
\frac{1}{\sqrt{N}}
\sum_{i=1}^{N}
\sum_{\Lambda=1}^{2^{M}}
\delta (\Lambda, \mbox{\boldmath $\lambda$}(A,Z : t))  
r_{\Lambda}^{i\tilde{c}_{i}(t)}.
\end{eqnarray}
We should notice that 
the above equation can be written 
by using the following relation
\begin{eqnarray}
r_{\Lambda}^{i\tilde{c}_{i}(t)} & = & 
w_{\Lambda}^{i}
+{\rm sgn}[q_{i}(t)] 
\eta_{\Lambda}^{i} ,\,\,\,
w_{\Lambda}^{i} = 
\frac{1}{2}
(
r_{\Lambda}^{i1}
+
r_{\Lambda}^{i2}
).
\end{eqnarray}
Then, we obtain the following 
coupled non-linear equations with respect to the total decision $A(t)$, 
the difference of pay-off values for two strategies $q_{i}(t)$ and 
the update equation of the price $p(t)$: 
\begin{eqnarray}
A(t) & = & 
\frac{1}{\sqrt{N}}
\sum_{i=1}^{N}
\sum_{\Lambda=1}^{2^{M}}
\delta (\Lambda, 
\mbox{\boldmath $\lambda$} (A,Z : t))
\left\{
w_{\Lambda}^{i} 
+{\rm sgn}
[q_{i}(t)]
\eta_{\Lambda}^{i}
\right\} 
\label{eq:Al} \\
q_{i}(t+1) & = & q_{i}(t) - 
\frac{1}{\sqrt{N}}
\sum_{\Lambda=1}^{2^{M}}
\delta (\Lambda, \mbox{\boldmath $\lambda$} (A,Z : t)) 
\eta_{\Lambda}^{i} A(t) 
\label{eq:qi} \\
p(t+1) & = & 
p(t) +\beta \{
A(t) + \psi \, {\rm sgn}[A(t-1)]
\}.
\label{eq:xl}
\end{eqnarray}
The above rules (\ref{eq:qi})(\ref{eq:Al}) and (\ref{eq:xl}) 
are our basic equations to discuss the response of 
double-auction markets having stochastic Bid-Ask spread to instantaneous 
Selling-Buying signals. 
\subsection{Making of the Bid-Ask spread in the minority game}
To make the Bid-Ask spread in our minority game, 
we assume that 
the buying price $a_{it}$ and the selling price $b_{it}$ 
which are posted to the market by each trader $i$ at round (time) $t$ are updated 
according to the following rules.
\begin{eqnarray}
a_{it} & = & 
p (t) + \gamma_{a}\,g_{it} + \delta
\label{eq:ait}\\
b_{it} & = & 
p (t) + \gamma_{b}\,g_{it} - \delta
\label{eq:bit}
\end{eqnarray}
where $\gamma_{a},\gamma_{b}$ and $\delta$ are constants to be set so as to satisfy $a_{it}-b_{it} > 0$. 
$g_{it}$ is an uncorrelated Gaussian variable with mean $\langle g_{it} \rangle =0$ 
and covariance 
$\langle g_{it}g_{i^{'}t^{'}} \rangle =\delta_{t,t^{'}} \delta_{i,i^{'}}$ ({\it additive white Gaussian noise: AWGN}). 
In our simulations, we set 
$\gamma_{a}=\gamma_{b}=0.01, \delta = 0.049$. 
Then, the Bid-Ask spread at round $t$ is given by 
\begin{eqnarray}
S_{t} & = & 
{\min}\{a_{it} | a_{it} \in N_{+}\} -
{\max}\{b_{it} | b_{it} \in N_{-}\}
\end{eqnarray}
where  the groups taking  `buying' and `selling' decisions are 
refereed to as $N_{+}$ and 
$N_{-}$, respectively ($N \equiv N_{+}+N_{-}$). 

\begin{figure}[ht]
\begin{center}
\includegraphics[width=8cm]{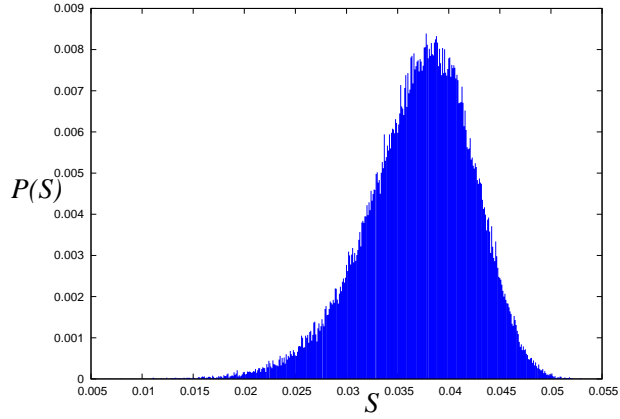}
\end{center}
\caption{\footnotesize 
Distribution of the spread $S=S_{t}$ generated by the minority game. }
\label{fig:fg_MGBA}
\end{figure}

In Fig. \ref{fig:fg_MGBA}, we plot the resulting distribution of the spread $S=S_{t}$. 
From  Fig.  \ref{fig:fg_MGBA}, 
we find that the stochastic Bid-Ask spread 
generated from the above modeling based on the minority game 
actually fluctuates and possesses a non-trivial shape of the distribution. 
\subsection{Results}
For the above set-up of the minority game, 
we evaluate two relevant statistics, 
namely, correlation function $C(t)$ (by (\ref{eq:empiricalC}))
and the response function $R(t)$ (by (\ref{eq:empiricalR})) to 
compare the results with the empirical evidence for the 
data with stochastic Bid-Ask spread.
\subsubsection{Auto-correlation  function}
\begin{figure}[ht]
\begin{center}
\includegraphics[width=6cm]{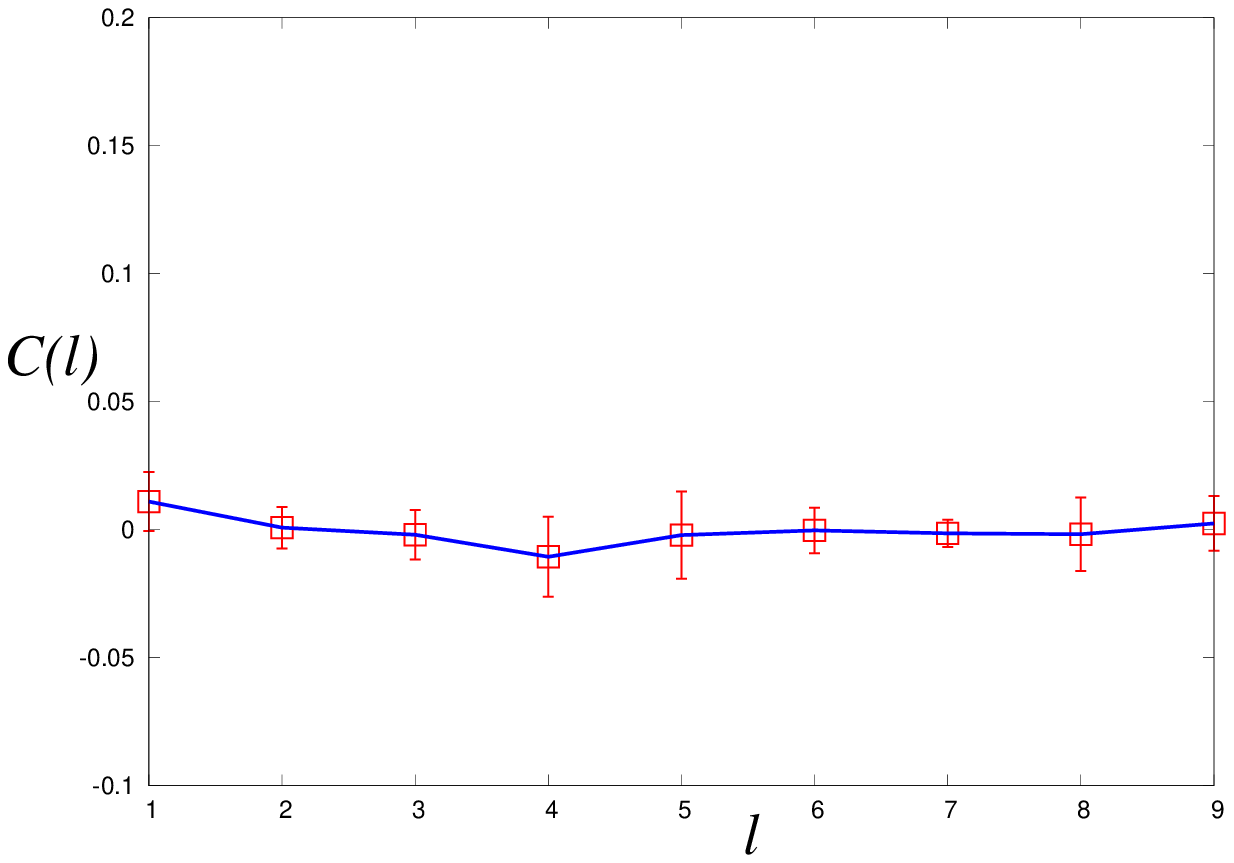}
\includegraphics[width=6cm]{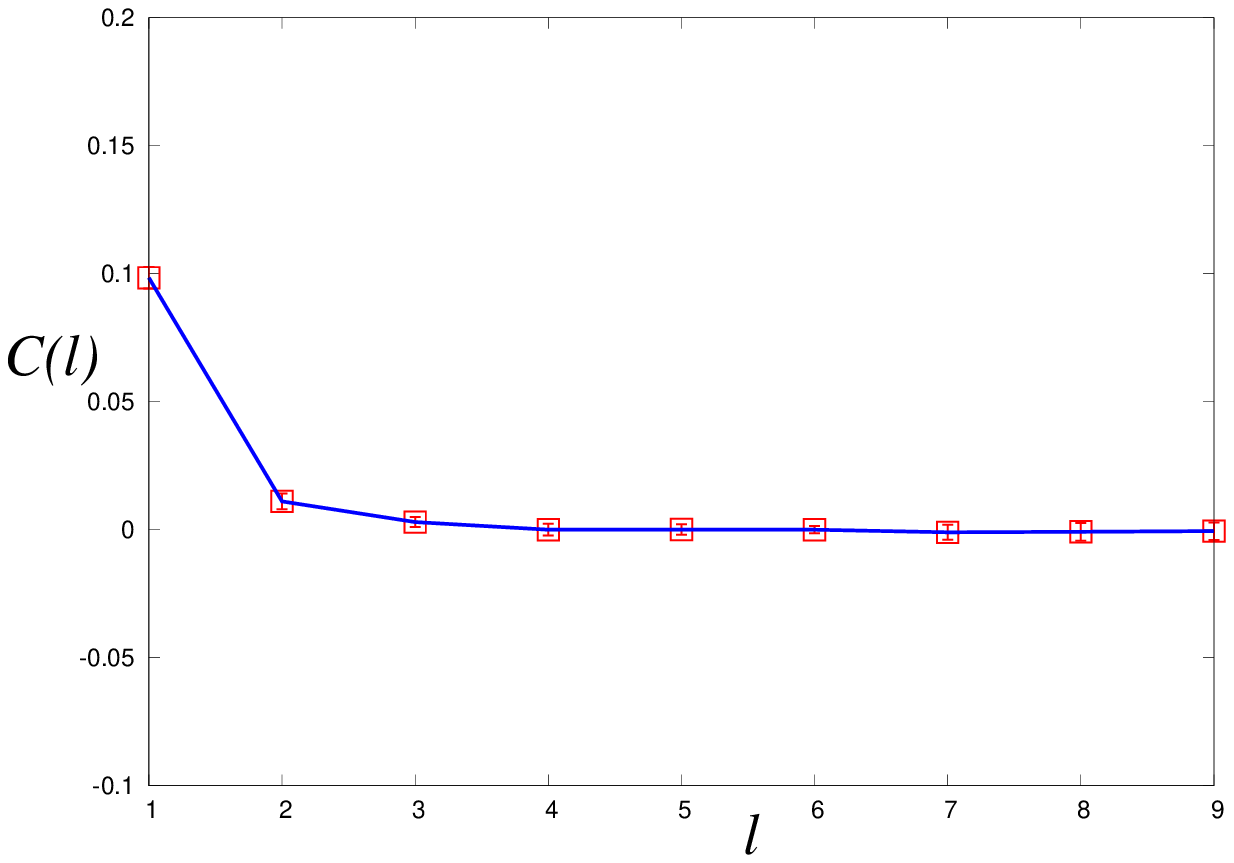}
\end{center}
\caption{\footnotesize
Auto-correlation function $C(l)$ in the minority game for 
$\psi=0$ (left). We set $N=1025, M=9, \beta=0.01$ and 
iterated the game $100010$ rounds. The $C(l)$ is zero for $l \geq 2$.
The right panel is auto-correlation function 
of the MRR model with $\rho=0.1$. The error-bars were calculated 
by 10-independent trials. 
}
 \label{fig:fg_C_l2}
\end{figure}

We first examine the effect of the bias term $\psi$ on the correlation function. 
The results for $\psi=0$ are shown in Fig. \ref{fig:fg_C_l2} (left). 
From this panel, we find that 
$C(l)=0$ for $l \geq 2$ and 
the result is apparently different from the result for 
the empirical data. 
Here we set 
$N=1025, M=9, \beta=0.01$ and 
iterated the game $t=100010$ rounds.   
The right panel of Fig. \ref{fig:fg_C_l2} shows 
the correlation function of the MRR model 
for $\rho=0.1$. 
The Selling-Buying signal 
\begin{eqnarray}
\epsilon_{t} & = & {\rm sgn}[A(t)]
\end{eqnarray}
is actually correlated automatically through the market history with 
length $M=9$, however, the correlation 
strength is very weak. 
Therefore, we need some other explicit 
correlation through the bias term $\psi$ which enhances the 
correlation by two-round back $A(t-1)$ from the present $t+1$. 
Hence, we here choose the non-zero bias term $\psi$ 
to reproduce the auto-correlation function 
as observed in the empirical data. 

\begin{figure}[ht]
\begin{center}
\includegraphics[width=6cm]{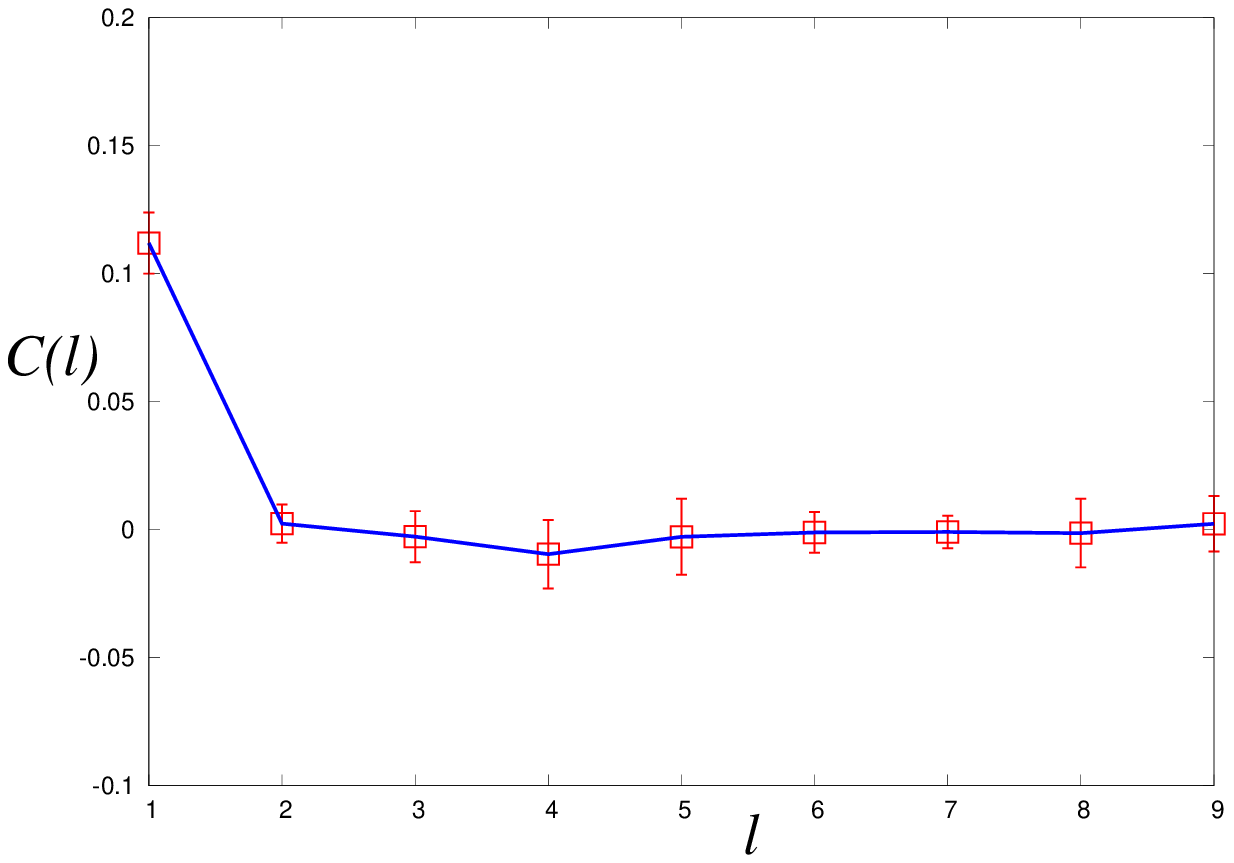}
\includegraphics[width=6cm]{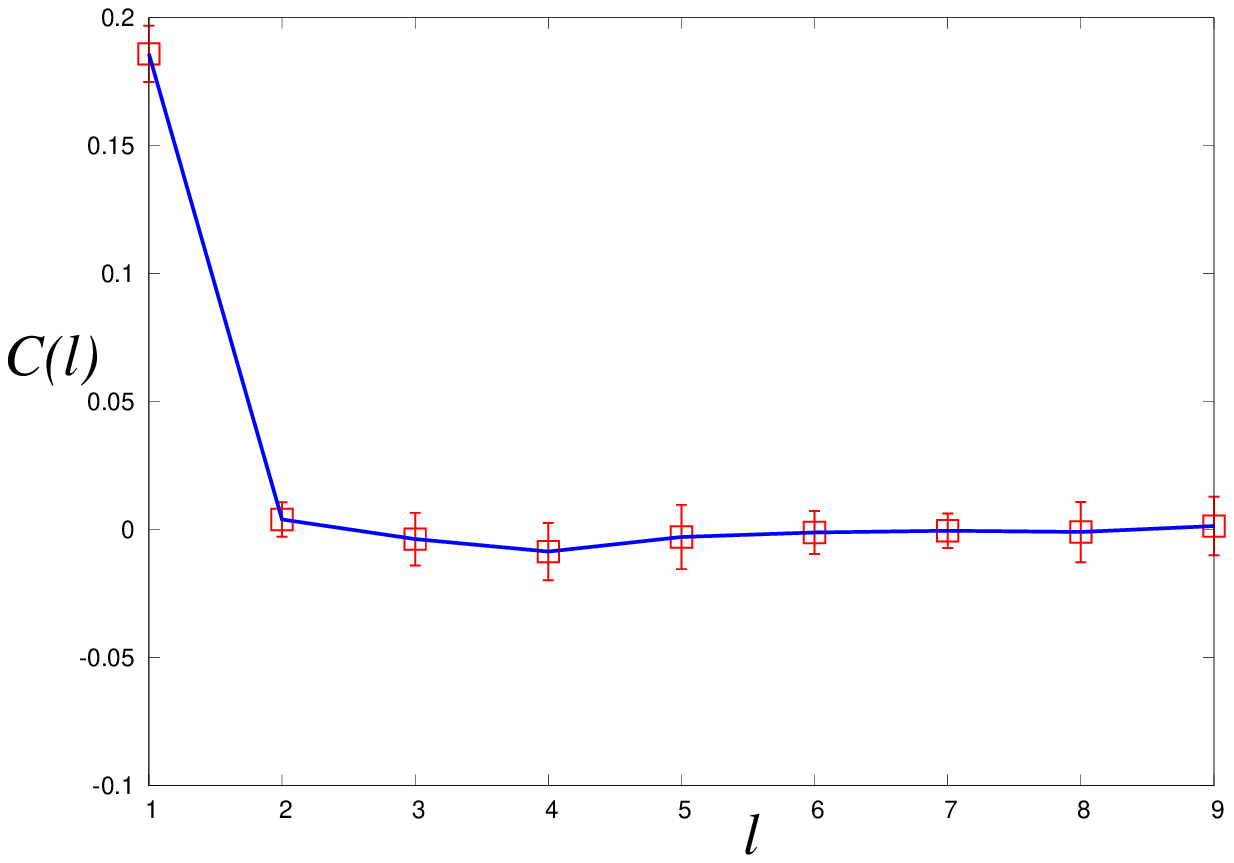}
\end{center}
\caption{\footnotesize
Auto-correlation function $C(l)$ in the 
minority game for 
$\psi=0.05$ (left) and $\psi=0.1$ (right). 
We set $N=1025, M=9, \beta=0.01$ and 
iterated the game for $100010$ rounds. 
The error-bars were calculated by 10-independent trials. 
}
 \label{fig:fg_C_l}
\end{figure}

We checked that the results are robust against 
the slight differences in the parameters 
appearing in the game such as $\beta, M, \delta$ {\it etc}. 
However, for only parameter $\psi$, we should be careful 
to choose the value. This is because 
from the definition of 
update rule of the price (\ref{eq:xl}), 
the effect of the $A(t)$ on 
the price change is relatively depressed by the large value of 
the $\psi$. Therefore, we should choose $\psi$ 
so as to make the value smaller than 
the standard deviation of 
the $A(t)$, namely, square root of the volatility as 
\begin{eqnarray}
\psi & < &  
\sqrt{
\frac{1}{T}
\sum_{t=1}^{T}
\{
A(t)-\overline{A(t)}
\}^{2}} \equiv \sigma_{A}.
\label{eq:cond_psi}
\end{eqnarray}
In our simulation, the square root of the volatility is estimated as $\sigma_{A}=0.44295$. 
In Fig. \ref{fig:fg_C_l}, we plot the 
correlation function for 
the case of $\psi=0.05$ (left panel) and 
$\psi=0.1$ (right). 
We should notice that 
these two choices of 
the bias term $\psi$ satisfy the condition (\ref{eq:cond_psi}).  
\mbox{}

From these panels, we find that the 
correlation function decreases as we observed 
in the same function for the empirical data sets.
\subsubsection{Response function}
\begin{figure}[ht]
\begin{center}
\includegraphics[width=6cm]{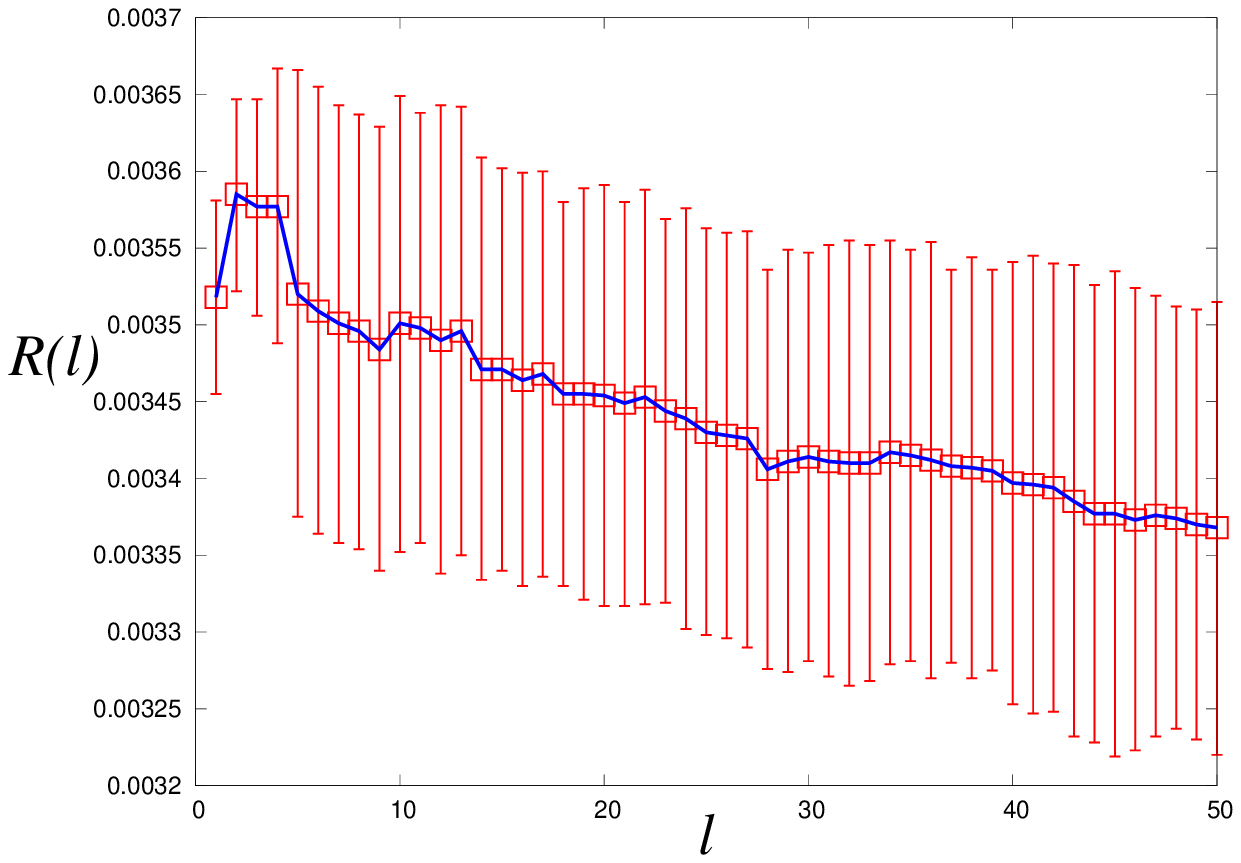}\\
\includegraphics[width=6cm]{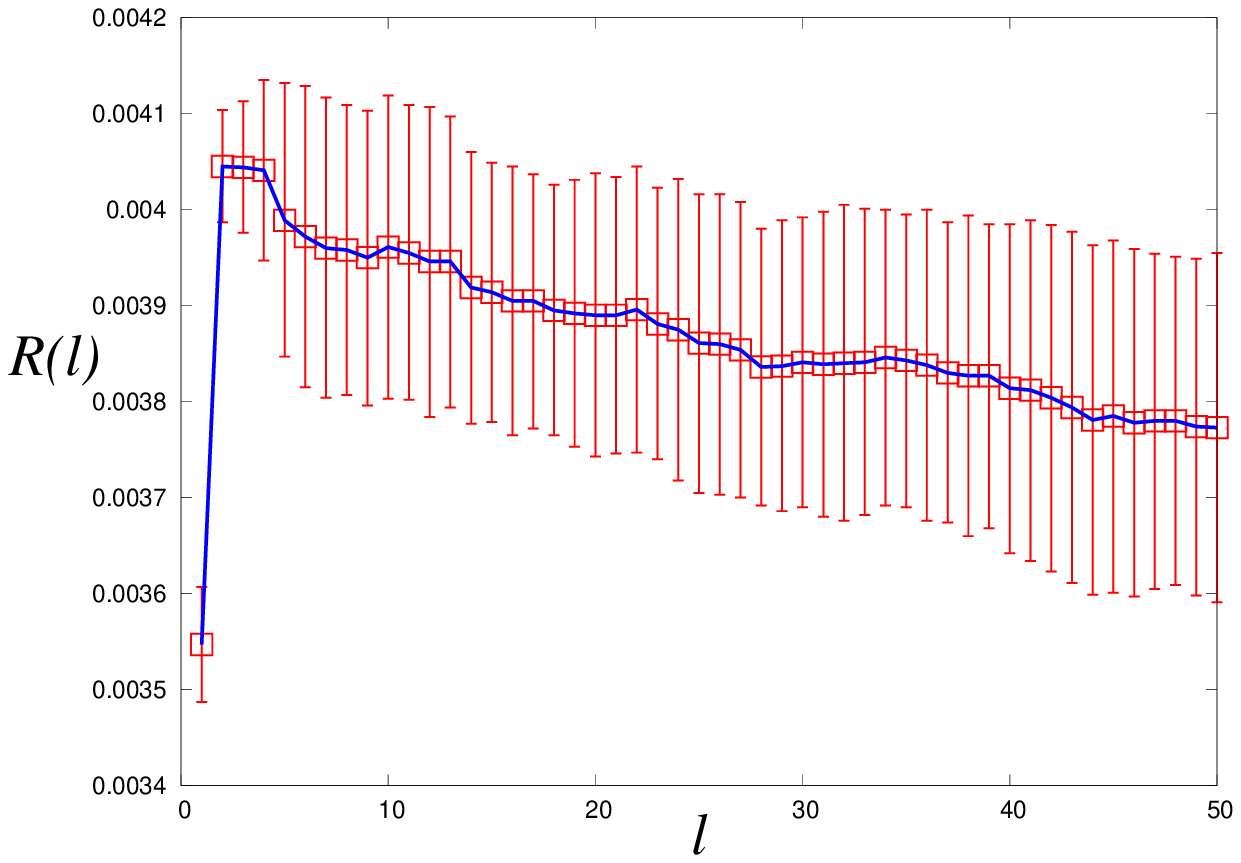}
\includegraphics[width=6cm]{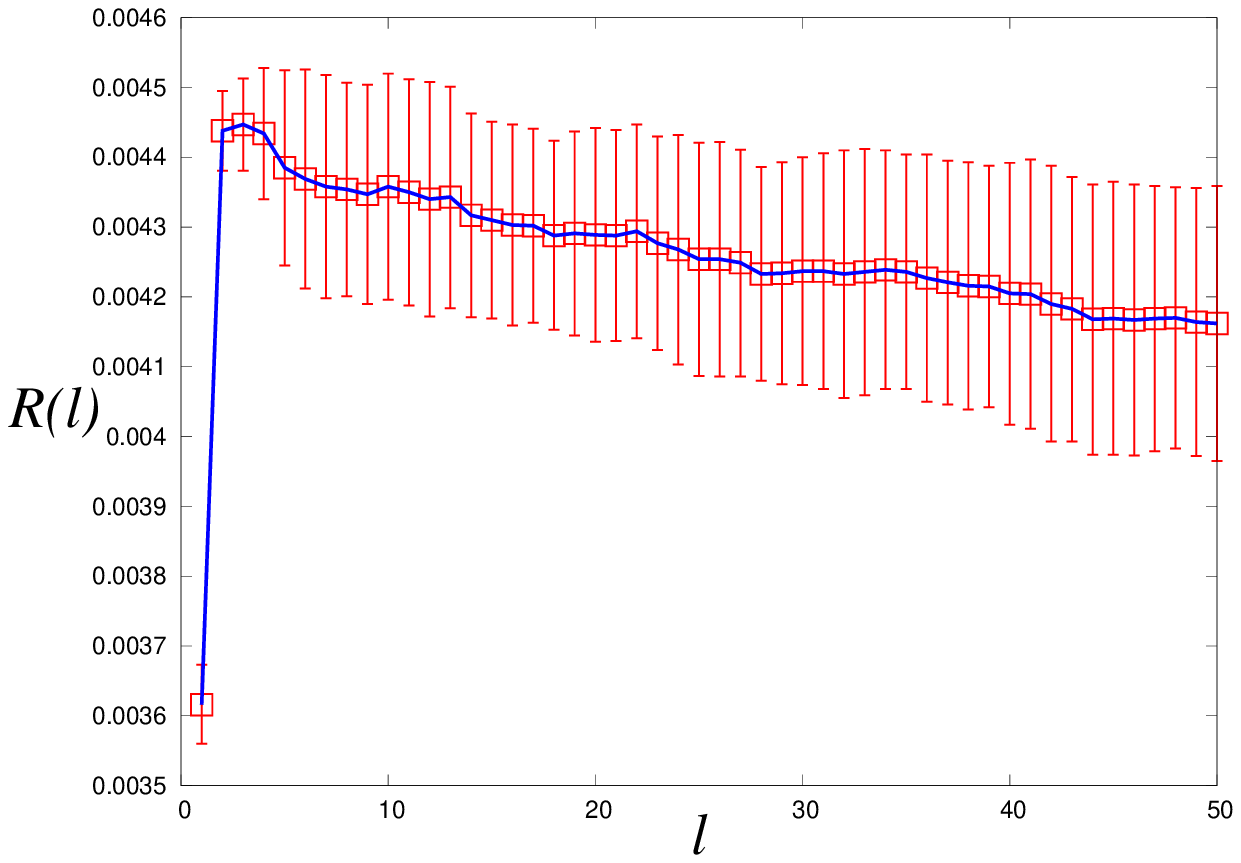}
\end{center}
\caption{\footnotesize
Response function $R(l)$ in the 
minority game for $\psi=0$ (upper), $\psi=0.05$ (lower left) and $\psi=0.1$ (lower right). 
We set 
$N=1025, M=9, \beta=0.01$ and 
iterated the game for $100010$ rounds. 
The error-bars were calculated by 10-independent trials. 
}
 \label{fig:fg_R_l}
\end{figure}

We next plot the response function 
in Fig. \ref{fig:fg_R_l} for 
$\psi=0.$(upper panel), $\psi=0.05$ (lower left) and 
$\psi=0.1$ (lower right). 
\begin{figure}[ht]
\begin{center}
\includegraphics[width=6cm]{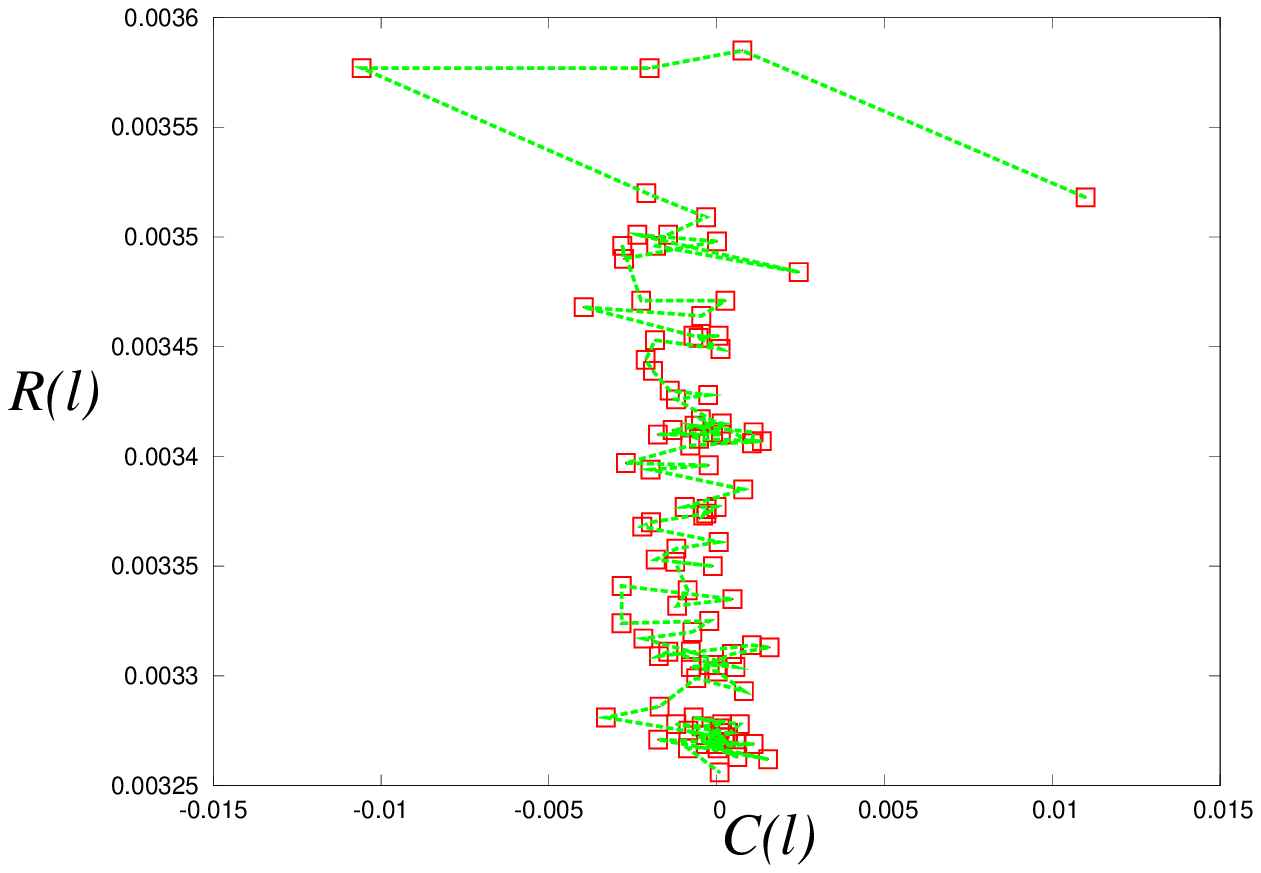}\\
\includegraphics[width=6cm]{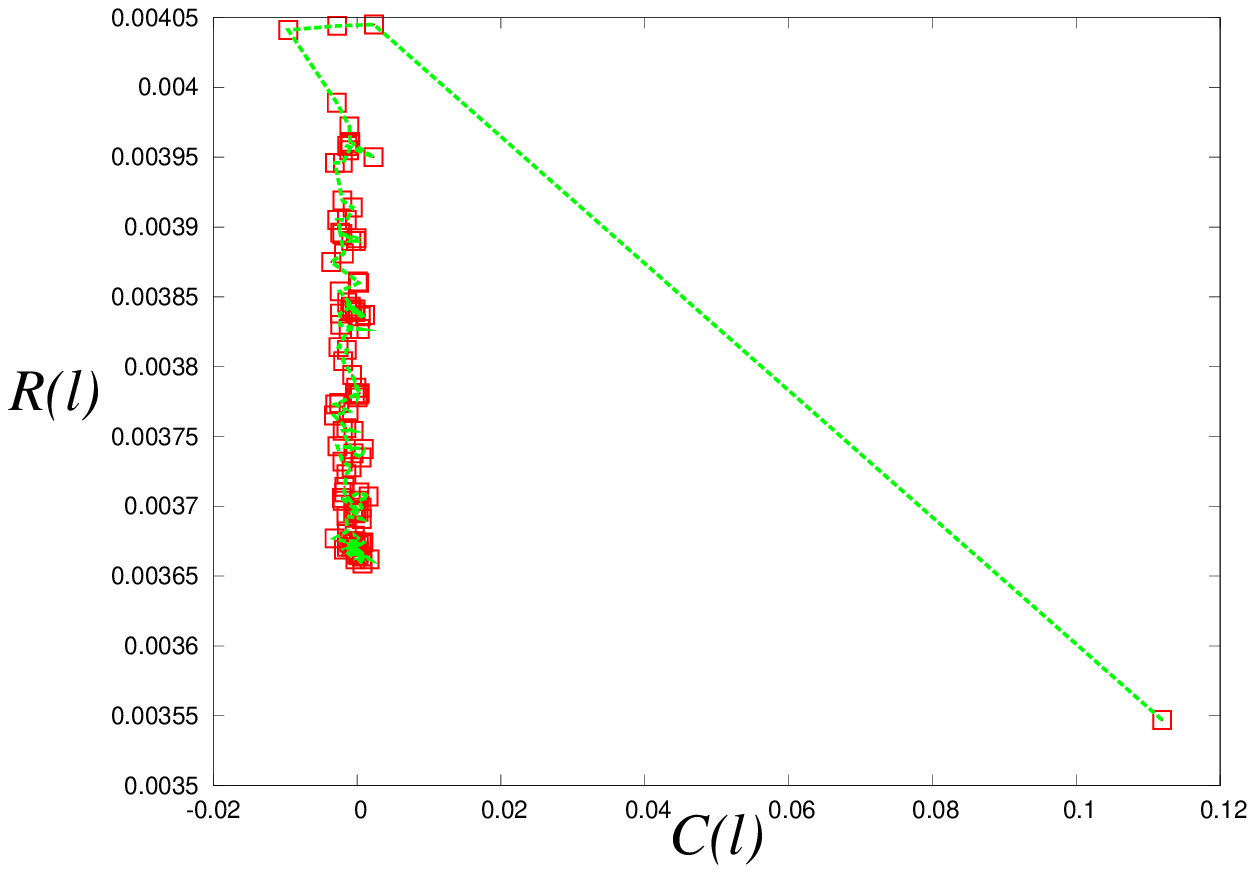}
\includegraphics[width=6cm]{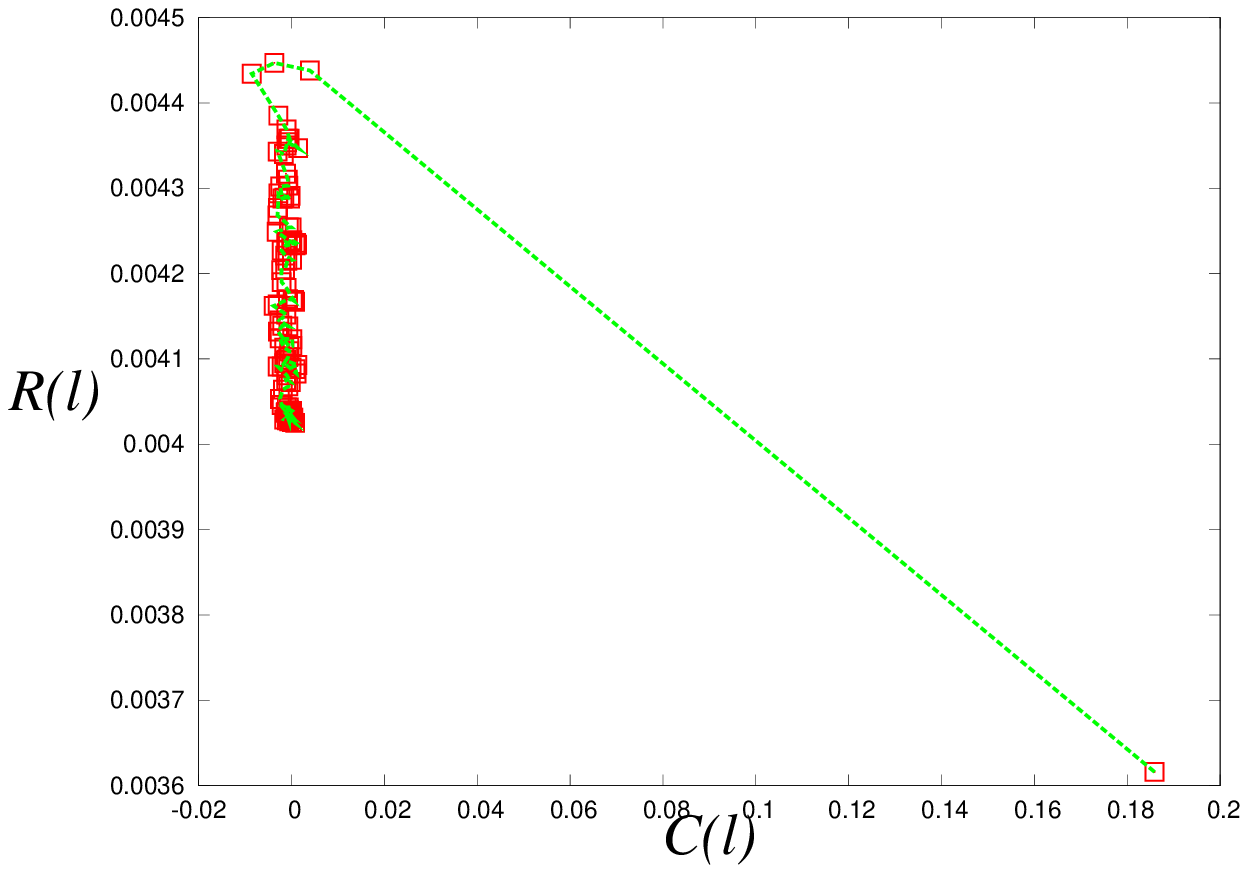}
\end{center}
\caption{\footnotesize
The relationship between $C(l)$ and $R(l)$ for 
minority game with $\psi=0$ (upper), $\psi=0.05$ (lower left) and $\psi=0.1$ (lower right). 
We set 
$N=1025, M=9, \beta=0.01$ and 
iterated the game for $100010$ rounds and 
we calculate $C(l),R(l)$ up to $l=100$. 
}
 \label{fig:fg_C_R}
\end{figure}
From these panels, we find that 
the behaviour of the response function 
is not monotonically increasing function 
leading up to the convergence to some constant value 
but `non-monotonic' as the response function of 
data sets having stochastic Bid-Ask spread 
(EUR/JPY, USD/JPY exchange rates) shows (see the lower panels of 
Fig. \ref{fig:fg_empirical}). 
\subsubsection{Relationship between the auto-correlation and response functions}
In Fig. \ref{fig:fg_C_R}, 
we plot the relationship between the auto-correlation and response functions. 
In the upper panel, we show the result for zero bias term $\psi=0$. 
we find that the curve is deviated from the 
linear relation $R(l) \propto -C(l)$. However, 
the shape is not `$\lambda$' as observed in the empirical evidence but `$T$-shape'. 
In the lower two panels are results for 
the non-zero 
bias term $\psi \neq 0$. 
We clearly find that the `$\lambda$-shapes' appears 
and the results are qualitatively similar to those of the empirical data. 
\section{Adaptive look-up tables}
\label{sec:adapt}
In the previous section, we fixed each decision component 
(buying: $+1$, selling:  $-1$) in 
their look-up tables before playing the game 
(in this sense, the decision components are `quenched variables' 
in the literature of disordered spin systems such as spin glasses).  
However, in this paper, a certain amount of traders update their 
decision components according to the macroscopic 
market history (they `learn' from the behaviour of markets) 
so as to be categorized into two-groups with a finite probability
(in this sense, the components are now regarded as `annealed variables').  
Namely, at each round of the game, if the number of sellers is smaller/greater than 
that of buyers, a fraction of  traders, what we call {\it optimistic group}/{\it pessimistic group}, 
is more likely to rewrite their own decision components from $-1$/$+1$ to $+1$/$-1$. 
In realistic trading, we might change our mind and rewrite the components 
of the look-up table according to the market history. 
Therefore, in this section 
we consider the case in which some amount of traders can 
rewrite their own table adaptively. 
\subsection{Adaptation using the latest market information}
We first consider the case in which 
each trader updates her/his own 
look-up table 
according to the latest information of the market. 
Some of the traders make their decisions  
as `buy' when the market is {\it seller's market}, namely, the signal of the latest market is `buying' 
(what we call {\it optimistic group}). 
On the other hand, 
they decide `sell' vice versa 
(they are referred to as {\it pessimistic group}). 
Namely, 
each trader rewrites the table according to 
the following algorithm. 
\\
\\
{\bf Adaptation algorithm using the latest market information} 
\begin{enumerate}
\item[(i)]
Fix (`quench') each component of the look-up table 
at the beginning of the game $t=-M$. 
\item[(ii)]
At each game round $t$ for $t > -M$, each trader rewrites her/his component 
$r_{\Lambda}^{i1}, r_{\Lambda}^{i2}$, 
where  
$\Lambda$ denotes the entry of market history for 
the information vector $\mbox{\boldmath $\lambda$}(A, Z: t)$,  
with probability $f_{1}$ as 
\[
r_{\Lambda}^{i1} =  {\rm sgn}[(A(t-1)],\,\,\,
r_{\Lambda}^{i2}  = {\rm sgn}[A(t-1)].
\]
\item[(iii)]
Each trader recovers her/his original (at the beginning 
of the game) look-up table with 
a probability $f_{2}$ at the next game round $t+1$.
\item[(iv)]
Repeat (ii) and (iii) until the game is over. 
\end{enumerate} 
Namely, 
a fraction $\sim Nf_{1}$ of the traders 
is categorized into the `optimistic group' if 
$A(t-1)>0$ ({\it seller's market}) and into the `pessimistic group' if $A(t-1)<0$ 
({\it buyer's market}). 
\begin{figure}[ht]
\begin{center}
\includegraphics[width=6cm]{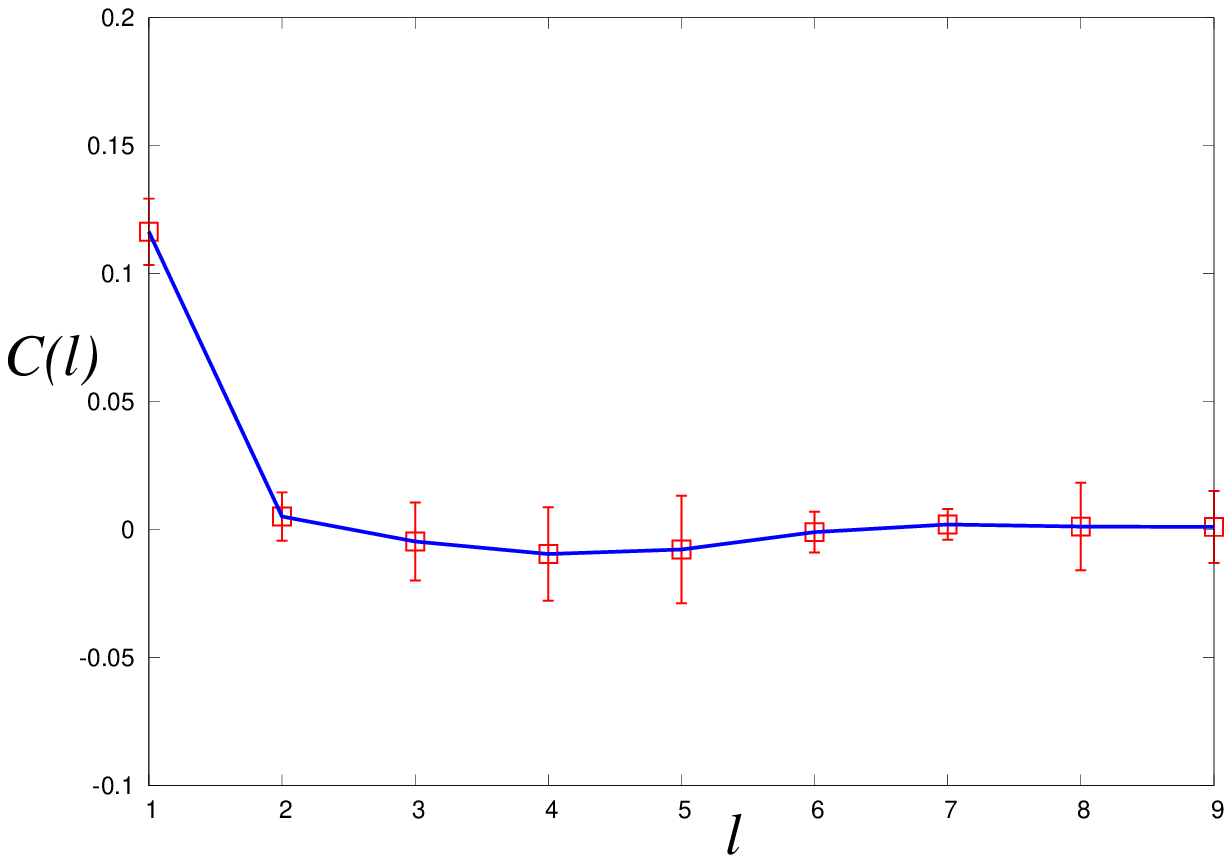} 
\includegraphics[width=6cm]{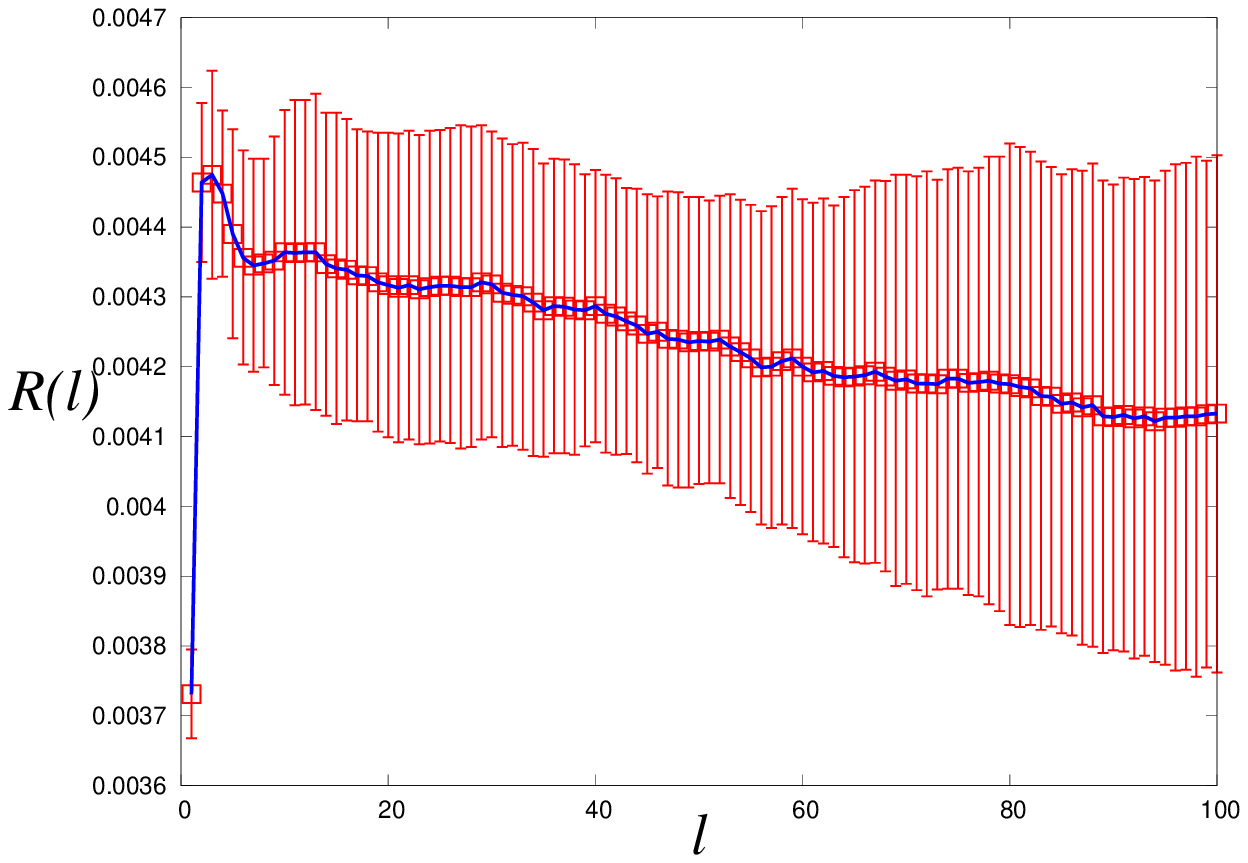}\\
\includegraphics[width=6cm]{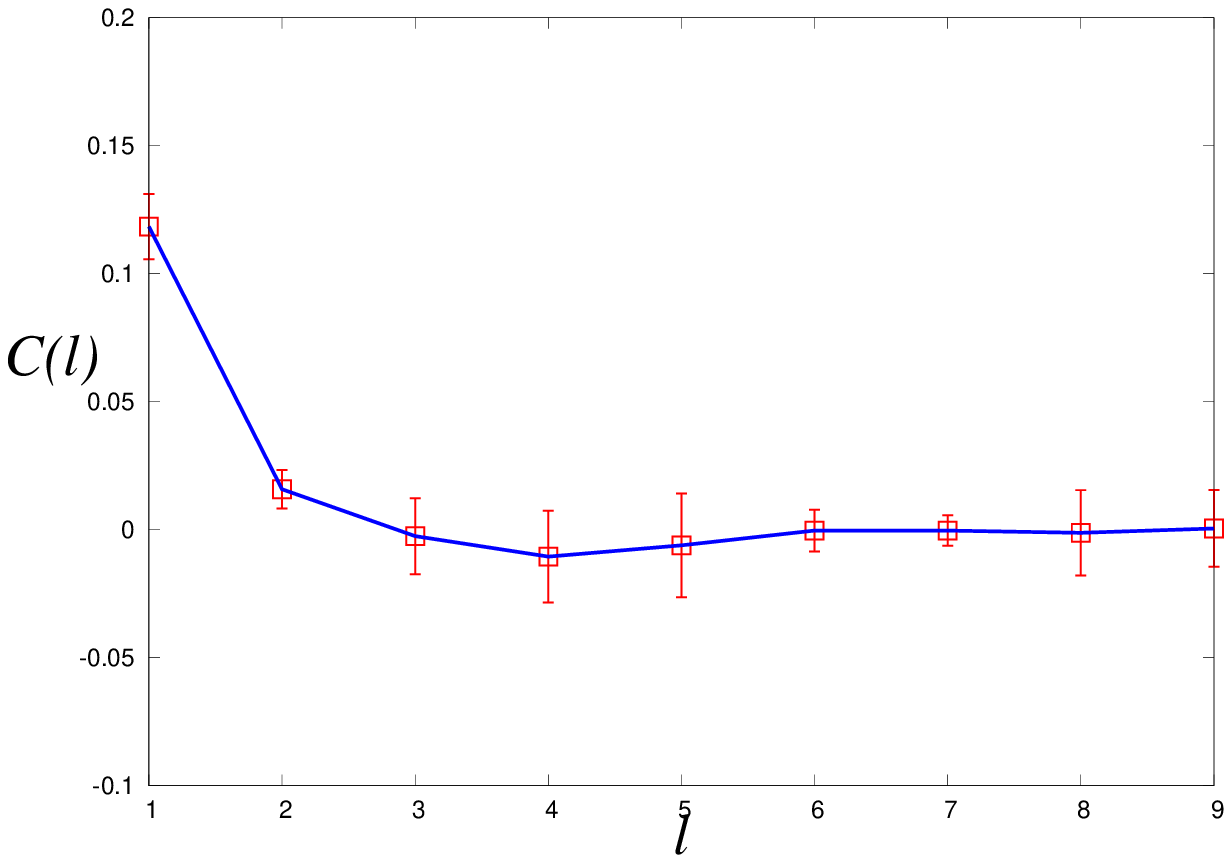} 
\includegraphics[width=6cm]{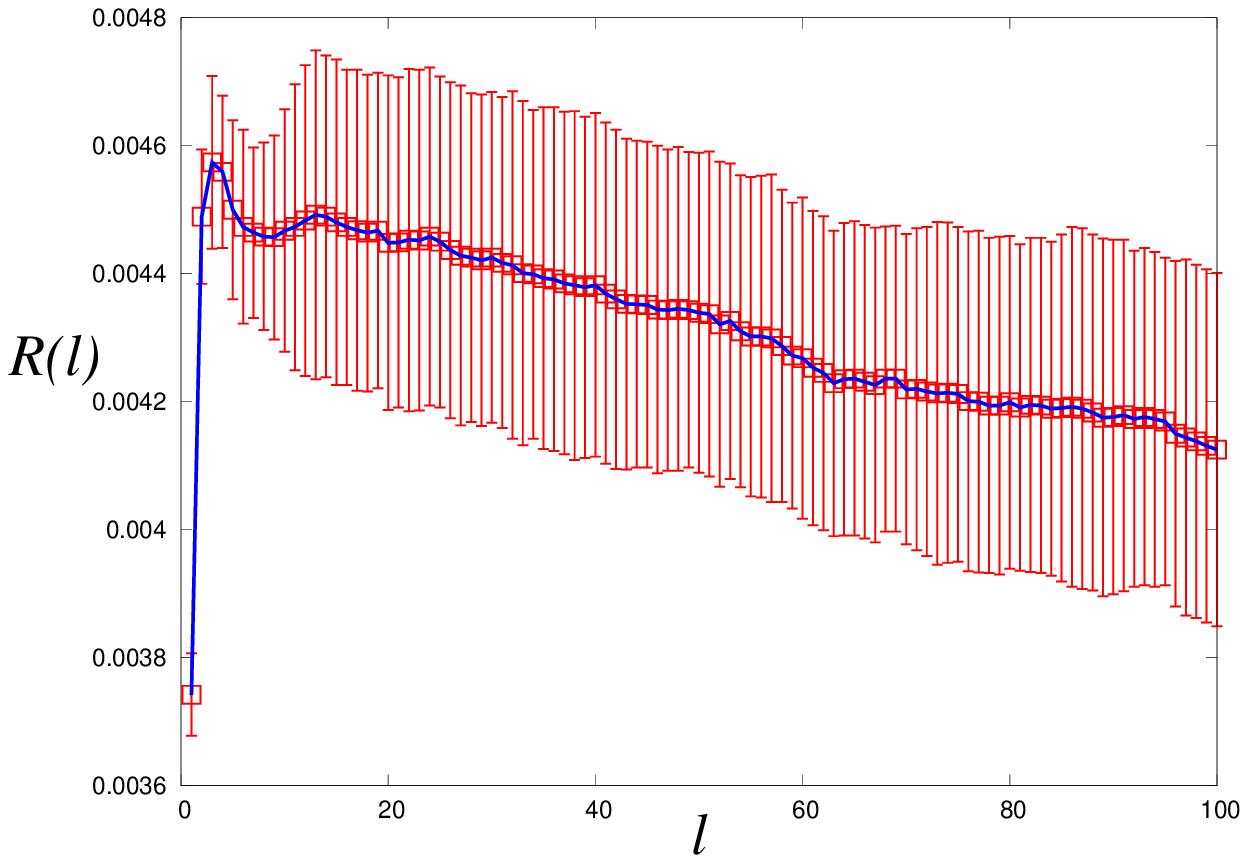} \\
\includegraphics[width=6cm]{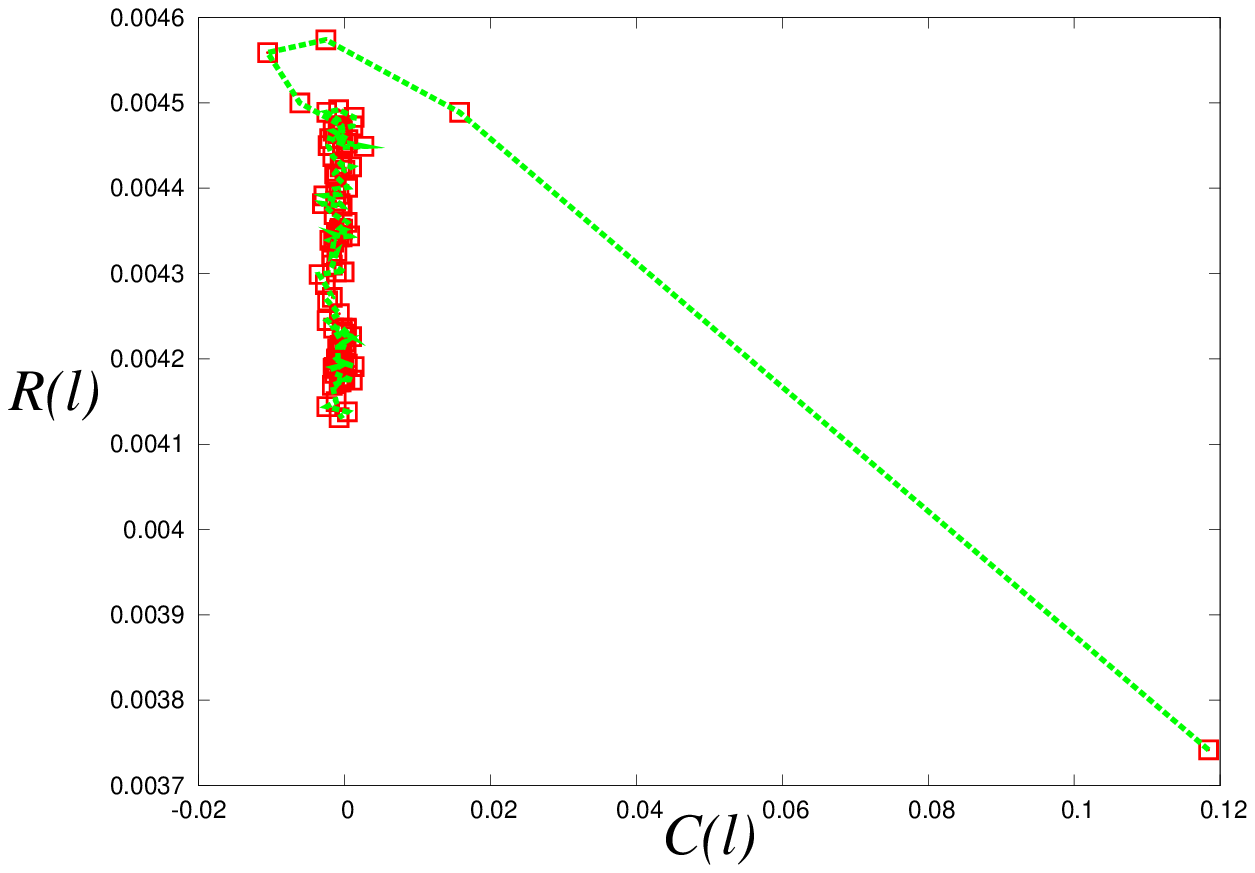} 
\includegraphics[width=6cm]{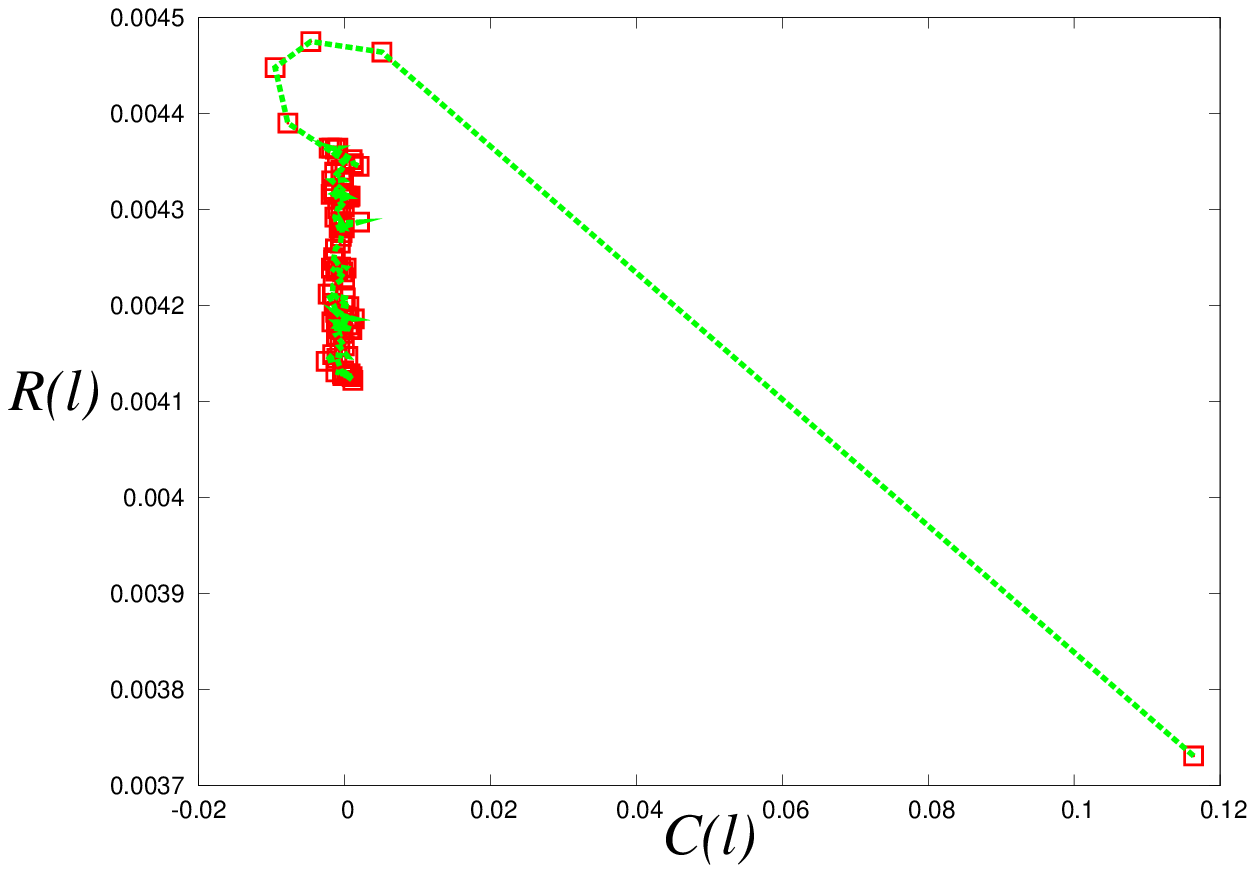}
\end{center}
\caption{\footnotesize 
Dynamical quantities $C(l)$ (upper left, middle left) and 
$R(l)$ (upper right, middle right) 
evaluated for the minority game with adaptive look-up tables.
We set  
$\psi=0, f_{1}=0.01$. $f_{2}=1$ (upper panels), $f_{2}=0.9$ (middle panels). 
The lower two panels show the relationship between 
the auto-correlation and response functions 
for $(f_{1},f_{2})=(0.01,1)$ (left) and $(f_{1},f_{2})=(0.01,0.9)$ (right).  
}
\label{fig:R_mgame_simu3}
\end{figure}
\subsubsection{Results}
In Fig. \ref{fig:R_mgame_simu3}, 
we show dynamical quantities $C(l)$ (upper left, middle left) and 
$R(l)$ (upper right, middle right) 
evaluated for the minority game with adaptive look-up tables.
We set  
$\psi=0, f_{1}=0.01$ and $f_{2}=1$ (upper panels), $f_{2}=0.9$ (middle panels). 
The lower two panels show the relationship between 
the auto-correlation and response functions 
for $(f_{1},f_{2})=(0.01,1)$ (left) and $(f_{1},f_{2})=(0.01,0.9)$ (right).  
From this figure, 
we find that the positive correlation $C(l)>0$  
for $l=1$ appears even if we set $\psi=0$ and 
non-monotonic behaviour of the response function 
is reproduced. 
As the result, `$\lambda$-shape' in 
the $C(l)$-$R(l)$ scatter plots is generated. 
From these results, we conclude that 
the adaptive modification of the look-up table 
by using the latest market information for 
each trader works well to explain the empirical evidence. 
\subsection{Adaptation by using the market history}
In the previous subsection, we succeeded in 
generating a positive finite auto-correlation 
by making use of the adaptive look-up table 
even if we set $\psi=0$. 
However, in this look-up table, each trader changes 
her/his decision from the latest information about the market. 
As the result, 
the auto-correlation function 
decays to zero for $l \geq 2$. 
To modify the weak correlation, we 
construct the adaptive look-up table by using 
the information about the market history with length $M$. 
As we mentioned, 
the information vector 
$\mbox{\boldmath $\lambda$}(A,Z : t)$ 
contains the 
useful information 
on the market.  Hence, we shall assume that each trader rewrites the component of her/his 
table as 
\begin{eqnarray}
r_{\Lambda}^{i1} = {\rm sgn}[\Omega_{\Lambda}(A,Z: t)],\,\,\, 
r_{\Lambda}^{i2}  = {\rm sgn}[\Omega_{\Lambda}(A,Z: t)]
\end{eqnarray} 
with probability 
\begin{eqnarray}
f_{3} & = & \frac{\alpha |\Omega_{\Lambda}(A,Z : t)|}{2^{M}} 
\end{eqnarray}
where we defined 
\begin{eqnarray}
\Omega_{\Lambda}(A,Z:t) & \equiv &  
\mbox{\boldmath $w$} \cdot 
\mbox{\boldmath $\lambda$}(A,Z:t) = 
\sum_{\tau=1}^{M}2^{M-\tau}{\rm sgn}[A(t-\tau)] \\
\mbox{\boldmath $w$}  & \equiv  & (2^{M-1},2^{M-2},\cdots,2^{0})
\end{eqnarray}
at each game round $t$. 
We should keep in mind that 
each trader recovers her/his original look-up table with 
a probability $f_{2}$ at the next game round. 
The $\Omega_{\Lambda}(A,Z:t)$ denotes 
cumulative weighted market status, 
namely, we assume that 
the importance of the market information 
decays as $2^{-\tau}$ in the history length $\tau$. 

For instance, 
for the information vector (let us define the 
entry by $\Lambda$) having $+1$ for 
all components:  
$\mbox{\boldmath $\lambda$}(A,Z: t) = 
({\rm sgn}[A(t-1)],\cdots, {\rm sgn}[A(t-M)])=
(+1,\cdots,+1)$, that is, 
the market remains as 
{\it seller's market} up to 
$M$-times back, 
we obtain 
$\Omega_{\Lambda} (A,Z: t) = 2^{M}-1>0$ and 
$Nf_{3}=N\alpha(1-2^{-M})$-traders 
rewrite their components as $r_{\Lambda}^{i1},r_{\Lambda}^{i2}=
{\rm sgn}[\Omega_{\Lambda}(A,Z,t)]=1$. 
As another example, 
when 
{\it seller's market} and {\it buyer's market} 
appears periodically as 
$\mbox{\boldmath $\lambda$}(A,Z: t) =
(-1,+1,-1,+1,\cdots)$, 
we have 
$\Omega_{\Lambda} (A,Z: t)=-\{2^{M}-(-1)^{M}/2\}/3<0$ 
and $Nf_{3}=N\alpha \{1-(-1)^{M}/2^{M+1}\}/3$-traders 
rewrite their component as 
$r_{\Lambda}^{i1},r_{\Lambda}^{i2}=
{\rm sgn}[\Omega_{\Lambda}(A,Z:t)]=-1$. 

Let us summarize the above procedure 
as the following algorithm. 
\\
\\
{\bf Adaptation algorithm using the market history} 
\begin{enumerate}
\item[(i)]
Fix (`quench') each component of the look-up table 
at the beginning of the game $t=-M$. 
\item[(ii)]
At each game round $t$ for $t > -M$, each trader rewrites her/his component 
$r_{\Lambda}^{i1}, r_{\Lambda}^{i2}$, 
where  
$\Lambda$ denotes the entry of market history for 
the information vector $\mbox{\boldmath $\lambda$}(A, Z: t)$,  
with probability as 
\[
r_{\Lambda}^{i1} =  {\rm sgn}[\Omega_{\Lambda}(A,Z:t)],\,\,\,
r_{\Lambda}^{i2}  = {\rm sgn}[\Omega_{\Lambda}(A,Z,t)]
\]
with probability 
\[
f_{3}=\frac{\alpha |\Omega_{\Lambda}(A,Z:t)|}{2^{M}}
\]
with 
$\Omega_{\Lambda}(A,Z:t)=\sum_{\tau=1}^{M}2^{M-\tau}{\rm sgn}[A(t-\tau)]$. 
\item[(iii)]
Each trader recovers her/his original (at the beginning 
of the game) look-up table with 
a probability $f_{2}$ at the next game round $t+1$.
\item[(iv)]
Repeat (ii) and (iii) until the game is over. 
\end{enumerate} 
\begin{figure}[ht]
\begin{center}
\includegraphics[width=6cm]{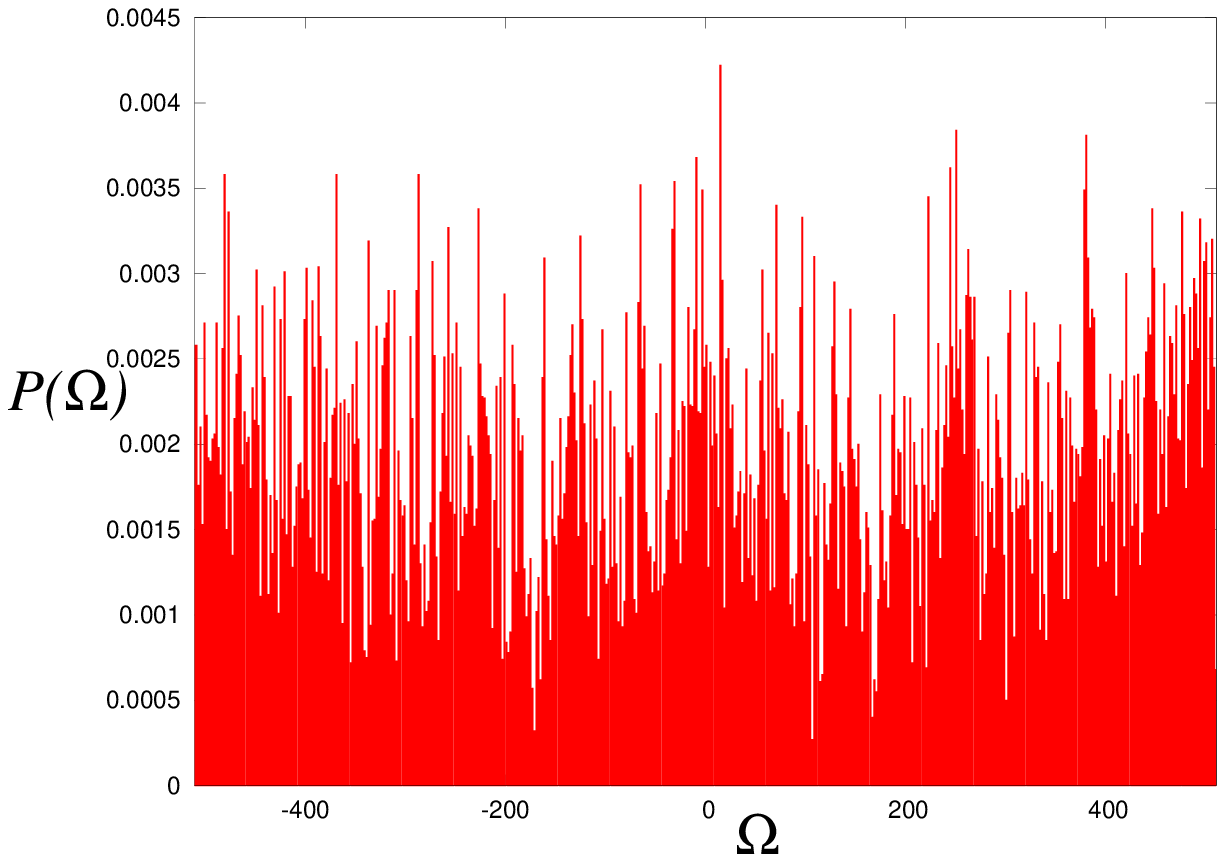}
\includegraphics[width=6cm]{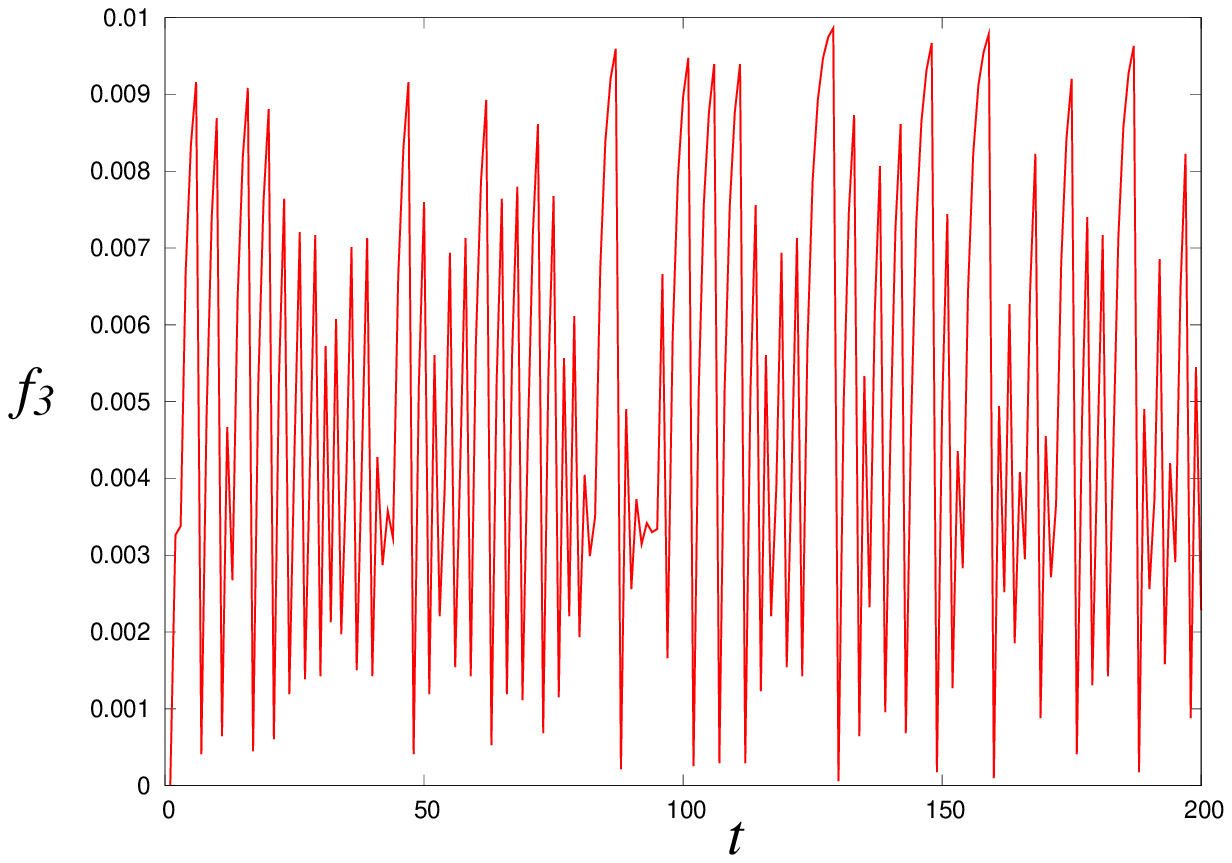}
\end{center}
\caption{\footnotesize 
The generated distribution 
$P(\Omega)$ 
of $\Omega_{\Lambda}(A,Z:t)=\Omega$ (left) and 
the typical time-evolution of 
the probability  
$f_{3}$ for  the first $200$-steps (right). 
}
 \label{fig:fg_Omega_f3}
\end{figure}
\mbox{}

We set $\alpha=0.01$ and 
select the same values as those of the previous section for the other parameters. 
\subsubsection{Results}
In Fig. \ref{fig:fg_Omega_f3}, 
we first plot the 
generated distribution 
$P(\Omega)$ 
of $\Omega_{\Lambda}(A,Z:t)=\Omega$ (left) and 
the typical time-evolution of 
the probability  $f_{3}$ for  the first $200$ steps (right).
It should be noted that 
$\Omega$ in 
the left panel is ranged from 
$\Omega_{\rm min}=\sum_{\tau=1}^{M}2^{M-\tau}(-1)=1-2^{9}=-511$ to 
$\Omega_{\rm max}=\sum_{\tau=1}^{M}2^{M-\tau}(+1)=2^{9}-1=511$ because 
we set the history length $M=9$ 
in the definition of the $\Omega_{\Lambda}(A,Z:t)$. 
We find from the right panel that 
the typical time-evolution of the 
probability $f_{3}$ obeys complicated dynamics.  

\begin{figure}[ht]
\begin{center}
\includegraphics[width=6cm]{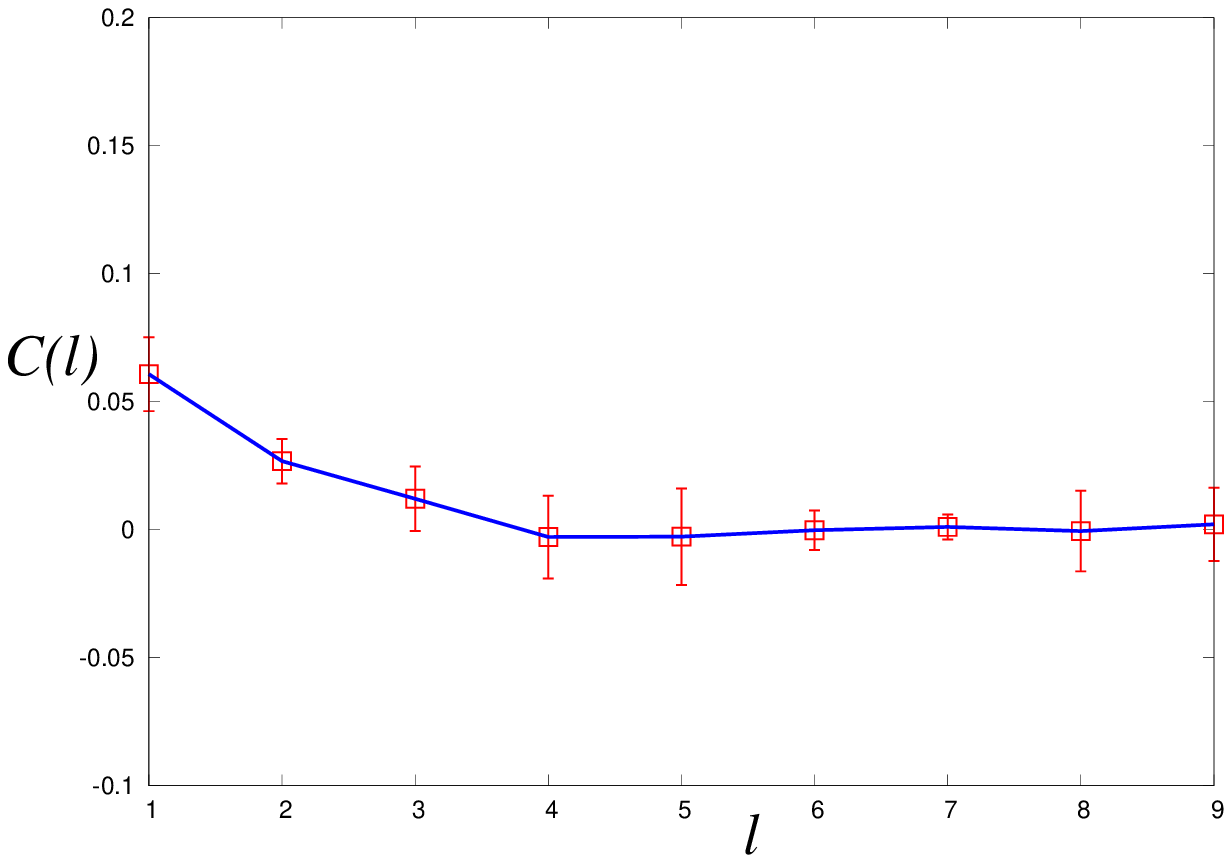} 
\includegraphics[width=6cm]{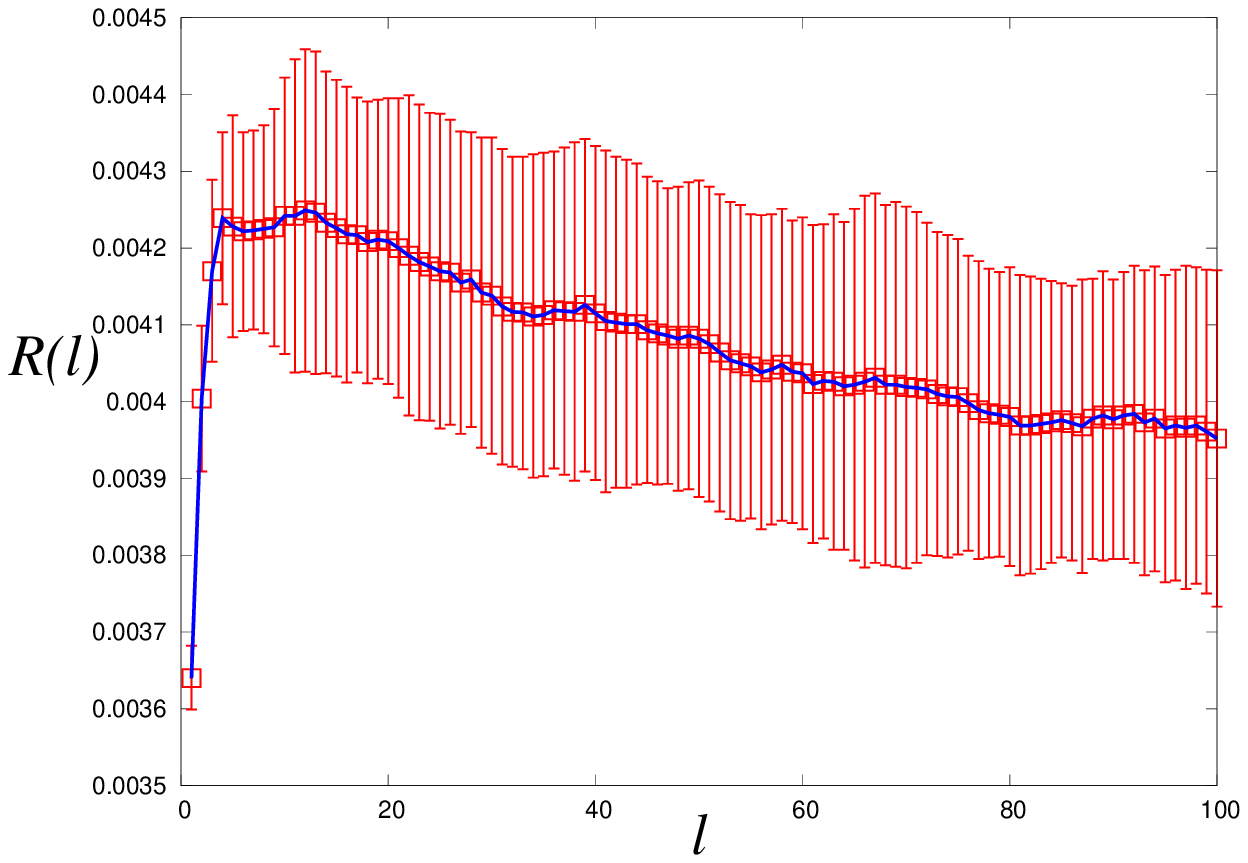}\\
\includegraphics[width=6cm]{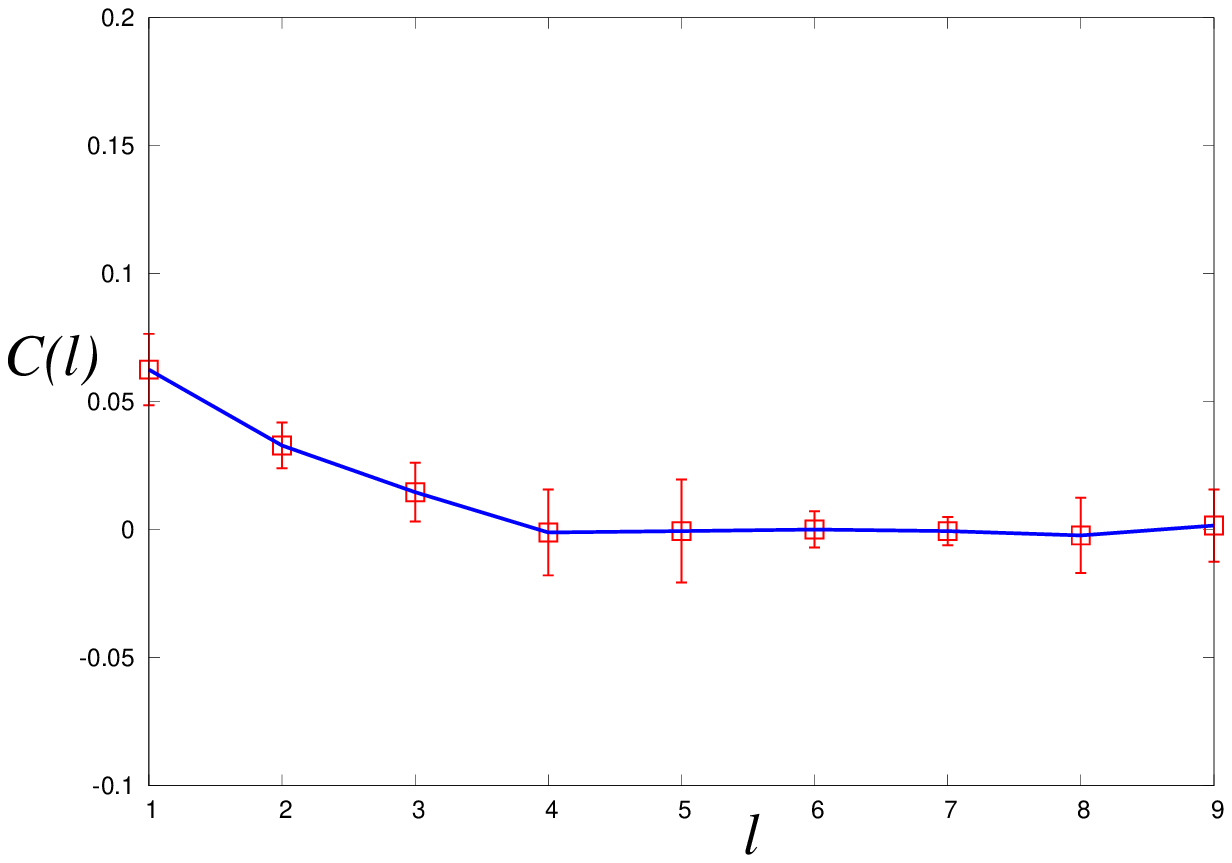} 
\includegraphics[width=6cm]{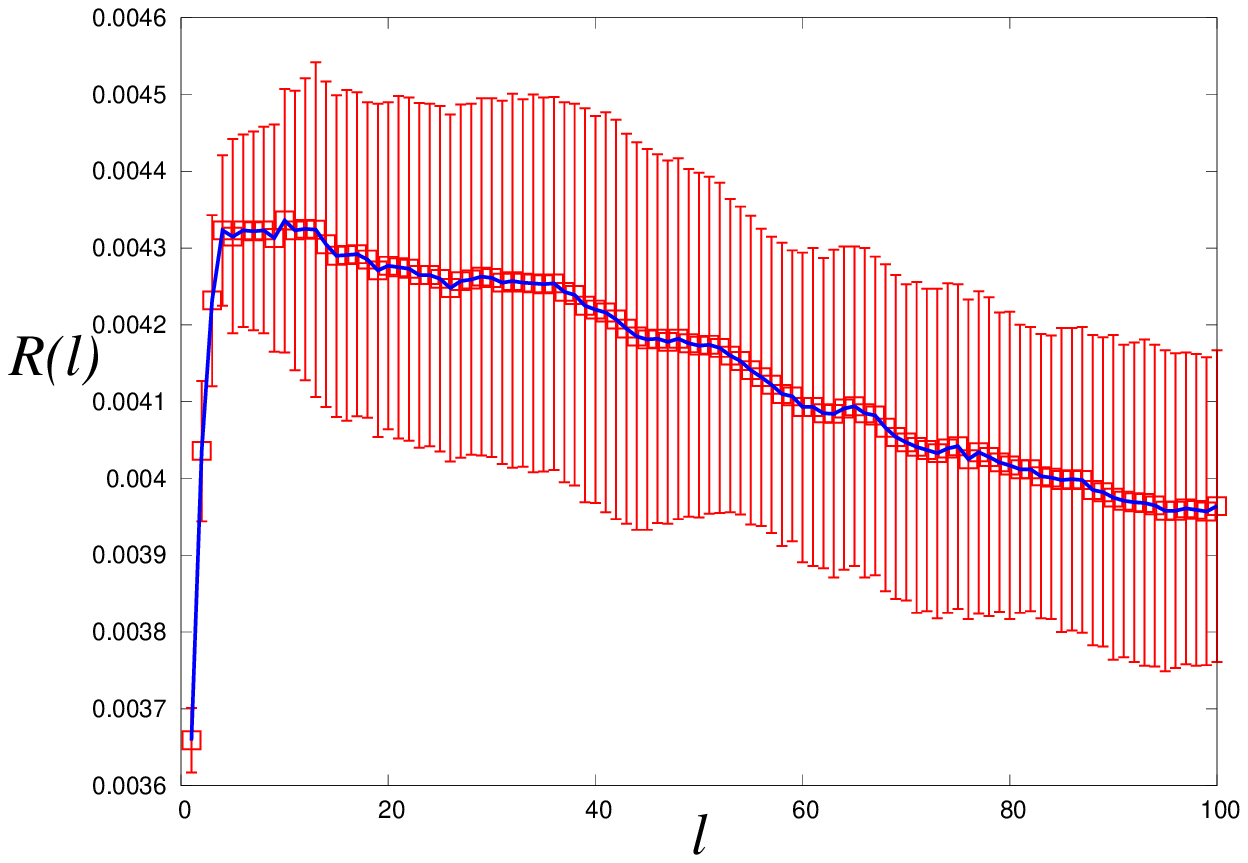} \\
\includegraphics[width=6cm]{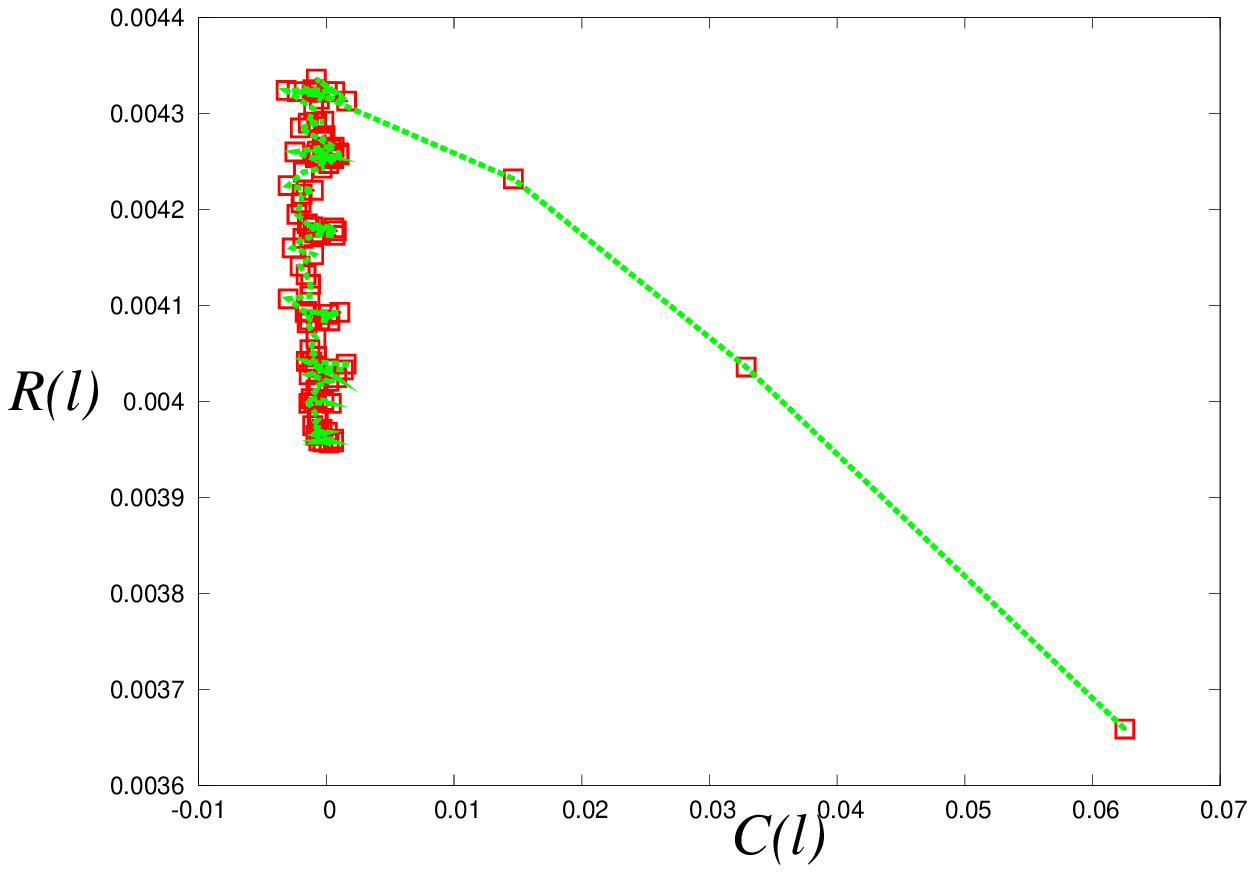} 
\includegraphics[width=6cm]{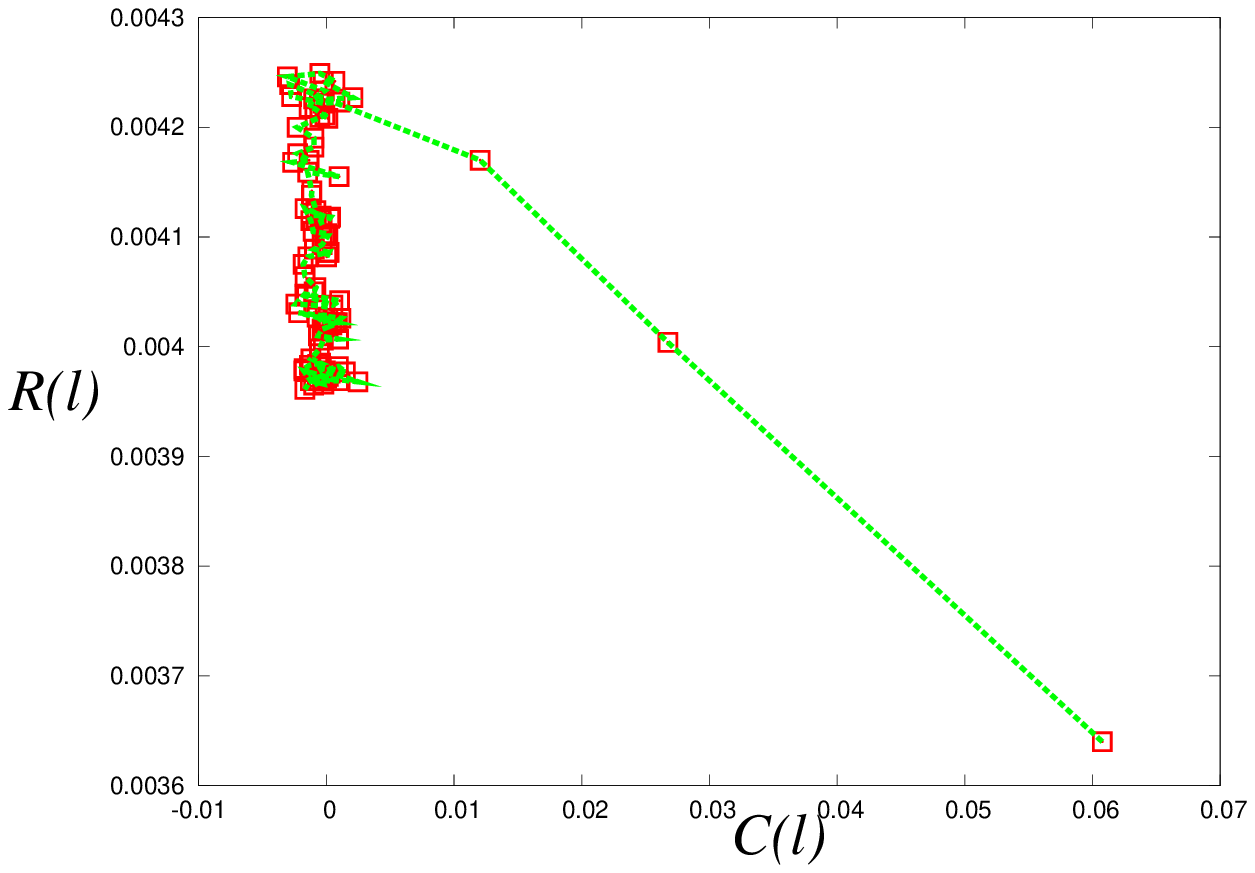}
\end{center}
\caption{\footnotesize 
Dynamical quantities $C(l)$ (upper left, middle left) and 
$R(l)$ (upper right, middle right) 
evaluated for the minority game with adaptive look-up tables 
for $M$ market history length.
We set  
$\psi=0$, $f_{2}=1$ (upper panels), $f_{2}=0.9$ (middle panels). 
The lower two panels show the relationship between 
the auto-correlation and response functions 
for $f_{2}=1$ (left) and $f_{2}=0.9$ (right).  
}
\label{fig:R_mgame_simu4}
\end{figure}

We next show the results for the macroscopic quantities 
in Fig. \ref{fig:R_mgame_simu4}. 
From this figure, we confirm that 
the `$\lambda$-shape' in $R(l)$-$C(l)$ 
scatter plots are much similar to 
the empirical evidence than the 
results in the previous subsection. 
\section{Discussion}
\label{sec:discuss}
Our general set-up presented in this paper is applicable to the analysis for the other quantities or 
for the other stochastic models.  Here we shall mention them briefly. 
\subsection{Waiting time statistics}
As we mentioned in section \ref{sec:Intro}, 
we can generate the duration between the price changes 
within the framework of our minority game. 

Let us introduce the maximum value of the spread 
$\bar{S}$ which is determined by market makers. Then, we might define 
a set of time points at which the price is updated as 
\begin{eqnarray}
\{t\} & \equiv & \{t_{k} | S_{t_{k}} =\min\{a_{it_{k}}|a_{it_{k}} \in N_{+}\}
-\max\{b_{it_{k}}|b_{it_{k}} \in N_{-}\} < \bar{S}\}
\end{eqnarray}
For these time points $t_{k}$, 
the duration between successive price changes is given by 
\begin{eqnarray}
\{\tau\} & \equiv & \{\tau_{k} | \tau_{k}=t_{k+1}-t_{k}\}
\end{eqnarray}
We can evaluate the distribution $P(\tau)$ and compare 
the results with well-known distributions, for instance, 
the Mittag-Leffler type \cite{Enrico,anowaiting,Enrico06}. 
\subsection{Mean-field models}
\begin{figure}[ht]
\begin{center}
\includegraphics[width=8cm]{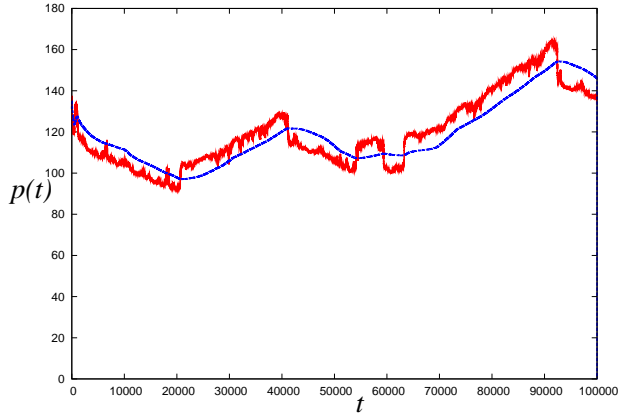}
\end{center}
\caption{\footnotesize 
Typical behaviour of time evolution of 
the price $p(t)$ (the solid line) and the moving average 
$\langle p(t) \rangle_{\tau}$ (the broken line) evaluated by 
(\ref{eq:sinha_s}) and (\ref{eq:sinha_p}). 
We set  $\mu=100,\tau=10000$.}
\label{fig:sinha}
\end{figure}

Recently, Vikram and Sinha \cite{Sinha} 
proposed a mean-field model to describe the collective behaviour of financial markets. 
In their model, each trader $i$ decides 
his (her) decision: 
$S_{i}(t)=+1$ (buy), 
$-1$ (sell) and $S_{i}(t)=0$ (no action) at time $t$ 
according to the following probability. 
\begin{eqnarray}
P[|S_{i}(t)|=1] = 1-P[S_{i}(t)=0] & = & 
{\exp}
\left(
-\mu 
{\Biggr |}
\log 
\frac{p (t)}{\langle p (t) \rangle_{\tau}}
{\Biggr |}
\right)
\label{eq:sinha_s}
\end{eqnarray}
where $p(t)$ stands for the price at time $t$ and 
the bracket $\langle \cdots \rangle_{\tau}$ means 
the moving average over the past $\tau$-time steps. 
The parameter $\mu$ is a parameter 
which controls the sensitivity of an agent to the magnitude 
of deviation of the price from its moving average $\langle p(t) \rangle$. 
For $\mu=0$, 
the system reduces to a binary decision model 
where every agent trades at 
all time instants. 
Namely, $P[S_{i}(t)=0]=0$ for $\mu=0$ means that 
the trader decides his (her) action, buy or sell, certainly. 
The traders who decide to trade at $t$ make his (her) action randomly, 
that is, $P[S_{i}(t)=1]=P[S_{i}=-1]=1/2$. 
The price at time $t+1$ is decided by the following 
recursion relation. 
\begin{eqnarray}
p (t+1) & = & 
\left(
\frac{1+A_{t}}{1-A_{t}}
\right) p (t)
\label{eq:sinha_p}
\end{eqnarray}
where $A_{t} \equiv (1/\sqrt{N})\sum_{i=1}^{N}S_{i}(t)$. 
In Fig.\ref{fig:sinha}, we plot the typical time-evolution of the price $p(t)$ and 
the moving average $\langle p(t) \rangle_{\tau}$ evaluated by 
(\ref{eq:sinha_s}) and (\ref{eq:sinha_p}). 

\mbox{}

\begin{figure}[ht]
\begin{center}
\includegraphics[width=4cm]{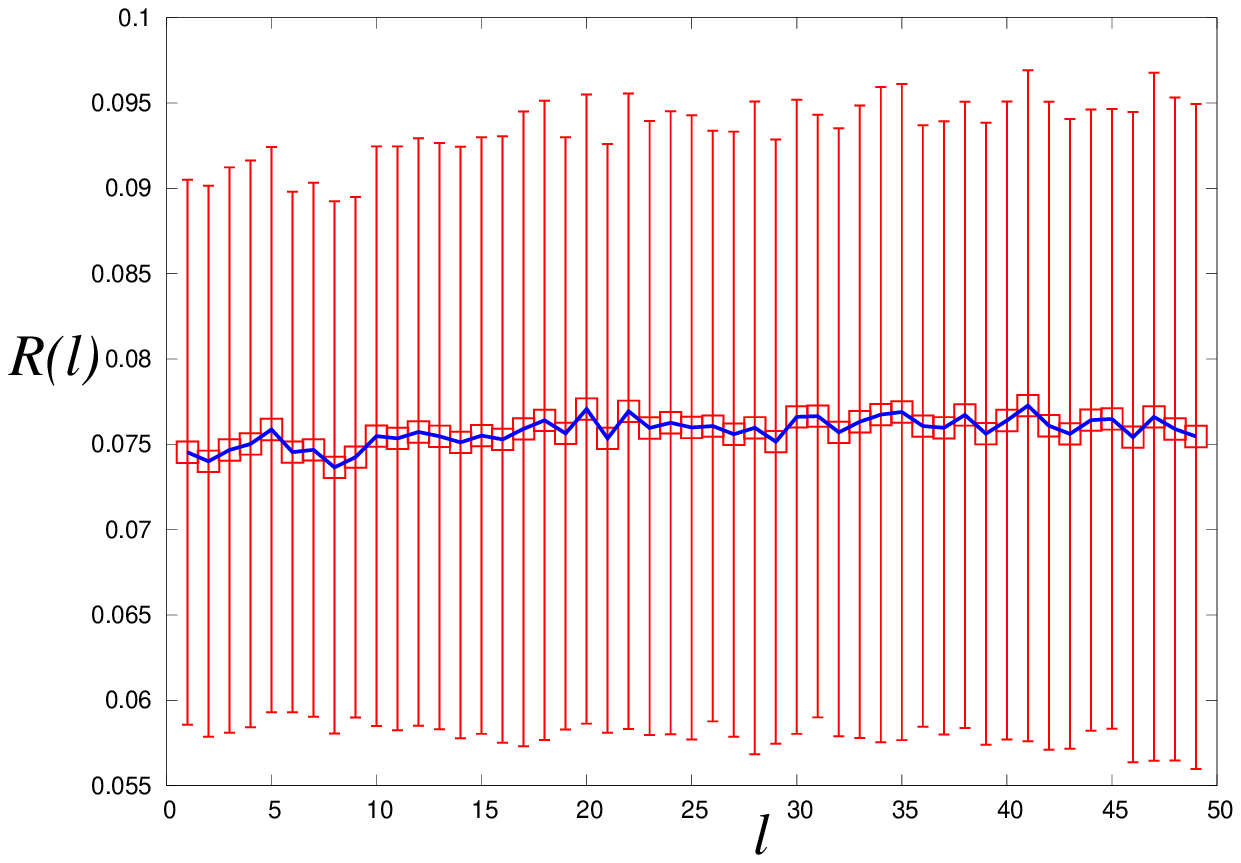}
\includegraphics[width=4cm]{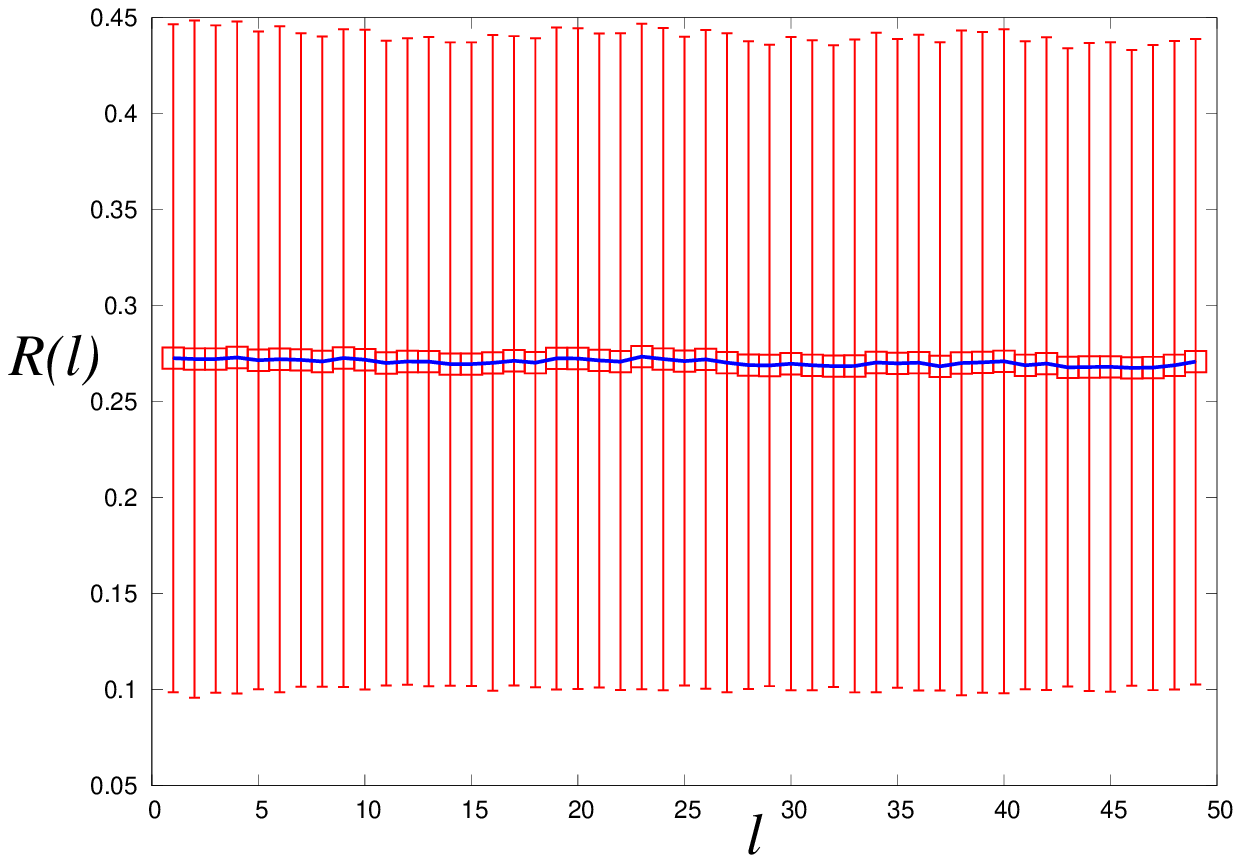}
\includegraphics[width=4cm]{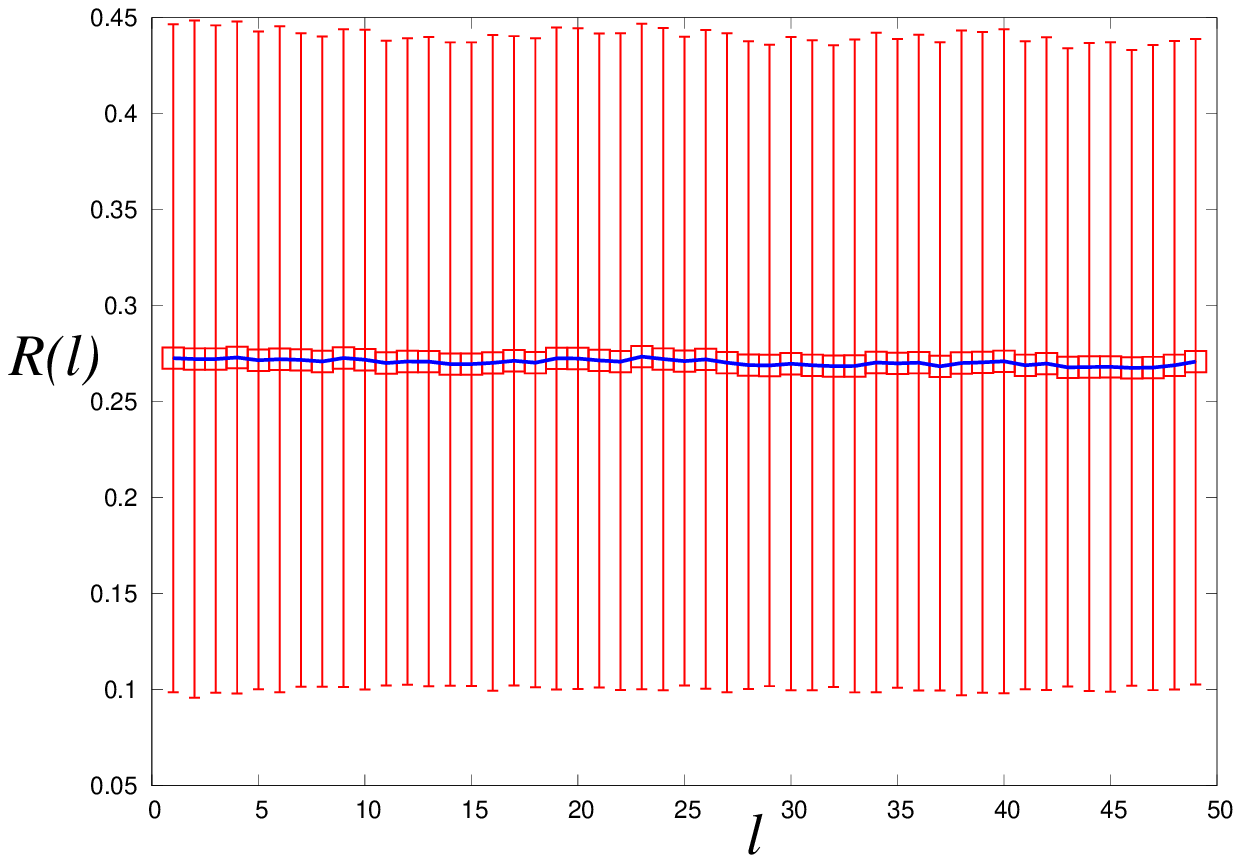}\\
\includegraphics[width=4cm]{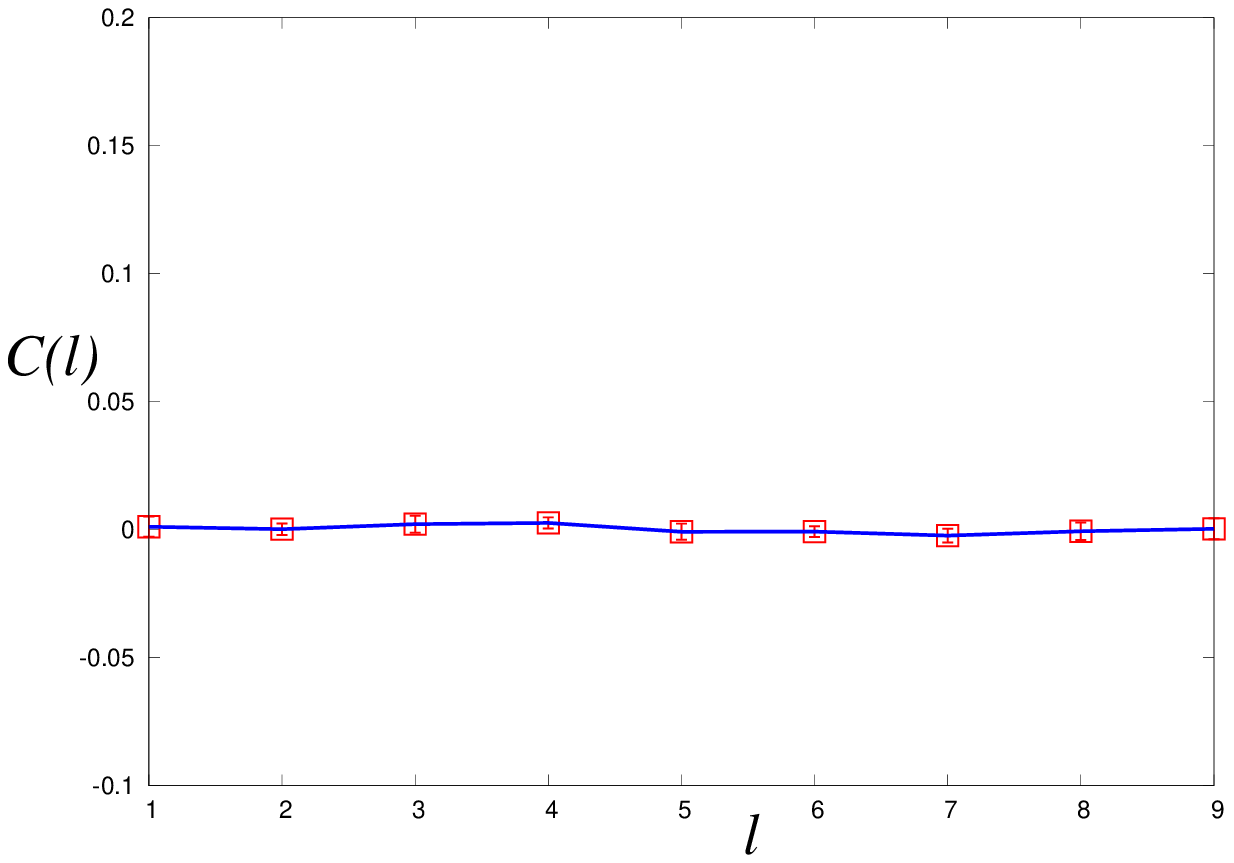}
\includegraphics[width=4cm]{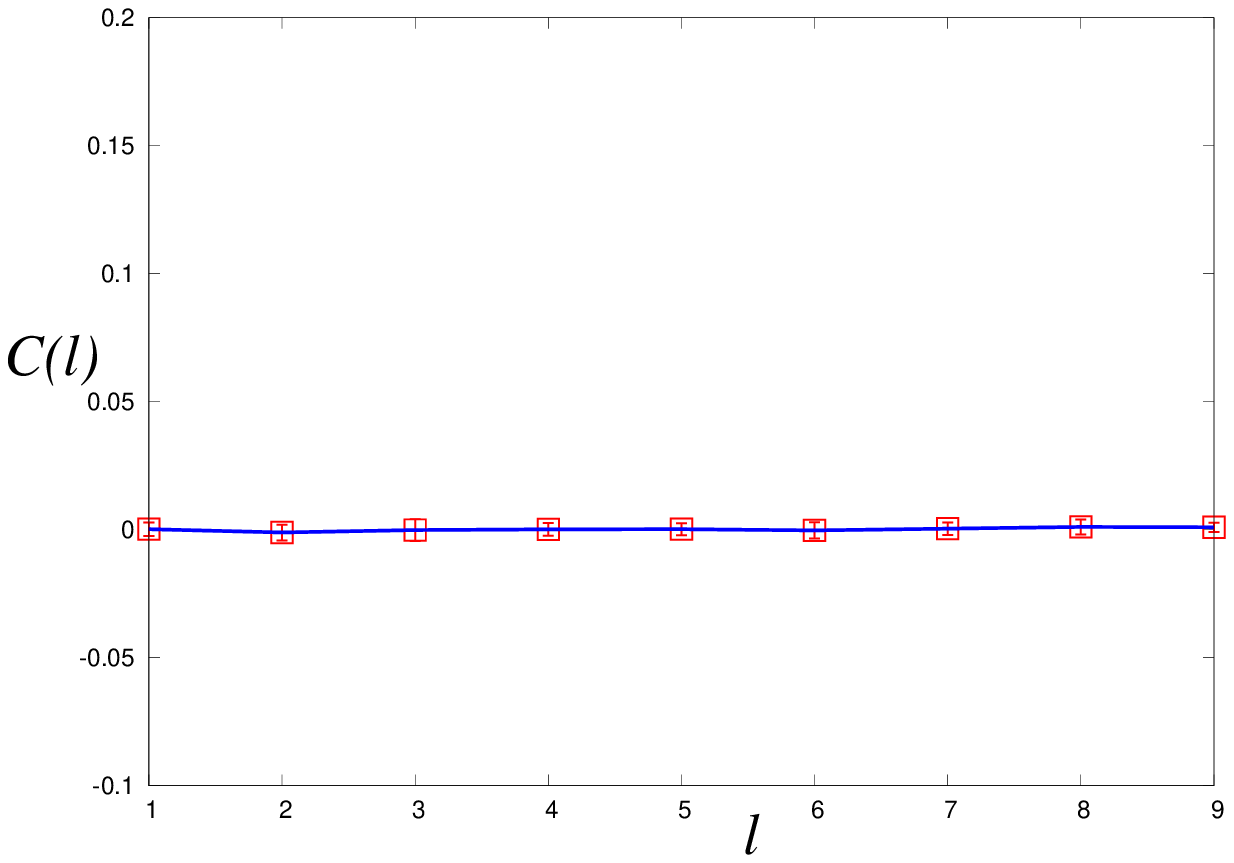}
\includegraphics[width=4cm]{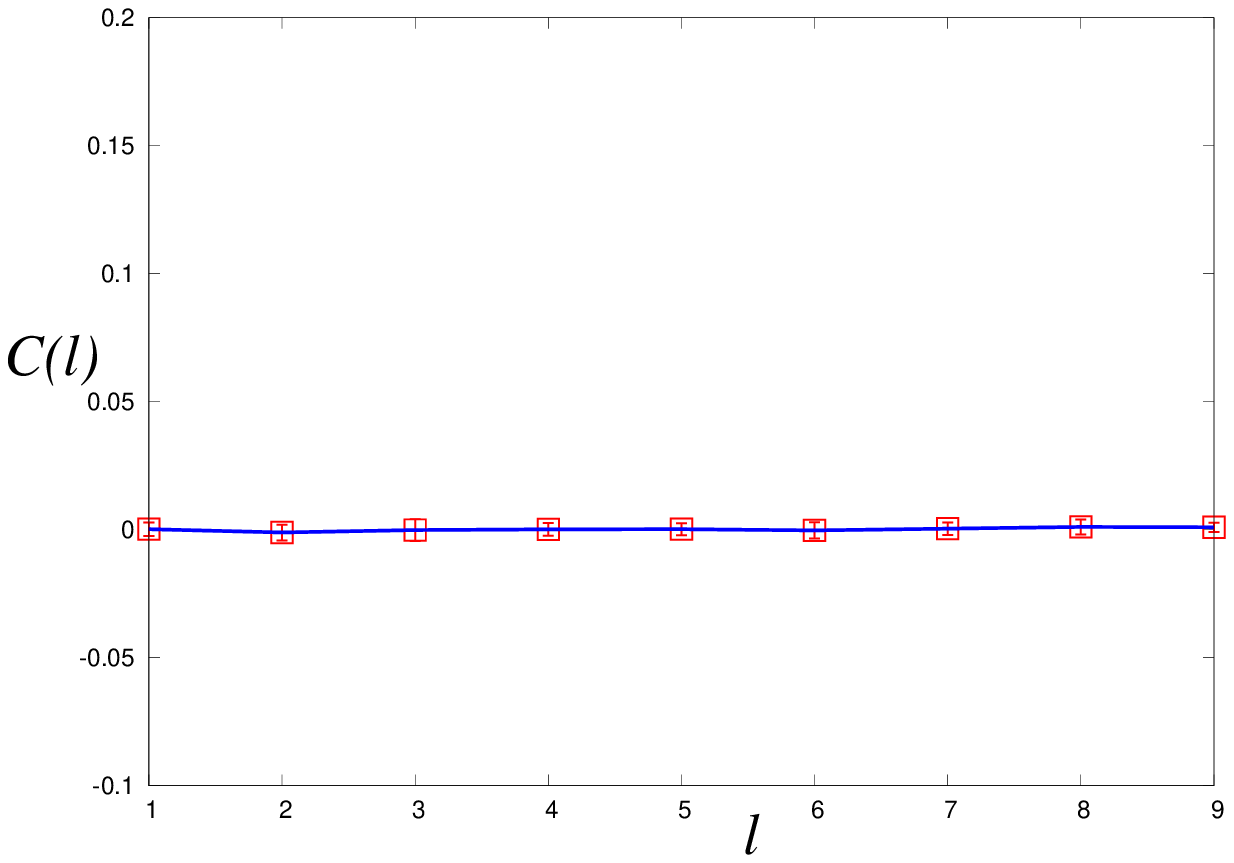}
\end{center}
\caption{\footnotesize 
Typical behaviour of the response function (upper panels) 
and the auto-correlation function (lower panels) 
for the Vikam-Sinha model. 
From the left to the right, we set 
$(\mu, \tau)=(100,10000),(10,10000),(100,1000)$. 
The error-bars were calculated by 10-independent trials.
}
\label{fig:sinha2}
\end{figure}

To construct the double-auction market for the above 
price change, we shall use 
the same definitions of buying $b_{it}$ and selling $a_{it}$ signals for each trader $i$ at time $t$ 
as (\ref{eq:ait})(\ref{eq:bit}) in our minority game modeling. 
Then, we calculate the response function and the 
auto-correlation function by using the above set-up and 
plot them in Fig. \ref{fig:sinha2}. 
We set the number of agents $N=20000$, 
the number of  iterations for the price change 
$T=100000$.  We also choose  
$\delta=0.02, \gamma_{a}=\gamma_{b}=0.001$. 
The initial condition on 
the price $p(t)$ is chosen from the Eur/JPY exchange rate in the empirical data. 
From this figure, we find that the behaviour is different from the empirical evidence. 
Especially, the auto-correlation function 
fluctuates with very large amplitudes. 
This result might come from the 
fact that in the Vikam-Sinha model, 
the buying-selling signal is chosen randomly.  
Therefore, we should modify 
the Vikram-Sinha model in order to explain the 
non-monotonic behaviour of the response function 
in double-auction markets with stochastic Bid-Ask spread. 
\section{Summary}
In this paper, statistical properties of double-auction markets with Bid-Ask 
spread were investigated 
through the response function. We first attempted 
to utilize 
the so-called {\it Madhavan-Richardson-Roomans 
model} (MRR for short) to simulate the stochastic process of 
the price-change in empirical data sets  (say, 
EUR/JPY or USD/JPY exchange rates) 
in which the Bid-Ask spread fluctuates in time.
We found that the MRR theory apparently does not  
simulate so much as the qualitative behaviour  (`non-monotonic' behaviour) 
of the response function calculated from the data.  
Especially, we were confirmed that 
the stochastic nature of the Bid-Ask spread causes 
apparent deviations from a linear relationship between the $R(l)$ and the 
auto-correlation function $C(l)$, namely, 
$R(l) \propto -C(l)$. 
To make the microscopic model of double-auction markets having stochastic Bid-Ask spread, 
we utilized the minority game with a finite market history length and 
found numerically that appropriate extension of the game shows quite similar 
behaviour of the response function to the empirical evidence. 
We also revealed that the minority game modeling with the 
 adaptive (`annealed') look-up table reproduces the 
 non-linear relationship $R(l) \propto -f(C(l))$ ($f(x)$ stands for a non-linear function 
 leading to `$\lambda$-shapes') 
more effectively than fixed (`quenched') look-up table does. 

Of course, there are still gaps between the theoretical prediction and 
the empirical evidence. 
We should modify our modeling of 
the double-auction market and 
figure out the micro-macro relationship in the market much more quantitatively.  
\section*{Acknowledgment}
We thank Enrico Scalas and Naoya Sazuka for useful comments on auction theory, especially 
for calling our attention to the reference \cite{Ohara}. 
We also thank Hikaru Hino for giving us useful information about the {\it MetaTrader4} \cite{Meta}. 
Our special thank goes to Fabrizio Lillo for 
valuable comments and his obliging advice on the double-auction markets. 
We would like to give an address 
of thanks to anonymous referees for 
critical reading of the manuscript and giving us 
useful comments.  
We were financially supported by 
Grant-in-Aid for Scientific Research (C) 
of Japan Society for 
the Promotion of Science, No. 22500195. 

\end{document}